\documentclass{tlp}

\usepackage{latexsym}
\usepackage{epsfig}

\newenvironment{des}{
\begin{list}
{$\bullet$}
{\topsep = 0 mm
\labelwidth = 3 mm
\labelsep = 2 mm
\parsep = 0 mm
\itemsep = \parskip 
\leftmargin = 5 mm}
}{\end{list}}

\def\blacksquare{\hbox{\vrule width 5pt height 5pt depth 0pt}}
\def\square{\hbox{\vrule\vbox{\hrule\phantom{o}\hrule}\vrule}}

\newtheorem{theorem}{Theorem}[section]
\newtheorem{lemma}[theorem]{Lemma}
\newtheorem{prop}[theorem]{Proposition}
\newtheorem{definition}[theorem]{Definition}

\newtheorem{exmp}[theorem]{Example}

\newtheorem{conj}[theorem]{Conjecture}

\newcommand{\reduce}{{\, \, \hookrightarrow_{{\cal P}} \, \,}}
\newcommand{\reducederi}{{\, \, \hookrightarrow^*_{{\cal P}} \, \,}}
\newcommand{\reducevar}{{\, \, \hookrightarrow_{{\it DVar}} \, \,}}
\newcommand{\reducevarderi}{{\, \, \hookrightarrow^*_{{\it DVar}} \, \,}}
\newcommand{\reduceap}{{\, \, \leadsto_{{\cal P}} \, \,}}
\newcommand{\tupla}{\bar{u}} 
\newcommand{\etupla}{\bar{e}} 
\newcommand{\eprimatupla}{\bar{e'}} 
\newcommand{\ttupla}{\bar{t}}
\newcommand{\tprimatupla}{\bar{t'}}
\newcommand{\sprimatupla}{\bar{s'}}  
\newcommand{\terminosprima}{{\it Term}_{\Sigma}(X)}

\newcommand{\terminospprima}{{\it Term}_{\Sigma_{\perp}}(X)}

\newcommand{\xtupla}{\bar{x}}
\newcommand{\stupla}{\bar{s}}
\newcommand{\calculo}{\mbox{LNCEC}}
\newcommand{\triangulo}{\triangleleft}
\newcommand{\menor}{\prec}
\newcommand{\ecuacional}{\sqsupseteq}
\newcommand{\aproximado}{\ecuacional_{\cal C}}

\newcommand{\tipos}{T_{{\it TC}}({\it TVar})}
\newcommand{\terminos}{{\it Term}_{\Sigma}({\it DVar})}
\newcommand{\expresiones}{{\it Expr}_{\Sigma}({\it DVar})}
\newcommand{\terminosp}{{\it Term}_{\Sigma_{\perp}}({\it DVar})}
\newcommand{\expresionesp}{{\it Expr}_{\Sigma_{\perp}}({\it DVar})}
\newcommand{\mi}{\{  \hspace*{-0.55ex} [\,}
\newcommand{\md}{\,] \hspace*{-0.55ex} \}}
\newcommand{\equivalente}{\approx_{\cal C}}
\newcommand{\deduce}{\vdash_{\Sigma_{\perp}}}
\newcommand{\programa}{{\cal P}=\langle \Sigma,{\cal C}, {\cal R} \rangle}
\newcommand{\gpc}{{\it GORC}}
\newcommand{\bpc}{{\it BRC}}
\newcommand{\algebra}{{\cal A}}
\newcommand{\relacion}{:^{\algebra}}
\newcommand{\orden}{\sqsubseteq^{\algebra}}
\newcommand{\botom}{\perp^{\algebra}}
\newcommand{\modterm}{{\cal M}_{{\cal P}}(V)}
\newcommand{\C}{[ \! [}  
\newcommand{\J}{] \! ]}
\newcommand{\igual}{==}
\newcommand{\lazo}{==}
\newcommand{\modelo}{\models}
\newcommand{\dominio}{{\it Term}_{\Sigma_{\perp}}(X)/_{\equivalente}}

\begin{document}

\title[Functional Logic Progr. with Algebraic Types]
  {A General Framework for Lazy Functional Logic Programming
       with Algebraic Polymorphic Types\thanks{Our research has been 
                partially supported by the Spanish National Project
                TIC98-0445-C03-02 ``TREND''  and the
                Esprit BRA Working Group EP-22457 ``CCLII''.}}

\author[Puri Arenas-S\'anchez and Mario Rodr\'{\i}guez-Artalejo]
        {P. Arenas-S\'anchez and M. Rodr\'{\i}guez-Artalejo \\
        Dpto. de Sistemas Inform\'{a}ticos y Programaci\'{o}n \\
        Universidad Complutense de Madrid \\
        Avenida Complutense s/n, 28040 Madrid, Spain \\
        \email{\{puri,mario\}@sip.ucm.es}}

\maketitle

{\small This paper is a revised and substantially 
                extended presentation of the results 
                from \cite{AR97a,AR97b}}

\medskip
\medskip

\begin{abstract}
We propose a general framework for first-order functional  logic  programming,  supporting 
lazy  functions,  non-determinism  and  polymorphic   datatypes   whose   data 
constructors obey a set ${\cal C}$ of equational axioms. 
On  top  of  a  given ${\cal  C}$\/, we specify a  program  as  a  set  ${\cal  R}$  of  
${\cal C}$-based  conditional rewriting rules for defined functions.  We  argue  that 
equational logic does not supply  the  proper  semantics  for  such  programs. 
Therefore, we present an alternative logic which  includes  ${\cal   C}$-based 
rewriting calculi and a notion of model. We get soundness and completeness for 
${\cal C}$-based rewriting w.r.t. models, existence of  free  models  for  all 
programs, and type preservation results. As operational semantics, we
develop a sound and  complete  procedure 
for goal solving, which is based on the combination  of  lazy  narrowing  with 
unification modulo ${\cal C}$. Our framework is quite expressive  for 
many purposes, as e.g. solving action and change problems, or realizing the 
GAMMA computation model. 

\medskip


\end{abstract}

\section{Introduction} 
\label{introduccion}

The interest in multiparadigm declarative programming has grown up during
the last decade, giving rise to different approaches to the integration
of functions into logic programming; see \cite{Han94b} for a good survey.
Declarative programming, in the wide sense, should have a firm foundation in logic. 
Therefore, we are especially interested in  approaches which provide a logical 
semantics for programs. Several early proposals, as e.g \cite{JLM84,GM87,Hol89},
focused on the idea of using {\em equational logic} as a basis for a 
semantically clean integration of functions and predicates. In  these
approaches, programs are built from {\em conditional rewrite rules}
(oriented conditional equations, see e.g.  \cite{DJ90}), and one obtains an 
analogon of the well-known least Herbrand model for 
pure logic programs  \cite{Apt90}, where the Herbrand universe is replaced by its 
quotient modulo the least congruence induced by an equational theory. 
Goals become systems of equations, and {\em narrowing} 
(a natural combination of rewriting and unification,  originally proposed as 
a theorem proving tool \cite{Fay79,Hul80}) 
can be used as a goal solving mechanism \cite{DO90}.

Unfortunately, equational logic has a drawback from the viewpoint of the 
semantics of {\em lazy functional languages}, such as Haskell  \cite{PH97,Bir98}.
These languages allow {\em non-strict} functions, that may return a  
result even if the values of some arguments are not known, or are known only 
partially. For instance, the function {\sf head} that returns the first element of 
a list,  does not need to know the rest of the list; and the function {\sf fst}  that 
returns the first component of an ordered pair, does not need to know at all
the value of the second component.  In a lazy functional language, expressions can 
sometimes denote infinite data structures (for instance, the list of all odd numbers),
and their values are computed gradually by means of a lazy reduction strategy 
(see  \cite{Bir98}) which delays the evaluation of function arguments until they are needed. 
In general, the identity between two expressions which have the same {\em infinite} 
value cannot be proved in equational logic. Consider, for instance, 
the following equations, which can be seen as a functional program defining 
the functions {\sf oddNumbers} and ${\sf oddNumbers}^*$ .  List constructors 
are written in Prolog notation.

\medskip

\( \begin{array}{llllllll}
\mbox{\sf oddNumbers}  \approx \mbox{\sf oddsFrom}(1)  \\
\mbox{\sf oddsFrom}(n)  \approx  {[n \mid \mbox{\sf oddsFrom}(n+2)]} 
\end{array} \)

\medskip

\( \begin{array}{llllllll}
\mbox{\sf oddNumbers}^*  \approx  \mbox{\sf oddsFrom}^*(1)  \\
\mbox{\sf oddsFrom}^*(n)  \approx  {[(2*n)-1 \mid   \mbox{\sf oddsFrom}^*(n+1)]}  
\end{array} \)

\medskip

According to the semantics of lazy functional languages, the two expressions 
{\sf oddNumbers} and ${\sf oddNumbers}^*$ have the same value, namely the infinite 
list of all odd positive integers. There is nothing unnatural in this. However, 
the equation {\sf oddNumbers} $\approx$ ${\sf oddNumbers}^*$ cannot be deduced 
in equational logic from the equations in the program. As a consequence, we cannot 
claim that the 
semantics of a program is characterized by  deducibility from the program, viewed 
as a theory in equational logic. In contrast to this, the semantics of pure 
logic programs can be characterized in terms of deducibility from the program
(a set of Horn clauses) in Horn logic, a very simple fragment of intuitionistic predicate logic. 
This claim is true both for the least Herbrand model semantics \cite{Apt90} 
as well as for the $\mathcal{C}$-semantics \cite{FLMP93}, which corresponds to the natural 
generalization of the least Herbrand model to the Herbrand universe consisting 
of open terms, with variables. To have a natural characterization of program 
semantics in terms of logical deducibility  realizes  the  ideal  of declarative 
programming, and helps to provide useful techniques for proving the semantic 
adequateness of program execution mechanisms.

Since identities between expressions with a common  infinite value can be 
unprovable in equational logic, using them in goals or conditions of conditional 
rewrite rules leads to some incompleteness results for conditional narrowing \cite{MH94}. 
In order to avoid this problem, the designers of the lazy 
functional+logic language K-LEAF \cite{GLMP91} proposed to distinguish between 
equalities $e \approx e'$ in the sense of equational logic (also known as 
{\em algebraic equalities}) and {\em strict equalities}, written as $e == e'$, 
intended to mean that expressions $e$ and $e'$ have a common value that is 
{\em finite} and {\em total}, in the sense of the theory of semantic domains used for 
the denotational semantics of programming languages \cite{Sco82,GS90}. 
Strict equality has been also adopted by other lazy functional logic languages, 
for instance BABEL \cite{MR92}. Typically, lazy functional logic languages with strict 
equality rely on a {\em constructor discipline}. Operation symbols are 
classified in two categories: {\em defined functions}, whose behaviour depends 
on the rewrite rules given in a program, and free {\em data constructors}, 
which are used to represent computed values. More precisely, {\em data terms} built 
from data constructors without any occurrence of defined function symbols,
always denote finite and total values. Moreover, different data terms always 
denote different values (this is why data constructors are called {\em free}).
Thanks to the constructor discipline,  strict equations $e == e'$ 
can be proved by reducing  both expressions $e$ and $e'$  to a common data term $t$.
In languages such as K-LEAF and BABEL, strict equality logically entails 
algebraic equality (but not vice versa). Regarding goal solving, it is known that
{\em lazy narrowing}  (a non-strict version of narrowing  originally proposed  in 
 \cite{Red85}) can provide a sound and complete operational semantics.

Unfortunately, data terms and strict equality do not fully characterize 
the semantics of a lazy language.
For expressions  such as {\sf oddNumbers} whose value is infinite,  there is
no data term $t$ that represents that value. However,  by introducing the 
special data constant $\bot$ which denotes the {\em undefined value},  it is
possible to build {\em partial data terms} $t$ which represent finite 
approximations of $e$'s value. For instance, the partial data term 
$[1, 3, 5, 7 ~|~ \bot]$ denotes a list formed by the first four odd numbers, 
followed by an undefined rest, which is a finite approximation of the value of 
{\sf oddNumbers}.  Now, imagine {\em approximation statements} of the form $e \rightarrow t$, 
intended to mean: ``$t$ denotes a finite approximation of $e$'s value''. Then, a 
logic with the ability to deduce such statements from a program could be used as logical 
framework for lazy functional logic programming. This approach has been recently 
developed in  \cite{GHLR96,GHLR99} under the name 
{\em \underline{C}onstructor-based \underline{R}e\underline{W}riting  \underline{L}ogic} 
(shortly, CRWL). In CRWL,  the semantic value of any expression $e$ can be 
characterized by the (possibly infinite) set of all approximation statements $e \rightarrow t$ 
that can be deduced from the program. Moreover,  strict equations $e == e'$ 
can be proved by proving two statements $e \rightarrow t$, $e' \rightarrow t$
for some common {\em total} data term $t$ (without occurrences of $\bot$).
In fact, CRWL does not enforce the restriction that such a $t$ must be unique.
For instance, if we assume constant constructors 0, 1 and 2, the following is 
allowed as a legal CRWL program:

\medskip

\( \begin{array}{llll}
{\sf coin} \rightarrow 0 & 
\ \ \ {\sf double}(x) \rightarrow {\sf plus}(x,x) & 
\ \ \ {\sf plus}(0,0) \rightarrow 0 & \ \ \ {\sf plus}(1,0) \rightarrow 1  \\
{\sf coin} \rightarrow 1 & &  \ \ \ {\sf plus}(0,1) \rightarrow 1 
& \ \ \ {\sf plus}(1,1) \rightarrow 2
\end{array} \)

\medskip

Given this program,  ${\sf coin} == {\sf coin}$ can be proved in two different 
ways, corresponding to the approximation statements ${\sf coin} \rightarrow 0$
and ${\sf coin} \rightarrow 1$. More generally, CRWL interprets $e == e'$ as a
{\em joinability statement}, meaning that $e$ and $e'$ admit some common total 
value, not necessarily unique. As illustrated by {\sf coin}, the rewrite rules in a CRWL 
program are not required to be confluent, and defined functions can be 
non-deterministic. The combination of non-deterministic functions and lazy 
evaluation turns out to be a very useful programming technique. However, 
in the presence of non-determinism,  neither joinability statement $e == e'$ nor
algebraic equations  $e \approx e'$ do entail that both expressions $e$ and 
$e'$ have the same semantic value. For instance, in CRWL the joinability
statement ${\sf coin} == 1$
can be deduced from the program above. Also, if we would view the rewrite rules 
of the program as equations, equational logic would allow us to deduce ${\sf coin} \approx 1$.
Since $1$ is not the denotation of {\sf coin} in the intended non-deterministic
semantics, we must use CRWL in place of equational logic, if we want to 
characterize the behaviour of programs.

From the various alternatives known for the semantics of non-determinism, CRWL 
has chosen {\em call-time choice} (see  \cite{Hus92,Hus93}), whose intuitive
meaning  is to fix a choice 
for the values of the arguments of a function,  before executing the function call.
This semantics does not force functions to be strict, because the values chosen 
for the argument expressions can be partial approximations (even $\bot$). 
Note that, according to this semantics, the possible total values of {\sf double}({\sf coin})
are 0 and 2, but not 1. For this reason,  CRWL uses lazy narrowing with sharing 
to obtain a sound and complete goal solving mechanism. Details can be found in 
 \cite{GHLR99}, along with  model theoretic semantics and a 
deeper motivation of the interest of CRWL as a framework for  declarative programming.

Extensions of CRWL dealing with modularity \cite{MP97} and higher-order 
programming \cite{GHR97} have been investigated.
The aim of the present paper is to extend CRWL in a different direction,
by introducing {\em algebraic polymorphic datatypes}.
The result will be a  more expressive framework ACRWL 
({\em \underline{A}lgebraic
\underline{C}onstructor-based \underline{R}e\underline{W}riting  \underline{L}ogic} ). 
More precisely, ACRWL will include user-defined polymorphic datatypes similar
to those used in modern functional languages such as Haskell
(see e.g. \cite{PH97}), but with a novel point: The data
constructors will be not necessarily free\footnote{Note that user-defined datatypes are
also called ``algebraic'' in Haskell. In spite of this terminology,
Haskell's data constructors are free.};
instead, we will allow
to specify a set ${\cal C}$ of equational axioms to control the 
constructors' behaviour. For instance, in our framework we can define a
datatype for polymorphic sets as follows:

\begin{center}

\( \begin{array}{lllll}
{\bf datatypes} &  {\bf constructors} & {\bf equations}\\
\ \ \ \ {\it Set}(\alpha) &  \ \ \ \ \{\ \}: \rightarrow {\it Set}(\alpha) & \ \ 
\ \ \{x \mid \{y \mid {\it zs}\}\}  \approx  \{y \mid \{x \mid {\it zs}\}\} \\
                     &  \ \ \ \ \{\cdot \mid \cdot\} : (\alpha,{\it Set}(\alpha)) 
\rightarrow {\it Set}(\alpha) 
                     & \ \ \ \ \{x \mid \{x\mid{\it zs}\}\}   \approx  \{x\mid{\it 
                     zs}\} 
\end{array} \)

\end{center}

\noindent
where the set constructors $\{\ \}$ (to build an empty set)
and $\{\cdot \mid \cdot\}$ (to add an element to a set) are controlled
by the two given equations. By omitting the second one, we can
obtain a data type for polymorphic multisets.

Data constructors with associated equations will be called {\em equational} 
or {\em algebraic} in the rest of this paper.
Algebraic data constructors play an important role in several recent proposals
for extended logic programming and multiparadigm declarative
programming; see 
e.g. \cite{JP89, Jay92, Leg94, DOPR91, DR93, DOPR96, DPR96, Mes92, Mes93,
CDE+99, KKV95, BKK+96, DFI+98, Llo99, MM95, HS90, GHSST92}. 
Some of these works do not consider functions, or lazy evaluation, while some 
others only allow some particular algebraic data constructors (most often sets 
and/or multisets). In a  higher-order language, sets and multisets can be
represented as functions, rather than using data constructors.
The advantages of each representation are problem dependent; see e.g. \cite{Llo99}.

We are also aware of some related work on functional logic languages with {\em free} polymorphic types \cite{Han90b,AGG96,AG97},  where the model theoretic semantics is more complex 
than the one we will develop, and algebraic data constructors are not considered. 
But, as far as we know, declarative programming with lazy functions 
and a general notion of algebraic polymorphic datatype, has not been 
investigated previously.  We view a program as a
set of ${\cal C}$-based conditional rewrite rules to define
the behaviour of lazy functions on top of a given set
${\cal C}$ of equational axioms for data constructors. Both constructors
and defined functions have polymorphic principal types. As
in CRWL  \cite{GHLR99}, defined functions are non-strict and possibly non-deterministic.  
For instance, a non-deterministic  function
which selects an arbitrary element from a non-empty set can be
defined by a single rewrite rule:

\vspace*{-0.10cm}

\[ \begin{array}{llll}
{\sf select} : {\it Set}(\alpha) \rightarrow \alpha \\
\ \ \ {\it select}(\{x\mid{\it xs}\}) \rightarrow x
\end{array} \]

\vspace*{-0.10cm}
 
\noindent
Now, due to the equational axioms for the set constructor, a goal such as 
${\it select}(\{a,$ $b,c\}) \lazo x$, where $x$ is a 
variable and $a,b,c$ are pairwise distinct constants, 
has three possible answers, namely  $x=a$, $x=b$ and $x=c$. 
A Prolog-like sequential implementation would  be expected to deliver the three answers one 
after the other, by using a backtracking mechanism.

We present  declarative and operational semantics for ACRWL programs. 
With respect to the declarative semantics, we have followed the lines of CRWL  \cite{GHLR99}, 
but with two major modifications. Firstly, our
models are algebras with two carriers (for data
and types, respectively), inspired by the polymorphically order-sorted
algebras from \cite{Smo89}. Secondly, the constructor-based
rewriting calculi from \cite{GHLR99} have been modified to incorporate
a set ${\cal C}$ of equational axioms for constructors while respecting 
the intended behaviour of lazy evaluation. To achieve this aim, we give
an {\em inequational calculus} which interprets each equational axiom
in ${\cal C}$ as a scheme for generating inequalities between {\em partial
data terms} (built from constructors and a bottom symbol $\perp$\/). For
instance, the equation $\{x \mid\{x\mid{\it zs}\}\} \approx \{x\mid{\it zs}\}$
 for sets will be
regarded as a scheme to generate all the inequalities $\{s\mid\{s\mid{\it r}\}\}  
\ecuacional  \{s\mid{\it r}\}$ and
 $\{t\mid{\it r}\}  \ecuacional  \{t \mid \{t\mid{\it r}\}\}$,
where $t$\/, ${\it r}$ are {\em partial} data terms, and $s$ is a {\em total}
data term (without occurrences of $\perp$\/). Inequalities are
thought of as defining an approximation ordering. The need to
deal with equations from ${\cal C}$ in this special way will be
justified in Sections \ref{terminos} and \ref{calculos}.

Regarding the operational semantics, we provide a lazy narrowing calculus, 
na\-med $\calculo$, for goal solving. In contrast to the narrowing calculus 
from  \cite{GHLR99} and other related approaches based on free data 
constructors, $\calculo$ must work modulo the equational axioms ${\cal C}$ which
control the algebraic data constructors in ACRWL programs. 
In fact, we have borrowed ideas from 
several previous works, such as \cite{GHLR99,techanus,JK91,Soc94}.
 The main novelty w.r.t. \cite{GHLR99,techanus} is the treatment of {\em algebraic polymorphic types}. 
Our lazy narrowing calculus   provides {\em mutation} rules
(in the line of \cite{JK91})
for applying equational axioms in ${\cal C}$\/.

Our goal solving calculus $\calculo$ is presented as a system of goal 
transformations. Thanks to the combination of lazy narrowing and 
${\cal C}$-based mutations, it can cope with infinite data structures and 
algebraic constructors simultaneously.
For instance, assume that we extend the little program above by adding
the datatype {\it Nat}, the constructors
${\it Zero} : \rightarrow {\it Nat}$ and ${\it Suc}: {\it Nat} \rightarrow 
{\it Nat}$\/, and the defining rule: 

 \vspace*{-0.10cm}

\[ \begin{array}{llll}
{\sf gen\_set\_nat}: {\it Nat} \rightarrow {\it Set}({\it Nat}) \\
\ \ \ {\it gen\_set\_nat}(n)
\rightarrow \{n\mid{\it gen\_set\_nat}({\it Suc}(n))\}
\end{array} \]

\vspace*{-0.10cm}

\noindent
Then,  the goal
${\it Suc}({\it Suc}($ ${\it Suc}({\it Zero})))
\lazo {\it select}({\it gen\_set\_nat}({\it Zero}))$ 
can be solved by $\calculo$. 
More generally,  we can prove 
soundness and completeness of $\calculo$ w.r.t. the declarative 
semantics.  Exactly as in the case of logic programming, 
the completeness result does not imply the absence of failing
computations and/or infinite computations in the search space.
The completeness proof splits the 
goal solving process in two phases, like in \cite{techanus}.
The first phase allows to transform a goal
into a quasi-solved goal only containing variables, whereas
the second phase transforms a quasi-solved goal into a solved goal
representing a computed answer for the initial goal.

Our theoretical results show that the ACRWL framework
provides a firm basis for a very expressive combination of  declarative programming 
features. Nevertheless, we are aware of the difficulty of filling the gap 
between the present theory and an efficiently implemented programming language.
In particular, the $\calculo$ calculus is far from being adequate as a 
description of a directly implementable computation strategy; its shortcomings 
will be discussed in sections \ref{goalsolving} and \ref{completeness} below.
In the absence of algebraic constructors,  the {\em needed narrowing} strategy 
\cite{AEH94,LLR93} can be used to alleviate the problem; see the discussion in 
 \cite{GHLR99}, section 8. As far as we know, no 
analogon of needed narrowing is available for rewrite systems based on algebraic 
constructors. In spite of this, we believe that there is hope of obtaining 
reasonably efficient implementations for some particular instances of ACRWL.
Especially, we have in mind the case of multisets, which (combined with other 
free data types) can be used for  many interesting applications, including {\em 
action and change problems}  \cite{MM95} and the GAMMA programming model 
\cite{BM90,BM93}.  A first proposal for implementing ACRWL, restricted to 
multisets and free data types, can be found in \cite{ALR98}\footnote{In fact, the 
language whose implementation is discussed in \cite{ALR98} has also some 
constraint solving capabilities.}.

The rest of the paper is organized as follows:
Section \ref{terminos} sets the basic formalism, defining
polymorphic signatures, expressions and equational axioms for data 
constructors, along with the calculus needed to deduce approximation inequalities from
them. In Section \ref{programas} we present ACRWL programs, given by ${\cal C}$-based
rewrite rules for defining lazy functions on top of a given set
 ${\cal C}$ of equational axioms. Some simple programming examples, dealing 
 with action and change problems and the GAMMA computation model, are included  
 here. The behaviour of ACRWL as a logic is given by rewriting calculi, which 
 are presented in Section \ref{calculos}, along with some type preservation 
 results. Section \ref{modelodeterminos} deals
with model theory, showing the existence of free models for ACRWL programs,
as well as soundness and completeness results for the rewriting calculi w.r.t.
models. Section \ref{goalsolving} presents
our goal solving calculus $\calculo$, whose main properties (namely, soundness, 
completeness and type preservation) are proved in 
Section \ref{completeness}.
Finally, some topics for future research are pointed in the concluding Section \ref{conclusiones}. 
In order to improve readability, many proofs have been moved to an Appendix.

\section{Signatures, Types, Expressions and Equations}
\label{terminos}

We assume a countable set ${\it TVar}$ of {\em type variables} $\alpha, 
\beta$\/, etc, and a countable ranked alphabet ${\it TC}=\bigcup_{n \geq 0} {\it TC}^n$ 
of {\em type constructors} $K,K'$, etc. {\em Polymorphic types} 
$\tau, \tau', \ldots \in \tipos$ are  built as
$\tau ::= \alpha \mid K(\tau_1, \ldots,\tau_n)$\/, where 
$\alpha \in {\it TVar}$, $K \in  {\it TC}^n$, $\tau_{i} \in \tipos$, $1 \leq i \leq n$\/.
The set of type variables occurring in $\tau$ is written $tvar(\tau)$.

We define a {\em polymorphic signature}
$\Sigma$ over ${\it TC}$ as a triple $\langle {\it TC},{\it DC},{\it FS} 
\rangle$\/, where ${\it DC}$ is a set of type declarations for {\em  data  constructors}, 
of the form $c:(\tau_1, \ldots, \tau_n) \rightarrow \tau$\/ with
$\bigcup_{i=1}^n  tvar(\tau_i)  \subseteq  tvar(\tau)$\/   (so-called   {\em 
transparency} property), and 
${\it FS}$ is a set of type declarations for {\em 
defined function symbols}, of the form
$f:(\tau_1, \ldots, \tau_n) \rightarrow \tau$\/.
In the following, we will say that 
$h: (\tau_1, \ldots, \tau_n) \rightarrow \tau
\in {\it DC} \cup {\it FS}$ is a {\em transparent} type declaration
iff $\bigcup_{i=1}^n tvar(\tau_i) \subseteq tvar(\tau)$.

We  require  that  $\Sigma$ 
does not include multiple type declarations for the  same  symbol.  The  types 
given by declarations in ${\it DC} \cup {\it FS}$ are  called  {\em  principal 
types}. We will write $h \in {\it DC}^n \cup {\it FS}^n$ to indicate the arity 
of a symbol according to its type declaration. In   the    following,    ${\it 
DC}_{\perp}$ will denote ${\it DC}$ extended  by  a  new  declaration  $\perp: 
\rightarrow \alpha$\/. The bottom constant constructor $\perp$ is intended  to 
represent an undefined value. Analogously, $\Sigma_{\perp}$  will  denote  the 
result of replacing ${\it DC}$ by ${\it DC}_{\perp}$ in $\Sigma$\/.

Assuming another countable set ${\it DVar}$ of {\em data variables} $x,y, $\/ 
etc, we build {\em total expressions} $e,r, \ldots \in \expresiones$ as
$e ::= x \mid h(e_1, \ldots, e_n)$\/, where $x \in {\it DVar}$, 
$h \in {\it DC}^n \cup {\it FS}^n$, $e_{i} \in \expresiones$, $1 \leq 
i \leq n$.\/ The set $\expresionesp$\/ of {\em partial expressions} is defined in the  same 
way, but using ${\it DC}_{\perp}$ in place of ${\it DC}$. {\em Total 
data terms} 
$\terminos \subseteq \expresiones$\/ and {\em partial data terms} 
$\terminosp \subseteq \expresionesp$\/ are built by using data variables  and  data 
constructors only. In the sequel, we reserve $t, s$, to denote possibly 
partial data terms, and we write $dvar(e)$ for the set of all data variables occurring 
in an expression $e$.

We define {\em type substitutions} $\sigma_{t} \in {\it TSub}$\/ as mappings from 
${\it TVar}$ to $\tipos$, and possibly {\em partial data 
substitutions} $\sigma_{d} \in 
{\it DSub}_{\perp}$\/ as mappings from ${\it DVar}$ to $\terminosp$. {\em Total 
data substitutions} $\sigma_{d} \in {\it DSub}$\/ are mappings from ${\it DVar}$ 
to $\terminos$. Pairs $\sigma=(\sigma_{t},\sigma_{d})$, with $\sigma_{t} \in TSub$ and $\sigma_{d} \in 
DSub_{\perp}$\/ are called {\em substitutions}. We will use  postfix  notation 
for the result of applying substitutions to types and expressions. We will say 
that $\sigma_{d} \in {\it DSub}_{\perp}$\/ is {\em  safe} for a data term $t$ if 
$\sigma_{d}(x)$ is a total term  for  every  variable  $x$  having  more  than  one 
occurrence in $t$. The notions of {\em  instance},  {\em  renaming}  and  {\em 
variant} have the usual definitions; see e.g. \cite{DJ90,Apt90}.

In the sequel,  given $A \subseteq {\it TVar}$
(respect. $X \subseteq {\it DVar}$) and
$\sigma_{t}, \sigma_{t}' \in {\it TSub}$ (respect.
$\sigma_{d}, \sigma_{d}' \in {\it DSub}_{\perp}$),
the notation $\sigma_{t} = \sigma_{t}'[A]$ (respect. $\sigma_{d}
=\sigma_{d}'[X]$) means that $\alpha\sigma_{t}
= \alpha\sigma_{t}'$ (respect.
$x\sigma_{d} = x\sigma_{d}'$), for
all $\alpha \in A$ (respect. for all $x \in X$). Similarly, the 
notation $\sigma_{t} = \sigma'_{t}[\backslash A]$ (respect. 
$\sigma_{d} =\sigma_{d}'[\backslash X]$)
 means that  $\sigma_{t}=\sigma'_{t}[{\it TVar}-A]$ (respect.
$\sigma_{d}=\sigma'_{d}[{\it DVar}-X]$)\/.

An {\em environment}  is  defined  as  any  set  $V$  of  type-annotated  
data 
variables $x:\tau$\/, such that $V$ does not include two different annotations 
for  the   same   variable.   The   set   of   {\em   well-typed   expressions} 
w.r.t. an environment $V$ is defined as ${\it Expr}_{\Sigma_{\perp}}(V) = 
\bigcup_{\tau \in \tipos} {\it Expr}^{\tau}_{\Sigma_{\perp}}(V)$, where 
$e \in {\it Expr}^{\tau}_{\Sigma_{\perp}}(V)$\/ holds iff the type judgment 
$V \deduce e:\tau$ is derivable by  means  of  the  following  type  inference 
rules:

\begin{des}

\item $V \deduce x:\tau$ if $x:\tau \in V$\/;
\item $V \deduce h(e_1, \ldots ,e_n):\tau$ if
    $V \deduce e_i:\tau_i$,  $1 \leq i \leq n$, 
where $h: (\tau_1, \ldots ,\tau_n) \rightarrow \tau$
is an instance of the unique declared principal type associated to
$h$ in ${\it DC}_{\perp} \cup {\it FS}$\/.

\end{des}

\noindent 
${\it Expr}^{\tau}_{\Sigma_{\perp}}(V)$\/ has subsets
${\it Expr}^{\tau}_{\Sigma}(V), 
{\it Term}^{\tau}_{\Sigma_{\perp}}(V), 
{\it Term}^{\tau}_{\Sigma}(V)$\/
that are defined in the natural way.

Note that, given any 
environment $V$\/, it holds that $V \deduce \perp: \tau$\/, for all $\tau \in 
\tipos$\/, but the type of an expression $e$ containing $\perp$ 
depends clearly on the   expression $e$. As an example, consider the 
following type declarations for data constructors:

\vspace*{-0.10cm}

\[ [\ ]: \rightarrow {\it  List}(\alpha),  \ [\cdot \mid \cdot ] : (\alpha, {\it List}(\alpha)) 
\rightarrow {\it List}(\alpha) \mbox{ and } {\it Zero}: \rightarrow {\it 
Nat} \]

\vspace*{-0.10cm}

\noindent
Then, given the environment $V = \{x:{\it Nat}\}$, it holds that $V 
\deduce [\perp]: {\it List}(\tau)$, for any $\tau \in \tipos$, 
$V \deduce [{\it Zero},\perp]: {\it List}({\it Nat})$ and $V \deduce 
[x \mid \perp]: {\it List(Nat)}$.

Remark that using well-known 
techniques \cite{Mil78,DM82}, it is easy to prove that every  
well-typed expression has a  most  general  principal  type,  
which  is  unique  up  to renaming.

The following definitions introduce equational axioms for data 
constructors.

\begin{definition}[Equational axiom]
An equational axiom is any logical statement of the form
$s \approx t$\/, where $s$ and $t$ are total data terms (i.e. $s,t 
\in \terminos$). An equational axiom $s \approx t$ is called:

\begin{des}
\item {\em regular} iff $dvar(s) = dvar(t)$\/;
\item {\em non-collapsing} iff neither $s$ nor $t$ is a variable;
\item {\em strongly regular} iff  it is regular and non-collapsing.
\end{des}

A finite set ${\cal C}$ of equational axioms is called
(strongly) regular iff  every 
axiom in ${\cal C}$ is (strongly) regular. \hfill \square

\end{definition}

Notice that a strongly regular equational axiom has the form
$c(t_1, \ldots, t_n) \approx d(s_1, \ldots, s_m)$\/, where
$\bigcup_{i=1}^n dvar(t_i)= \bigcup_{j=1}^m dvar(s_j)$\/,
whereas a collapsing regular equational axiom has
the structure $c(t_1, \ldots,$ $ t_n) \approx x$ or
$x \approx c(t_{1}, \ldots, t_{n})$\/, where
$\bigcup_{i=1}^n dvar(t_i)= \{x\}$\/. In the sequel, when we speak of 
an equation $s \approx t$\/, we mean (by an abuse of language) $s 
\approx t$ or $t \approx s$\/.
By convention, we assume that no equational axiom in ${\cal C}$
is a trivial identity $t \approx t$\/.

In the rest of the paper we focus on strongly regular
equations, because strong regularity is needed for our current
type preservation results; see Theorem \ref{conservacion} 
and Example \ref{r3} in Section \ref{calculos} below.

\begin{definition}[Well-typed strongly regular equation]
\label{ecuacion}
We say that a strongly regular equation
$c(t_1, \ldots, t_n) \approx d(s_1, \ldots, s_m)$\/ is {\em well-typed}
iff the principal type declarations for $c, d$\/ have variants
$c: (\tau_1, \ldots, \tau_n)  \rightarrow\tau$\/  and 
$d:  (\tau'_1, \ldots,\tau'_m)  \rightarrow  \tau$\/ such that 
$c(t_1, \ldots, t_n), d(s_1, \ldots, s_m) \in 
{\it Term}_{\Sigma}^{\tau}(V)$\/, for some environment $V$. 

A set ${\cal C}$\/ of strongly regular  axioms  is 
called {\em well-typed} iff each axiom in ${\cal C}$\/ is well-typed.
\hfill \square

\end{definition}

Since principal types of data constructors are transparent, 
the above definition implies that $t_i \in {\it 
Term}_{\Sigma}^{\tau_i}(V)$\/,
$1 \leq i \leq n$\/, and
$s_j \in {\it Term}_{\Sigma}^{\tau'_j}(V)$\/, $1 \leq j \leq m$\/.
In the following, we will  say that two data constructors
$c,d$ are {\em constructors of the same datatype} iff the declared principal types
for $c$ and $d$  admit variants $c: (\tau_1, \ldots, \tau_n) \rightarrow
\tau$ and $d: (\tau'_1, \ldots ,\tau'_m) \rightarrow \tau$\/,
respectively.

The following example  presents   different algebraic datatypes  and 
illustrates the expressiveness of strongly regular equations.
All equational axioms in the example 
are either strongly regular or collapsing and regular.  Furthermore,
all those being strongly regular  are well-typed in the sense of 
Definition \ref{ecuacion}.

\begin{exmp}[Equational axioms]
\label{tiposecuaciones}  

\begin{des}

\item[(1)]  Suppose that $\Sigma$ includes the following
            declarations:

\medskip

\noindent
\( \begin{array}{lllllll}
\textbf{datatypes}  \\
\ \ {\it Set}/1, {\it Mset}/1,  {\it Nat}/0  
\end{array} \)

\medskip

\noindent
\( \begin{array}{lllllll}
 \textbf{constructors} \\
\ \ \ {\it Zero}: \rightarrow {\it Nat} & {\it Suc}: {\it Nat} \rightarrow 
{\it Nat} \\
 \ \ \  \{ \ \}: \rightarrow {\it Set}(\alpha)  &
            \{\cdot \mid \cdot \}:(\alpha,{\it Set}(\alpha)) 
            \rightarrow {\it Set}(\alpha) \\
 \ \ \ \mi \ \md: \rightarrow {\it Mset}(\alpha) &
            \mi \cdot \mid \cdot \md: (\alpha,{\it Mset}(\alpha)) 
            \rightarrow {\it Mset}(\alpha) 
\end{array} \)

\medskip

\noindent
Then, the following equational axioms for the set   ($\{ \cdot 
\mid \cdot \}$) and multiset ($\mi \cdot \mid \cdot \md$)
constructor

 \medskip

\( \begin{array}{lllllllll}
\textbf{equations} \\
\ \ \ \{x \mid \{y \mid{\it zs}\}\}  \approx  \{y \mid\{x\mid{\it 
zs}\}\}  & \ \ \ \  \mi x \mid 
\mi y \mid {\it zs} \md \md \approx  \mi y \mid 
\mi x \mid {\it zs} \md \md  \\
\ \ \ \{x\mid\{x\mid{\it zs}\}\}   \approx  \{x\mid{\it zs}\}
\end{array} \)

\medskip

\noindent
are strongly regular.

\item[(2)] Suppose now that $\Sigma$ contains the  datatypes  
 ${\it USet}/1$ and ${\it UMset}/1$, together with the following data 
 constructor type declarations:

 \medskip

\noindent
\textbf{constructors} 

\smallskip 

{\small
\noindent
\( \begin{array}{llll}
\{  \ \}: \rightarrow {\it USet}(\alpha)  &
\mi \ \md: \rightarrow {\it UMset}(\alpha) \\
\{\cdot\}: \alpha \rightarrow {\it USet}(\alpha)   &
\mi \cdot \md: \alpha \rightarrow {\it UMset}(\alpha)  \\
\ \cup :({\it USet}(\alpha),{\it USet}(\alpha)) 
            \rightarrow {\it USet}(\alpha) & 
\ \uplus:  ({\it UMset}(\alpha),{\it UMset}(\alpha)) 
            \rightarrow {\it UMset}(\alpha)
\end{array} \)
}
 
\medskip

\noindent 
For the data constructors  $\cup$ and $\uplus$, let us consider the 
following equations:

  \medskip

\( \begin{array}{lllllllll}
\textbf{equations} \\
\mbox{(a)} \ \ \  ({\it xs} \cup {\it ys}) \cup {\it zs} \approx {\it xs} \cup ({\it ys} 
\cup {\it zs}) &  \ \ \ ({\it xs} \uplus {\it ys}) \uplus {\it zs}
 \approx {\it xs} \uplus ({\it ys} \uplus {\it zs})  \\
\mbox{(b)} \ \ \ {\it xs} \cup {\it ys} \approx {\it ys} \cup {\it xs} &
\ \ \ {\it xs} \uplus {\it ys} \approx {\it ys} \uplus {\it 
xs} \\
\mbox{(c)} \ \ \ {\it xs} \cup \{ \ \} \approx {\it xs} &
\ \ \ {\it xs} \uplus  \mi \ \md \approx {\it xs} \\
\mbox{(d)}\ \ \ {\it xs} \cup {\it xs} \approx {\it xs}
 
\end{array} \)

\medskip

The above declaration constitutes an alternative to point (1)  for
specifying sets (respect. multisets) using singletons and set union $\cup$ 
(respect. multiset union $\uplus$) as data constructors. For instance, we can build the 
set  $\{a,b\}$ (respect. the multiset $\mi a,b \md$), where $a$,$b$ 
are two constant symbols,  as a  union of singletons  
$\{a\} \cup \{b\}$  (respect. $\mi a \md \uplus \mi b \md$).

The equational axioms in lines   (a), (b)  are strongly regular,
whereas those in lines  (c), (d) are regular and collapsing.  Since 
our type preservation and semantic results are based on strongly regular 
equational axioms, we can not use the specifications for sets and multisets 
given in this item. However, this is not a  serious lack of expressiveness, since in 
fact the specifications given in item (1) are very adequate for 
programming languages, as discussed in \cite{DOPR91,DR93,DOPR96,DPR96}.

Note that the operator $\cup$ (respect. $\uplus$) does not exactly 
correspond to the ACI1 (respect. AC1) operator used in the  theory of 
E-unification \cite{JK91}, where there is no explicit distinction 
between elements and sets (respect. multisets) but unitary sets 
(respect. multisets) are identified with elements.

\item[(3)] We conclude this example by presenting another datatype 
for polymorphic lists generated by means of unitary lists and a 
concatenation operator (associative and with neuter element $[ \ ]$).
Again, the traditional treatment of an A1 operator in  $E$-unification 
does not correspond exactly to that we are going to present.
In order to build the list datatype we consider the type constructor
${\it CList}/1$ together with the following data constructors:

 \medskip

\noindent
\( \begin{array}{llll}
[ \ ]: \rightarrow {\it CList}(\alpha) &
\ [\cdot] : \alpha \rightarrow {\it CList}(\alpha) &
\ \otimes: ({\it CList}(\alpha),{\it CList}(\alpha)) 
\rightarrow {\it CList}(\alpha)
\end{array} \)

\medskip

The equations which control the behaviour of  the data constructor
$\otimes$ are the following:

 \medskip

\noindent
\( \begin{array}{llll}
\ \ \ ({\it xs} \otimes {\it ys}) \otimes {\it zs} \approx 
        {\it xs} \otimes ({\it ys} \otimes {\it zs})  &
\ \ \ \ \ \ {\it xs} \otimes [ \ ] \approx {\it xs} & 
\ \  \ \ \ \ [\ ]  \otimes {\it xs} \approx {\it xs} 
\end{array} \)

\medskip

\noindent
where the first equation is strongly regular but the other two ones 
are regular and collapsing. \hfill \blacksquare

\end{des}

\end{exmp}

In subsequent examples, we will use  abbreviations  such as 
$\{x,y\mid{\it zs}\}$\/, $\{x,y\}$\/, and $\{x\}$ for the terms
$\{x\mid\{y\mid{\it zs}\}\}$\/, $\{x\mid\{y\mid\{\ \}\}\}$\/
and $\{x\mid\{\ \}\}$\/, respectively. We will use  similar notations 
for multisets and lists.

\begin{definition}[Algebraic and free data constructors]
Let ${\cal C}$ be a finite set of equational axioms and
$\Sigma$ a polymorphic signature.
$c \in {\it DC}^n$ is free iff ${\cal C}$ contains no equation
of the form $c(t_1, \ldots, t_n) \approx s$\/. Otherwise, we say that
$c$ is an algebraic (or equational) data constructor. \hfill \square
\end{definition}

As explained in the introduction, we must interpret equational axioms 
as schemes for generating approximation inequalities. This is achieved 
by the following  inequational calculus:

\begin{definition}[Inequational calculus] \label{calin} 
Given a set ${\cal C}$\/ of equational axioms, the
inequational calculus associated to $\cal C$  is defined
by the following inference rules:

 \smallskip

\noindent
\( \begin{array}{lllllllllll}
\textsf{(B)} \ \ \textsf{ Bottom:} &
\left.
 \begin{array}{c}
\\
 \hline   
 t \ecuacional \perp
\end{array} \right.
&
\ \ \textsf{(RF)} \ \ \textsf{ Reflexivity:} &
\left.
 \begin{array}{cc}
\\
 \hline   
 t \ecuacional t
\end{array} \right.
\end{array} \)
 
\smallskip

\noindent
\( \begin{array}{llllllll}
\textsf{(TR)} 
\ \ \textsf{ Transitivity:} &
\left.
 \begin{array}{cc}
 t \ecuacional t', t' \ecuacional t'' \\
 \hline
 t \ecuacional t''
\end{array} \right. 
\end{array} \)

\noindent
\( \begin{array}{llllllll}
 \textsf{(MN)} \ \  \textsf{ Monotonicity:} &
\left.
\begin{array}{cc}
t_1 \ecuacional s_1, \ldots, t_n \ecuacional s_n   \\
 \hline
c(t_1, \ldots, t_n) \ecuacional c(s_1, \ldots, s_n)
\end{array} \right.
\end{array} \)

\noindent
\( \begin{array}{llll}
\textsf{(IN)} \ \ \textsf{${\cal C}$-Inequation:} &
 \left.
\begin{array}{ccc}
 \\
 \hline
s \ecuacional t
\end{array}  \right. & \mbox{if $s \ecuacional t \in
                  {[{\cal C}]}_{\ecuacional}$}
\end{array}\)

\medskip

\noindent
where $t,t',t'',c(t_1, \ldots, t_n),
c(s_1, \ldots, s_n) \in \terminosp$\/, and:

\smallskip

\noindent
\( \begin{array}{llll}
{[{\cal C}]}_{\ecuacional} = \{ s\sigma_{d} \ecuacional
  t \sigma_{d}, t \sigma_{d}' \ecuacional s \sigma_{d}'
& \mid & s \approx t \in {\cal C}\/, \sigma_{d},\sigma_{d}' \in {\it DSub}_{\perp}, \\
& & \mbox{ $\sigma_{d}$ and $\sigma_{d}'$ are safe for $s$ and
         $t$ respectively} \}
\end{array} \) 

\hfill \square

\end{definition}

In the rest of the paper,  the  notation $s  \aproximado  t$  will denote  
the  formal derivability of $s \ecuacional t$ using the 
above  inequational  calculus for  ${\cal  C}$\/.  Moreover,  we  will write 
$s \equivalente t$ iff $s \aproximado t$ and $t \aproximado  s$\/.
Thinking of partial data terms as approximations of data, 
$s \aproximado t$\/ can be read as 
``$t$ approximates $s$\/''. Note  that  the 
formulation of  the rule ${\cal C}$-{\sf Inequation} forbids to use the axiom 
$\{x,x\mid{\it  zs}\} \approx  \{x\mid{\it zs}\}$ 
from Example \ref{tiposecuaciones} (1) to derive the inequality 
$\{\perp,\perp\}  \aproximado  \{\perp\}$,  which   would   have   undesirable 
consequences (see Example \ref{conjuntos} in Sect. \ref{calculos} below).

The next proposition states some simple properties of $\aproximado$ and
$\equivalente$, which follow easily from the form of the inference 
rules in the inequational calculus.

 \newpage

\begin{prop}[Properties of $\aproximado$ and $\equivalente$]
\label{caracteristicas}
 Let ${\cal C}$ be a finite set of equational axioms. Then:

 \begin{des}
 \item[(a)] $\aproximado$ is the least precongruence over $\terminosp$
            that contains ${[{\cal C}]}_{\ecuacional}$\/;

 \item[(b)] $\equivalente$ is the least congruence over $\terminosp$
            that contains ${[{\cal C}]}_{\ecuacional}$\/;

 \item[(c)] If ${\cal C}$ is regular then for any $s,t \in \terminosp$\/:
   If $s \aproximado t$ and $t$ is a total data term, then $s$ is also 
   a total data term and $s \equivalente t$\/. \hfill \square

\end{des}
\end{prop}

 Note that (c) may fail for non-regular equational axioms. For example, if
 ${\cal C}$ includes the axiom $c(x) \approx d(y)$ and $t \in \terminos$
 then $c(\perp) \aproximado d(t)$\/.

\section{Defining Rules and Programs}
\label{programas}

In this section  we introduce ACRWL programs, and we present some simple 
programming examples to illustrate the expressiveness of our framework. 
An ACRWL  program consists of some set ${\cal C}$  of equational axioms for
data constructors, together with  constructor-based rewrite rules
for defined functions. More precisely, assuming a principal type
declaration $f: (\tau_1, \ldots, \tau_n) \rightarrow \tau \in
{\it FS}$\/, a {\em defining rule} for $f$ must have the form:

 \vspace*{-0.10cm}

\[ f(t_1, \ldots, t_n) \rightarrow r \Leftarrow 
    a_1 \lazo b_1, \ldots, a_m \lazo b_m \]
    
\vspace*{-0.10cm}

\noindent
where the $n$-tuple $(t_{1}, \ldots, t_{n})$ is {\em linear} (i.e. without multiple
occurrences of variables), $t_i \in \terminos$\/, 
$1 \leq i \leq n$\/, and $a_j,b_j, r \in \expresiones$,
$1 \leq j \leq m$\/. {\em Joinability conditions}
$a_j \lazo b_j$ are intended to hold if and only if $a_j,b_j$ can
be reduced to some common {\em total} data term $t \in \terminos$\/, 
as in \cite{GHLR99}. A formal definition will be given below.

A defining rule is called {\em regular} if and only if all variables
occurring in $r$ occur also in the left-hand side.
Extra variables in the conditions are allowed, as well as
the unconditional case $m=0$\/.

\begin{definition}[Programs]
A program is a triple $\programa$\/, where
$\Sigma$ is a polymorphic signature, ${\cal C}$ is a finite
set of equational axioms for constructors in $\Sigma$,
and ${\cal R}$ is a finite set of defining rules for defined
functions symbols in $\Sigma$\/. 

We will say that
a program ${\cal P}$ is {\em strongly regular} if and only if
${\cal C}$ is strongly regular and all rules in ${\cal R}$ are regular.
\hfill \square

\end{definition}

Programs are intended to solve {\em goals} composed  of 
joinability conditions; i.e. goals  will  have  the  same form as conditions 
for defining rules.  Some of our subsequent results will refer to well-typed programs.
Let us introduce this notion.

\begin{definition}[Well-typed strongly regular program]
 \label{tipos} 

\begin{des}

\item A Joinability condition $e \lazo e'$ is well-typed w.r.t. an 
environment $V$ iff $e,e' \in {\it Expr}^{\tau}_{\Sigma_{\perp}}(V)$, 
for some $\tau \in \tipos$\/;

\item A regular defining rule $f(t_1, \ldots,$ $ t_n) \rightarrow r
     \Leftarrow C$ for a defined function symbol $f:(\tau_{1}, 
     \ldots, \tau_{n}) \rightarrow \tau$ is well-typed if there exists an
     environment $V$ such that $t_i \in {\it Term}^{\tau_i}_{\Sigma}(V)$,
     $1 \leq i \leq n$, $r \in {\it Expr}^{\tau}_{\Sigma}(V)$\/,
     and for all $e \lazo e' \in C$\/, $e \lazo e'$ is well-typed 
     w.r.t. $V$;
       
\item A strongly regular program $\programa$ is  well-typed,
     if all equations in ${\cal C}$ and all rules in ${\cal R}$ are
     well-typed.\hfill \square  
\end{des}
\end{definition}

Note that, according to the previous definition,  the left-hand sides of rewrite 
rules in a well-typed program must conform to the principal type of the 
corresponding function symbol, rather than being a more particular instance. 
Therefore, given $\textsf{append}: ({\it List}(\alpha),{\it List}(\alpha)) \rightarrow
{\it List}(\alpha)$\/, a defining rule such as 
$append({[{\it Zero}\mid{\it xs}]}$ $, {\it ys})
\rightarrow
    {[{\it Zero}\mid append({\it xs},{\it ys})]}$ would be ill-typed,
since the type of ${[{\it Zero}\mid{\it xs}]}$ is too particular
($List({\it Nat})$ instead of ${\it List}(\alpha)$)\/.
For technical convenience, 
we are assuming that the principal types of functions are declared as part of a 
program's signature. This assumption, however, is not essential in practice. 
Type reconstruction algorithms based on \cite{Mil78,DM82} can be used
to infer principal types for functions,  going out from the declared principal 
types of data constructors and the rewrite rules in the program.

 The expressive power of algebraic data constructors allows to write short and 
 clear ACRWL programs for many kinds of problems. We will now illustrate this by 
 means of two examples, dealing with typical applications of the datatype multiset. 
 The reader is also referred to \cite{GHLR99} for more programming examples
 in the CRWL framework (with free data constructors),  and to \cite{Han94b} for 
 the general advantages of functional logic programming.

\subsection{Planning Problems}
\label{planning}

Planning problems are a particular case of {\em action and change problems},
where one is interested in finding actions that will transform a given initial 
situation into a final situation  which satisfies some desired property.  When 
attempting  to solve action and change problems in classical predicate logic, 
one meets the so-called {\em frame problem}, roughly meaning that all the 
properties of a situation that are not affected by the application of 
an action, must be explicitly asserted within the logical formula which 
formalizes the effect of the action. This gives rise to a combinatorial 
explosion when trying to use automated deduction techniques (resolution, say)
to solve action and change problems.

It is known that various non-classical logics can be used to solve action and 
change problems declaratively, while avoiding the frame problem; see 
e.g. \cite{MM95}. One of the known approaches is based on the representation of 
situations as multisets of facts. Assuming such a representation, actions can 
be conveniently specified as multiset transformations. In general, an action 
will be applicable to those situations which include certain facts. The 
effect of the action will be to take away the facts which enable its 
application, and to add some other facts to the new situation. The
frame problem is avoided, because the rest of the facts is carried along implicitly.

Following these ideas, H\"olldobler and his group have developed an approach 
to planning based on {\em equational logic programs} \cite{HS90,GHSST92}. 
In equational logic programming,  programs consist of Horn clauses with algebraic data 
constructors, in addition to free data constructors. SLD resolution uses 
unification modulo the equational theory of the algebraic constructors present 
in the program. More precisely, for the case of planning problems, H\"olldobler
and his co-workers propose to use a binary associative-commutative constructor $\circ$ 
(written in infix notation) to represent situations as multisets of facts 
${\sf fact_1  \circ  \ldots  \circ fact_n}$\/, and a ternary 
predicate $\textsf{execPlan(InitialSit,Plan,FinalSit)}$ to model the 
transformation of an initial situation into a final situation by the execution 
of a plan.

In ACRWL we can follow the same 
idea  using multisets of facts to represent situations, 
and a non-deterministic function 
$\textsf{execPlan}: ({\it List}({\it Action}), {\it  Mset}({\it  Fact})) 
\rightarrow {\it  Mset}({\it   Fact})$ 
 to  represent  the  effect  of  plan 
execution.  In general, when dealing with search problems, ACRWL gives the 
freedom to use either predicates or non-determi\-nistic functions. 
As shown in \cite{GHLR99}, the use of non-determi\-nistic functions can bring 
advantages, when combined with the effect of lazy evaluation.

As a concrete illustration, we will show a particular ACRWL 
program which solves a very simple planning problem, adapted from \cite{GHSST92}.
More complicated planning problems, as well as other kinds of action and change 
problems, could be treated analogously.

\begin{exmp} \label{accion} 
The following typical
blocksworld problem consists in finding a plan
for transforming situation (A) into situation (B) (see figure
below) by means of a robot's hand. The possible facts are:

\begin{des}
\item $O(b_1,b_2)$\/: block $b_1$ is over block $b_2$\/;
\item $C(b)$\/: block $b$ is clear (i.e. there is no block over it);
\item $T(b)$\/: block $b$ is over the table;
\item $H(b)$\/: the robot's hand holds block $b$;
\item $E$\/: the robot's hand is empty.
\end{des}

The available actions are ${\it Pickup(b)}$\/,
${\it Unstack(b_1,b_2)}$\/, ${\it Putdown(b)}$, 
${\it Stack}(b_1,$ $b_2)$. Their behaviour can be easily
deduced from the definition of the function \textsf{execAction}
below.

\begin{center}
\epsfig{file=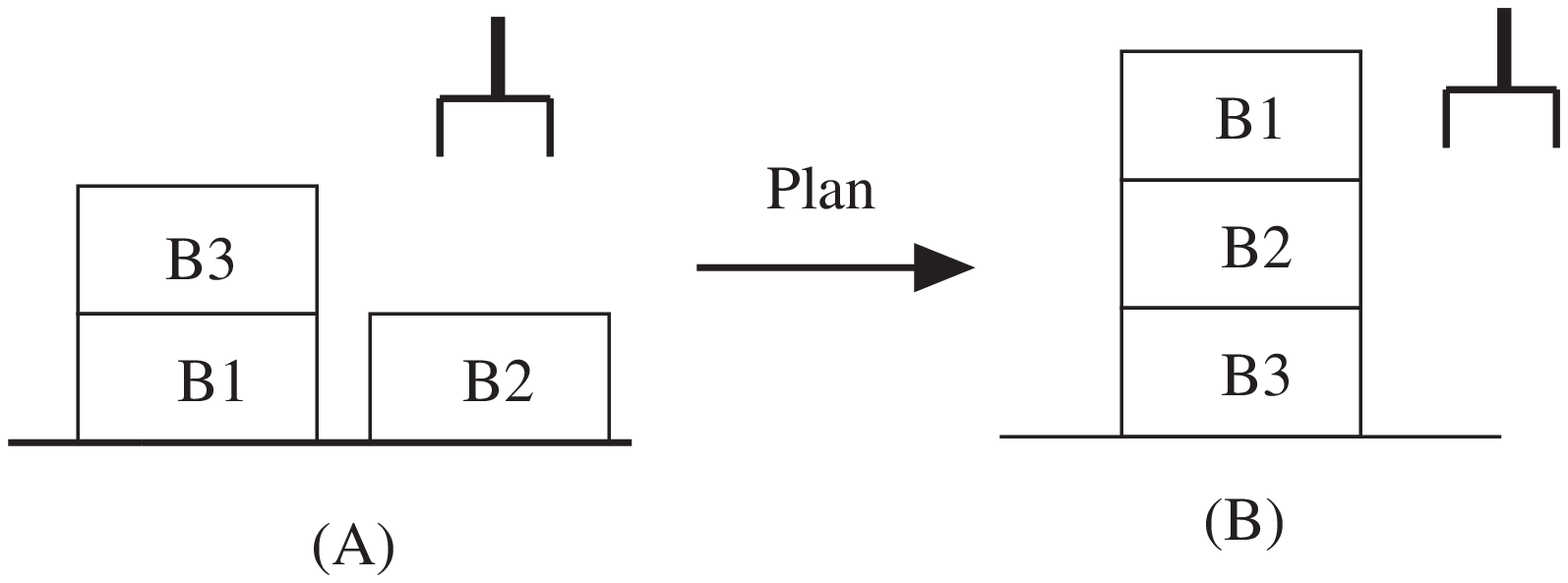,width=7cm, height = 3cm}
\end{center}
 
The problem of finding a plan for transforming situation (A) into (B) can be 
described in our framework by means of the following simple program:

 \medskip

\noindent
{\bf datatypes} ~~~${\it Block}/0,{\it Fact}/0,{\it Action}/0,{\it Mset}/1,{\it 
List}/1$

\smallskip

\newpage 

\noindent
{\bf constructors}

\smallskip

\noindent
\( \begin{array}{llll}
\left.
\noindent
\begin{array}{llll}
B_1,B_2,B_3: \rightarrow {\it Block} \\
C,{\it T},{\it H}: {\it Block} \rightarrow {\it Fact} \\
{\it O}: ({\it Block},{\it Block}) \rightarrow {\it Fact} \\
{\it E}: \rightarrow {\it Fact}\\
{\it Pickup},{\it Putdown}: {\it Block} \rightarrow {\it Action}\\
{\it Stack},{\it Unstack}: ({\it Block},{\it Block}) \rightarrow {\it Action}
\end{array} \right.
&
\left.
\noindent
\begin{array}{llll}
\mi \ \md: \rightarrow {\it Mset}(\alpha) \\
\mi \cdot \mid \cdot \md: (\alpha, {\it Mset}(\alpha))
              \rightarrow {\it Mset}(\alpha) \\
{[ \ ]}: \rightarrow {\it List}(\alpha) \\
{[ \cdot \mid \cdot ]}: (\alpha, {\it List}(\alpha))
              \rightarrow {\it List}(\alpha) \\
\end{array} \right.
\end{array} \)

\medskip

\noindent
{\bf equations}

\smallskip

 \( \begin{array}{llll}
      \mi x,y\mid{\it xs} \md \approx \mi y,x\mid{\it xs} \md
 \end{array} \)

\medskip

\noindent
{\bf functions}

\smallskip

\noindent
\( \begin{array}{lllllll}
\textsf{execPlan}: ({\it List}({\it Action}),
        {\it Mset}({\it Fact})) \rightarrow {\it Mset}({\it Fact}) \\
\  {\it execPlan}({[\ ]}, sit) \rightarrow sit \\
\  {\it execPlan}({[act\mid ract]},sit) \rightarrow
            {\it execPlan}(ract,{\it execAction}(act,sit))
\end{array} \)

\smallskip

\noindent
\( \begin{array}{llll}
\textsf{execAction}: ({\it Action},{\it Mset}({\it Fact})) \rightarrow
        {\it Mset}({\it Fact}) \\
\  {\it execAction}({\it Pickup}(v), \mi {\it C}(v_1),{\it T}(v_2),{\it 
E}\mid{\it facts} \md)
            \rightarrow \mi {\it H}(v)\mid{\it facts} \md \\
\ \ \ \ \ \ \ \ \ \ \ \ \Leftarrow v \lazo v_1, v \lazo v_2 \\
\  {\it execAction}({\it Unstack}(v,w), \mi {\it C}(v_1),{\it O}(v_2,w_1),
             {\it E}\mid{\it facts} \md) \\
\ \ \ \ \  \rightarrow
    \mi {\it H}(v),{\it C}(w)\mid{\it facts} \md \\
\ \ \ \ \ \ \ \ \ \ \ \ \Leftarrow v \lazo v_1, v \lazo v_2, w_1 \lazo w \\
\  {\it execAction}({\it Putdown}(v), \mi {\it H}(v_1)\mid{\it facts} \md) 
\rightarrow
     \mi {\it T}(v),{\it C}(v),E\mid{\it facts} \md \\
\ \ \ \ \ \ \ \ \ \ \ \ \Leftarrow v \lazo v_1 \\
\ {\it execAction}({\it Stack}(v,w), \mi {\it H}(v_1),{\it C}(w_1)\mid{\it 
facts} \md)
    \rightarrow
     \mi {\it O}(v,w),{\it C}(v),
                  {\it E}\mid{\it facts} \md \\
\ \ \ \ \ \ \ \ \ \ \ \ \Leftarrow v \lazo v_1, w \lazo w_1
\end{array} \)

\medskip

\noindent
The appropriate goal for 
getting a plan solving
the planning problem at hand is:

 \medskip

\(\begin{array}{llll}
{\it execPlan}({\it plan},\mi C(B_2),C(B_3),O(B_3,B_1),T(B_2),T(B_1),E
      \md)   
  \lazo \\
\ \ \ \ \ \ \ \ \ \ \ \ \ \ \ \ \ \ \ \mi C(B_1),O(B_1,B_2),O(B_2,B_3),T(B_3), E \md.
\end{array} \)

\medskip

\noindent
The  completeness result proved in Theorem \ref{completitud2} 
in Section \ref{goalsolving} ensures that the answer: 

\medskip

\noindent
\( \begin{array}{llll}
{\it plan} & = & [{\it Unstack}(B_3,B_1),{\it Putdown}(B_3),{\it Pickup}(B_2), \\
          &    & \ {\it Stack}(B_2,B_3), {\it Pickup}(B_1),{\it
            Stack}(B_1,B_2)]
\end{array} \)

\medskip

\noindent
can be computed by the lazy narrowing calculus $\calculo$ presented in 
Section \ref{goalsolving}. Of course, other possible plans for solving the
same planning problem can be also computed. However, in spite of the theoretical 
completeness result, $\calculo$ has many shortcomings from the viewpoint of 
practical computation. This will be discussed in sections \ref{goalsolving} 
and \ref{completeness}. \hfill \blacksquare

\end{exmp}

\subsection{The GAMMA Programming Model}
\label{gamma}

The \underline{G}eneral \underline{A}bstract \underline{M}odel for \underline{M}ultiset 
M\underline{a}nipulation (GAMMA, for short) has been proposed by Ban\^{a}tre and 
Le M\'{e}tayer \cite{BM90,BM93}, aiming at problem solving at a high level of 
abstraction. Two important motivations behind the GAMMA approach are to 
avoid unfortunate sequential biases during program design, and to facilitate the 
application of systematic program derivation methods.

The basic data structure in GAMMA is the multiset, which allows to describe 
compound data without any form of constraint or hierarchy between its 
components.  Other more conventional data structures can be encoded as 
multisets; see \cite{BM93}. A GAMMA computation proceeds as a series of 
transformations, carrying a multiset from some initial situation (representing the input) to 
some final situation (representing the output). More precisely, a GAMMA program 
is a set of pairs $(R,A)$ where $R$ (called {\em reaction condition}) is a 
boolean function of some given arity $n$, and $A$ (called {\em action}) is a 
function of the same arity $n$, returning a multiset. The  behaviour of a GAMMA program as 
a multiset transformer is as follows: given a multiset ${\it xs}$, GAMMA looks for some n-tuple 
$\overline{x}$ of elements from ${\it xs}$ (ignoring the order, but not ignoring the repetitions), 
such that $R(\overline{x})$ holds for some pair $(R,A)$  in the program. If no 
such tuple can be found, the computation halts giving ${\it xs}$ as result. Otherwise, 
the components of $\overline{x}$ are removed from ${\it xs}$, the remaining elements 
are added to $A(\overline{x})$ (in the sense of multiset union), and GAMMA 
transformation is iterated from the new multiset. The notation 
$\Gamma_{P}({\it xs})$ 
is used to indicate the final result obtained by iterating the GAMMA 
transformation w.r.t. program $P$ and starting with the multiset ${\it 
xs}$. 
In fact, $\Gamma_{P}$ is almost always a non-deterministic function (unless 
$P$ is a trivial program). Ban\^{a}tre and Le M\'{e}tayer assume that one of the 
possible results will be chosen, non-deterministically, at each GAMMA 
iteration\footnote{They also assume that several action-reaction pairs $(R,A)$ could be 
performed simultaneously by a parallel implementation.}.

GAMMA is a quite powerful computation model. In particular, the approach to 
planning problems discussed in the previous subsection, can be understood as  
an instance of GAMMA computation.  An interesting collection of GAMMA 
programs is presented in \cite{BM93},  showing a variety of programming styles.
For example, in order to compute shortest paths between all pairs of vertices 
in a weighted directed graph, we can represent the graph as a multiset of edges.
Assume that an edge of cost $c$ going from the vertex $u$ to the vertex $v$ is 
encoded as $(u,v,c)$. Then, \cite{BM93} gives the following GAMMA program for
solving the shortest path problem:

\medskip

\(\begin{array}{lcl}
\textsf{shortestPaths}({\it graph}) & = & \Gamma_{(R,A)}({\it graph}) \textbf{ where} \\
R((u,v,c),(u,w,d),(w,v,e)) & = & c > d+e \\
A((u,v,c),(u,w,d),(w,v,e)) & = & \mi (u,v,d+e),(u,w,d),(w,v,e) \md 
\end{array}\) 

\medskip

The idea behind this program is simple: Each time one finds an edge in the 
graph whose cost is greater than the cost of a path of length 2 going through 
some intermediate node, the cost of the edge is rewritten to the smaller cost of 
the path. As soon as this local transformation cannot be applied any longer, 
the cost $c$ of each edge $(u,v,c)$ will be that of a path of minimal cost going 
from $u$ to $v$ in the original graph.

Now, we will  present an ACRWL program which can be viewed as a
translation of the former GAMMA program. More generally, any GAMMA program $P$
could be translated into an ACRWL program based on the algebraic datatype 
multiset, whose rewrite rules would define the function $\Gamma_{P}$, as well as an
auxiliary boolean function ${\it irreducible}_{P}$ which tests the GAMMA termination 
condition (namely, that no action-reaction pair $(R,A) \in P$ is applicable).
Unfortunately,  ${\it irreducible}_{P}$ tends to be complex and inefficient in most 
cases.  After all, GAMMA is not intended as a programming language in the 
conventional sense, but rather as a convenient intermediate language between 
specifications and programs; see \cite{BM93}, pg. 108. For our particular 
example, we will use the more suggestive names {\sf minimizePaths} and 
{\sf minimal} in place of $\Gamma_{P}$ and ${\it irreducible}_{P}$,
 respectively.
 
 \begin{exmp} We will use lists and multisets, as defined in Example \ref{accion}, 
 as well as boolean values, given by free constructors 
 ${\it True}, {\it False}: \rightarrow {\it Bool}$.  Moreover, we will 
 assume the existence of the datatypes $\textsf{Node}$ and $\textsf{Cost}$,  
 together with  infix binary operations 
 $+~:~({\it Cost,Cost}) \rightarrow {\it Cost}$ and
 $>,\not =, \leq ~:~({\it Cost,Cost}) \rightarrow {\it Bool}$,
 intended to add and compare costs, respectively.
 In order to represent edges and graphs, we introduce the following data 
 constructors: 
 
  \vspace*{-0.10cm}
 
 \[ E: ({\it Node,Node,Cost}) \rightarrow {\it Edge}
  \ \ \ \ \ G: {\it Mset(Edge)} \rightarrow {\it Graph} \] 
 
 \vspace*{-0.10cm}

The function \textsf{minimizePaths} is defined as follows:

  \medskip

\noindent
\( \begin{array}{llll}
\textsf{minimizePaths}: {\it Graph} \rightarrow {\it Graph} \\
\ \ \ {\it minimizePaths}(G(\mi  E(u,v,c),E(u_{1},w,d),E(w_{1},v_{1},e) | 
{\it rest }\md)) \rightarrow \\
   \ \ \ \ \ \  G(\mi  E(u,v,d+e),E(u_{1},w,d),E(w_{1},v_{1},e) | 
            {\it rest } \md) \\
       \ \ \ \ \ \ \ \ \ \ \ \ \ \ \ \ \      \Leftarrow  u \lazo u_{1}, v \lazo v_{1}, w \lazo w_{1}, 
             (c > d + e) \lazo {\it True} \\
\ \ \ {\it  minimizePaths}({\it graph}) \rightarrow
       {\it graph} \Leftarrow {\it minimal(graph)} \lazo {\it True}
 
\end{array} \)

\medskip

Now, we need to define a function \textsf{minimal} which detects if 
a graph is irreducible, i.e., the first rule of \textsf{minimizePaths} 
can not be applied.   The ACRWL definition for this function is the 
following:

 \medskip

\noindent
\( \begin{array}{llll}
\textsf{minimal}: {\it Graph} \rightarrow {\it Bool} \\
\ \ \ {\it minimal}(G({\it graph})) \rightarrow 
{\it irreducible}({\it triples(graph)}) 
 \end{array} \)

\medskip

\noindent
where the function \textsf{triples} returns a list composed of all  
possible triples of edges coming from the graph, and the function 
\textsf{irreducible} checks that no one of them enables  a 
GAMMA reaction. To represent a triple of edges, we use the following data 
constructor:

 \vspace*{-0.10cm}

\[ T : (\alpha,\alpha,\alpha ) \rightarrow {\it Triple}(\alpha) \]

\vspace*{-0.10cm}
 
\medskip

\noindent
Next, the definition for \textsf{irreducible} is the following:

\medskip

\noindent
\( \begin{array}{llll}
\textsf{irreducible}: {\it List(Triple(Edge))} \rightarrow {\it Bool} \\
\ \ \ {\it irreducible}([ \ ]) \rightarrow  {\it True} \\
\ \ \ {\it irreducible}([T(e,e_{1},e_{2}) | {\it rest}]) 
     \rightarrow {\it True} \\
     \ \ \ \ \ \ \ \ \ \ \Leftarrow {\it reaction}(e,e_{1},e_{2}) 
     \lazo {\it False}, {\it irreducible}({\it rest}) \lazo {\it 
     True}  
 \end{array} \)

\noindent
where the function \textsf{reaction} is defined as follows:

 \medskip

\noindent
\( \begin{array}{llll}
\textsf{reaction}: {\it (Edge,Edge,Edge)} \rightarrow {\it Bool} \\
\ \ \ {\it reaction}(E(u,v,c),E(u_{1},w,d),E(w_{1},v_{1},d)) \rightarrow  
{\it False}  \\
\ \ \ \ \ \ \ \ \Leftarrow (u \not = u_{1} \vee v \not = v_{1} \vee w 
\not = w_{1} \vee c \leq d+e) \lazo {\it True} 
 \end{array} \)

\medskip

\noindent
Note that ``$\vee$'' represents the boolean disjunction defined as usual.
Finally,  let us see the definition of function \textsf{triples}.

 \medskip

\noindent
\( \begin{array}{llll}
\textsf{triples}: {\it  Mset(\alpha)} \rightarrow {\it 
List(Triple(\alpha))} \\
\  {\it  triples}(\mi \ \md) \rightarrow [ \ ] \\
\   {\it triples}(\mi x | {\it xs} \md) \rightarrow 
         {\it triples1}(x,{\it xs}) \mbox{++} {\it triples(xs)}
 \end{array} \)

\medskip

\noindent
\( \begin{array}{llll}
\textsf{triples1}: {\it  (\alpha, Mset(\alpha))} \rightarrow {\it 
List(Triple(\alpha))} \\
\   {\it  triples1}(x,\mi \ \md) \rightarrow [ \ ] \\
\  {\it triples1}(x,\mi y | {\it ys} \md) \rightarrow 
         {\it triples2}(x,y,{\it ys}) \mbox{++} {\it triples1(x,ys)}
 \end{array} \)
 
 \medskip

\noindent
\( \begin{array}{llll}
\textsf{triples2}: {\it  (\alpha, \alpha, Mset(\alpha))} \rightarrow 
{\it List(Triple(\alpha))} \\
\ {\it  triples2}(x,y,\mi \ \md) \rightarrow [ \ ] \\
\ {\it triples2}(x,y, \mi z | {\it zs} \md) \rightarrow  \\
\ \ {[T(x,y,z),T(x,z,y),T(y,x,z),T(y,z,x),T(z,x,y),T(z,y,x) | {\it triples2}(x,y,{\it zs})]}
 \end{array} \)

\medskip

\noindent
where $++$ is   the concatenation of lists, which is easy to define in ACRWL. 
\hfill \blacksquare
 
 \end{exmp}

\section{Rewriting Calculi}
\label{calculos}

In   this  section  we 
present  two constructor-based rewriting  calculi, named
{\em Basic Rewriting Calculus} ($\bpc$) and {\em Goal-Oriented Rewriting 
calculus} ($\gpc$) respectively, which   are intended   as   a 
proof-theoretical specification of programs'  semantics.
Although both rewriting calculi will be proved equivalent in 
Theorem \ref{equivalencia}, we have preferred to  
present both of them. The reason is that  $\bpc$ is closer to the intuition, 
while  the  goal-oriented format of the $\gpc$-like  calculus  is
useful as a basis for designing  the lazy narrowing calculus described 
in Section \ref{goalsolving}.

As  in \cite{GHLR99}, 
our calculi are designed to  derive  two  kinds of  statements:
{\em  reduction statements} $e \rightarrow e'$\/, intended to mean that $e$ can be  reduced  to 
$e'$, and {\em joinability statements} $e \lazo e'$, intended to mean that $e$ 
and $e'$ can be reduced to some common total data term. Reduction statements of 
the form $e \rightarrow t$, where $t$ is a possibly partial data term, will  
be  called {\em approximation statements}.

\begin{definition}[Rewriting calculi] \label{calculitos} 
For  a  given  program  $\programa$\/:

\smallskip

\noindent
$\bullet$ The  basic 
rewriting calculus ($\bpc$)  is  defined  as  follows:

\smallskip

\noindent
\(\begin{array}{lllllllll}
\textsf{(B)} \ \  \textsf{ Bottom:} &
\left.
 \begin{array}{cc}
\\
 \hline   
 e \rightarrow \perp \\
\end{array} \right.
&
\ \ \ \textsf{(RF)} \ \ \textsf{ Reflexivity:} &
\left.
 \begin{array}{cc}
\\
 \hline   
 e \rightarrow e
\end{array} \right.
\end{array} \)

\medskip
 
\noindent
\( \begin{array}{lll}
\textsf{(TR)} \ \ \textsf{ Transitivity:} &
 \left.
\begin{array}{cc}
 e \rightarrow e', e' \rightarrow e''   \\
 \hline
e \rightarrow e''
\end{array} \right. 
\end{array} \)

\medskip

\noindent
\(\begin{array}{lllllll}
\textsf{(MN)} \ \  \textsf{ Monotonicity:} &
\left.
\begin{array}{cc}
e_1 \rightarrow e'_1, \ldots , e_n \rightarrow e'_n   \\
 \hline
h(e_1, \ldots , e_n) \rightarrow h(e'_1, \ldots ,e'_n)
\end{array}   \right.
& \mbox{ if $h \in {\it DC}^{n} \cup {\it FS}^n$}
\end{array} \)

\medskip

\noindent
\( \begin{array}{llllllll}
\textsf{(R)}  \ \ \textsf{${\cal R}$-Reduction:} &
\left.
\begin{array}{cc}
 C   \\
 \hline
l \rightarrow r
\end{array} \right. & \mbox{if $l \rightarrow r \Leftarrow C
                        \in {[{\cal R}]}_{\rightarrow}$}
\end{array} \)

\medskip

\noindent
\( \begin{array}{llllllllll} 
\textsf{(MUT)}  \ \ \textsf{${\cal C}$-Mutation:} &
\left.
\begin{array}{cc}
\\
 \hline
s \rightarrow t
\end{array} \right.  & \mbox{if $s \ecuacional t \in {[C]}_{\ecuacional}$}
\end{array} \)

\medskip

\noindent
\( \begin{array}{llll}
\textsf{(J)} \textsf{ Join:} &
\left.
\begin{array}{cc}
e \rightarrow t,  e' \rightarrow t \\
 \hline
e \lazo e'
\end{array} \right. &  \mbox{if $t \in \terminos$
      is a {\em total} data term}
\end{array} \)

\medskip

\noindent
where $e,e',e'',h(e_1, \ldots, e_n),h(e'_1, \ldots, e'_n)
\in \expresionesp$\/, $[{\cal C}]_{\ecuacional}$ has been specified in
Definition \ref{calin} and
${[{\cal R}]}_{\rightarrow} =_{{\it Def}} \{ (l \rightarrow r
           \Leftarrow C)\sigma_{d} \ \mid 
              \  l \rightarrow r \Leftarrow C \in {\cal R}, \sigma_{d} \in
                DSub_{\perp}\}$\/.
                
\medskip
                
\noindent
$\bullet$ The  goal-oriented
rewriting calculus ($\gpc$)  is  defined  as  follows:

\smallskip

\noindent
\(\begin{array}{llllll}
\textsf{(B)} \ \  \textsf{ Bottom:} &
\left.
 \begin{array}{cc}
\\ 
\hline   
 e \rightarrow \perp
\end{array} \right. 
\end{array} \)

\medskip

\noindent
\(\begin{array}{llllll}
\textsf{(RR)}  \textsf{ Restricted Reflexivity:} &
\left.
 \begin{array}{cc}
\\
 \hline   
 x \rightarrow x
\end{array} \right.  
& \mbox{ if $x \in {\it DVar}$}
\end{array} \)

\medskip

\noindent
\( \begin{array}{llllllllll}
\textsf{(DC)} \ \  \textsf{ Decomposition:} &
 \left.
\begin{array}{cc}
e_1 \rightarrow t_1, \ldots , e_n \rightarrow t_n   \\
 \hline
c(e_1, \ldots , e_n) \rightarrow c(t_1, \ldots ,t_n)
\end{array} \right.
& \mbox{ if $c \in {\it DC}^n$}
\end{array} \)

\medskip

\noindent
\( \begin{array}{lllll}
\textsf{(OMUT)} \ \ \textsf{ Outer ${\cal C}$-Mutation:} &
\left.
\begin{array}{cc}
e_1 \rightarrow t_1, \ldots, e_n \rightarrow t_n, s \rightarrow t    \\
 \hline
 c(e_1, \ldots, e_n) \rightarrow t
\end{array} \right.
\end{array} \)

if $t \not = \perp$\/, $c(t_{1}, \ldots, t_{n}) \ecuacional s \in 
{[C]}_{\ecuacional}$

\medskip

\noindent
\( \begin{array}{lcllll}
\textsf{(OR)} \ \ \textsf{ Outer ${\cal R}$-Reduction:} &
\left.
\begin{array}{cc}
e_1 \rightarrow t_1, \ldots , e_n \rightarrow t_n, C, r \rightarrow t   \\
 \hline
f(e_1, \ldots ,e_n) \rightarrow t
\end{array} \right. 
\end{array} \)

\smallskip

if $t \not = \perp$,
                    $f(t_{1}, \ldots, t_{n}) \rightarrow r \Leftarrow C
                        \in {[{\cal R}]}_{\rightarrow}$ 

\smallskip

\noindent
\( \begin{array}{lcll}
\textsf{(J)} \ \ \textsf{ Join:} &
\left.
\begin{array}{cc}
e \rightarrow t',  e' \rightarrow t' \\
 \hline
e \lazo e'
\end{array} \right. & \mbox{ if $t' \in \terminos$
  is a {\em total} data term}
\end{array} \)

\medskip
 
\noindent
where $e,e',c(e_1,\ldots, e_n),f(e_1,\ldots, e_n)
\in \expresionesp$\/, $x \in {\it DVar}$, and
$t,c(t_1, \ldots,$ $ t_n) \in 
\terminosp$. \hfill \square

\end{definition}

Note  that   the construction   of   ${[{\cal R}]}_{\rightarrow}$ 
does not require $\sigma_{d}$ to be safe for $l$\/, in contrast to the 
construction of  ${[C]}_{\ecuacional}$ 
in the inequational calculus. This is because $l$ is known to be linear.

As in \cite{GHLR99}, neither of the two calculi specifies 
rewriting in the usual sense. The main reason is the presence of rule
\textsf{(B)} and the  formulation of rules \textsf{(R)} (respect. \textsf{(OR)}) and 
\textsf{(MUT)} (respect. \textsf{(OMUT)}). The need of the  
rule \textsf{(B)} is because of non-strict (also called lazy) functions, and shows that  
$e  \rightarrow  t$ is intended to mean  ``$t$ approximates $e$''.
The construction of ${[{\cal R}]}_{\rightarrow}$ and
 ${[C]}_{\ecuacional}$  reflects the ``call-time choice'' treatment of 
non-determinism. Our motivation to adopt these ideas has been explained in the 
Introduction. As we will see in Section \ref{goalsolving} 
(in particular, in Example \ref{sharing1}), our goal solving 
calculus incorporates sharing in order to ensure a sound realization of 
call-time choice.

As the main novelty w.r.t. \cite{GHLR99}, we  find  the  {\em  mutation} 
rules \textsf{(MUT)}  (respect. \textsf{(OMUT)}) to deal  with  equations between  
constructors.  Note that the use of such mutation rules can cause 
cycles. This is easy to see for the multiset equation $\mi x,y |{\it 
zs} \md \approx \mi y,x | {\it zs} \md$. From a theoretical point of 
view, this inconvenience can be avoided by rewriting with equivalence 
classes instead of terms and eliminating the mutation rules. However, 
we have preferred the current presentation of the rewriting calculi since it 
enables (as shown in Lemma \ref{progreso}) a very intuitive 
completeness proof for the lazy narrowing calculus in Section 
\ref{goalsolving}. Unfortunately, the presence of cycles in lazy narrowing 
derivations is also possible  and quite hard to  avoid in a general 
framework, where arbitrary algebraic data constructors are allowed.

Finally, we can also establish several differences between our rewriting calculi 
and another well-known approach to rewriting as logical deduction, namely 
Meseguer's {\em Rewriting Logic} \cite{Mes93}, which has been used as 
a basis for computation systems and languages such as 
Maude \cite{Mes93,CELM96,CDE+99}, Elan \cite{KKV95,BKK+96} and
CafeOBJ \cite{DFI+98}. As an analogy between  \cite{Mes93} 
and the calculi $\bpc$ and $\gpc$,  we have that in \cite{Mes93} rewriting 
is performed  modulo a set of  equations (as for instance, 
associativity and/or commutativity), which allow to establish term 
equivalences. As the main difference, note that our rewriting calculi 
allow to model expression evaluation in a language based on 
constructors, which includes non-strict functions (possibly 
non-deterministic). Thus, $\bpc$ and $\gpc$ can serve as a basis for 
declarative programming languages based on lazy evaluation. On the 
contrary, the logic  described in \cite{Mes93} was originally proposed 
as a semantic framework for the specification of  concurrent languages 
and systems, and as a framework in which to be able to specify 
other logics. Hence, \cite{Mes93} is not constructor-based and lacks 
of the rule \textsf{(B)}.  Finally, as we have 
commented before, we only consider instances over partial data terms 
as shown by the construction of the sets $[{\cal C}]_{\ecuacional}$ and 
$[{\cal R}]_{\rightarrow}$ (i.e., we adopt a ``call-time choice'' view), whereas 
in \cite{Mes93} arbitrary  instances are allowed (i.e., a ``run-time 
choice'' view is adopted).  As shown in \cite{GHLR99}, call-time choice is a 
good option from the programming viewpoint.

Remark that $\gpc$ means {\em Goal-Oriented Rewriting Calculus}. Such a name 
has been inherited from \cite{GHLR99}, where goal-oriented 
proofs have the property that the outermost syntactic structure of 
the statement to be proved determines the inference rule which must 
be applied at the last proof step; in this way, the structure of the 
proof is determined by the structure of the goal. In our case, proofs 
derived by using the rewriting calculus $\gpc$ are not totally goal-oriented 
due to the presence of algebraic constructors. More concretely, an 
approximation statement of the form $c(\etupla_{n}) \rightarrow t$, 
where $c$ is an algebraic data constructor, 
presents two alternatives given by rules \textsf{(DC)} and \textsf{(OMUT)}. 
However, we have preferred to maintain the name of goal-oriented because
$\gpc$ is really goal-oriented when algebraic data constructors are 
absent. Let us see a simple example:

\begin{exmp} Consider the program rules ${\it 
select}(\mi x \mid {\it xs} \md) \rightarrow x$ and $f \rightarrow \mi {\it 
Zero} \mid f \md$, where $\mi \cdot \mid \cdot \md$ is the multiset 
constructor defined in Example \ref{tiposecuaciones} (1). 
Consider the approximation statement $\varphi  \equiv {\it select}(f) \rightarrow 
{\it Zero}$. Now, let us look for a $\gpc$-proof for $\varphi$.

It is clear that the only $\gpc$-rule applicable to $\varphi$ is  
\textsf{(OR)}. 
Let us choose the instance of \textsf{select} given by the rewriting rule ${\it 
select}(\mi {\it Zero}\mid\perp\md) \rightarrow {\it Zero}$. Then we need 
to prove $\varphi_{1} \equiv f \rightarrow \mi {\it Zero}\mid\perp\md$ and $
\varphi_{2} \equiv {\it Zero} \rightarrow {\it Zero}$. $\varphi_{2}$ 
can be proved uniquely by rule   \textsf{(DC)}, whereas $\varphi_{1}$  
necessarily requires an application of  \textsf{(OR)}, i.e., we need to 
prove that $\varphi_{3} \equiv \mi {\it Zero} \mid f \md \rightarrow \mi {\it Zero} \mid 
\perp\md$.  Up to this point, the proof has been clearly 
goal-oriented. Now, $\varphi_{3}$ can be proved by using either  
\textsf{(DC)} or \textsf{(OMUT)}. Applying  \textsf{(DC)}, we would need to prove 
the approximation statements ${\it Zero} \rightarrow {\it Zero}$ 
and $f \rightarrow \perp$ which are trivially true by rules  \textsf{(DC)} and 
\textsf{(B)} respectively. If we apply \textsf{(OMUT)} we can 
also get a proof but with more inference steps. To this end, it is enough to 
take the following  instance of the commutativity equational axiom for multisets:
$\mi {\it Zero,Zero}\mid\perp \md \approx \mi {\it Zero,Zero}\mid\perp 
\md$. \hfill \blacksquare

\end{exmp}

The next result ensures that both calculi 
are essentially equivalent. Moreover, they are compatible with the
inequational calculus presented in Sect. \ref{terminos}. The complete 
proof can be found in Appendix \ref{demos}.

\begin{theorem}[Calculi equivalence]
\label{equivalencia}
Let $\programa$ be a program.

\begin{des}
\item[(a)] For strongly regular ${\cal C}$\/,
      $e,e' \in \expresionesp$ and $t \in \terminosp$\/:
      $e \rightarrow t$ (respect. $e \lazo e'$) is derivable in $\gpc$
      if and only if $e \rightarrow t$ (respect. $e \lazo e'$)
      is derivable in $\bpc$\/;

\item[(b)] For any $t,t' \in \terminosp$\/, $t \aproximado t'$ if and only if 
       $t \rightarrow t'$ is derivable in $\bpc$\/;

\item[(c)] If ${\cal C}$ is regular, then for any $s,t \in \terminosp$\/, 
       $s \lazo t$ is derivable
      in $\bpc$ if and only if $s \equivalente t$ and $s,t$ are total 
      data terms. \hfill \square

\end{des}
\end{theorem}

In  the  rest  of  the  paper,  when  we  write $e \rightarrow_{{\cal P}} t$ 
(respect. $e \lazo_{{\cal P}} e'$\/) we mean  that
$e \rightarrow t$ (respect. $e \lazo e'$\/)
is derivable from program $\cal P$ in $\bpc$  or $\gpc$\/.

At this point, we can give an example that justifies 
why we require  left-linear  defining rules 
and safe data substitutions for the construction of ${[C]}_{\ecuacional}$
in the inequational calculus.

\begin{exmp} \label{conjuntos} Let ${\cal P}$ be the program
obtained by extending Example \ref{tiposecuaciones} (1) with
the following  type declarations and defining rules for functions: 
 
 \vspace*{-0.50cm}

\begin{center}
\noindent
\( \begin{array}{lllll}
\left.
\begin{array}{llll}
\textsf{eq}: (\alpha,\alpha) \rightarrow {\it Bool} \\
eq(x,x) \rightarrow {\it True}
\end{array} 
\right.
& \left.
\begin{array}{lll}
\\
\ \ \ \textsf{unit,duo}: {\it Set}(\alpha) \rightarrow {\it Bool} \\
\ \ \ unit(\{x\}) \rightarrow {\it True} \\
\ \ \ duo(\{x,y\}) \rightarrow {\it True}
\end{array} \right.
&
\left.
\begin{array}{lll}
\ \ \ \textsf{om}: \rightarrow \alpha \\
\ \ \ om \rightarrow om
\end{array} \right.
\end{array} \)
\end{center}

\noindent
Note that the defining rule for \textsf{eq} is not left-linear and
thus illegal. If it were allowed, we would obtain
$eq(e,e') \rightarrow_{\cal P} {\it True}$ for arbitrary $e,e' \in 
\expresionesp$
(by using $e \rightarrow_{\cal P} \perp$, $e' \rightarrow_{\cal P}
\perp$ and
$eq(\perp,\perp) \rightarrow_{\cal P} {\it True}$).

On the other hand, if we would define $\aproximado$ in such 
a way that $\{\perp,\perp\} \aproximado \{\perp\}$ could be
derived as  an instance of the equation $\{x,x\mid{\it zs}\}
\approx \{x\mid{\it zs}\}$, we could use 
${\it True} \rightarrow_{\cal P} \perp$, ${\it False} \rightarrow_{\cal P} \perp$ and
$unit(\{\perp\}) \rightarrow_{\cal P} {\it True}$ for obtaining 
$unit(\{{\it True},{\it False}\})$
 $\rightarrow_{\cal P} {\it True}$, which is not expected as a reasonable consequence
from \textsf{unit}'s defining rule.

Finally, note that the inequational calculus permits
$\{\perp\} \aproximado \{\perp,\perp\}$. We can combine this
with $om \rightarrow_{\cal P} \perp$ and
$duo(\{\perp,\perp\}) \rightarrow_{\cal P} {\it True}$ to obtain
$duo(\{om\})$ $\rightarrow_{\cal P} {\it True}$, which does not
contradict our intuitive understanding of the program. \hfill \blacksquare

\end{exmp}

To conclude this section, we give a type preservation result.
We need some auxiliary lemmas and notation. Let $V$ be an environment,
$\{x_1, \ldots, x_m\}$ a set of data variables 
and $\tau_1, \ldots, \tau_m \in \tipos$\/.
$V[x_1:\tau_1, \ldots, x_m:\tau_m]$  denotes
the environment verifying the following conditions:

\begin{des}

\item for all $x_i$, $1 \leq i \leq m$,
      $x_i :\tau_i \in V[x_1:\tau_1, \ldots, x_m:\tau_m]$\/;
\item for all $x \in {\it DVar}$ such that $x \not \in \{x_1, \ldots, 
x_m\}$,
      $x:\tau \in V[x_1:\tau_1, \ldots, x_m:\tau_m]$ if and only
       if $x:\tau \in V$\/.
     \end{des}

Given $\sigma_{t} \in {\it TSub}$\/, we will write
$V\sigma_{t}$ to denote the environment
$\{x:\tau\sigma_{t} \ \mid \ x:\tau \in V\}$. Finally,
given $X \subseteq {\it DVar}$ and
two environments $V$ and $V'$, the notation
$V = V'[X]$ will mean that
for all $x \in X$,
$x:\tau \in V$ if and only if $x:\tau \in V'$\/.

Next, we  present four lemmas that can be easily proved by 
structural induction.

\begin{lemma}
\label{lema2}
Let $V$ be an environment and $e \in \expresionesp$. If $e$ has type 
$\tau$ in $V$, i.e., 
$e \in {\it Expr}_{\Sigma_{\perp}}^{\tau}(V)$, 
then $e$ has type $\tau\sigma_{t}$ in $V\sigma_{t}$, i.e., $e \in
{\it Expr}_{\Sigma_{\perp}}^{\tau\sigma_{t}}(V\sigma_{t})$,
for all $\sigma_{t} \in {\it TSub}$. \hfill \square
\end{lemma}

\begin{lemma}
\label{lema1}
Let $V$, $V_{0}$ be two environments and  $\sigma_{t} \in {\it TSub}$\/.
For all $e \in \expresionesp$ such that all function
symbols occurring in $e$ have a transparent principal type: If
$e \in {\it Expr}_{\Sigma_{\perp}}^{\tau}(V_0)
\cap {\it Expr}_{\Sigma_{\perp}}^{\tau\sigma_{t}}(V)$,
then
$V=V_0\sigma_{t}[{\it dvar(e)}]$. \hfill \square

\end{lemma}

\begin{lemma}
\label{lema4}
Consider $e \in \expresionesp$ and $\sigma_{d} \in {\it DSub}_{\perp}$\/
such that $\{x_1, \ldots,$  $x_m\} =
\{x \in dvar(e) \ \mid \ x\sigma_{d} \not = x\}$ and
$x_i\sigma_{d} = t'_i$\/, $1 \leq i \leq m$.
Let $V$ be an environment such that
$t'_i \in {\it Term}_{\Sigma_{\perp}}^{\tau'_i}(V)$, 
      $1 \leq i \leq m$, and
$V[x_1:\tau'_1, \ldots, x_m:\tau'_m] \deduce e:\tau'$.
Then $e\sigma_{d} \in {\it Expr}_{\Sigma_{\perp}}^{\tau'}(V)$. \hfill \square

\end{lemma}

\begin{lemma} \label{lema5}
Consider $e \in \expresionesp$  and
$\sigma_{d} \in {\it DSub}_{\perp}$ such that
$\{x_1, \ldots, $ $x_m\} =
\{x \in dvar(e) \ \mid \ x\sigma_{d} \not = x\}$ and
$x_i\sigma_{d} = t'_i$, $1 \leq i \leq m$\/. Assume
that $x_i$, $1 \leq i \leq m$, occurs at most once in $e$.
Let $V$ be an environment such that
$e\sigma_{d} \in {\it Expr}_{\Sigma_{\perp}}^{\tau'}(V)$.
Then, there exist $\tau'_i \in \tipos$, $1 \leq i \leq m$,
such that $t'_i \in {\it Term}_{\Sigma_{\perp}}^{\tau'_i}(V)$ 
and
$V[x_1:\tau'_1, \ldots, x_m:\tau'_m] \deduce e:\tau'$.
\hfill \square

\end{lemma}

The following two lemmas establish that the well-typedness of
program rules and equational axioms is preserved by type instances.

\begin{lemma}[Type preservation by type instances]
\label{lema3}

\begin{des}
\item[(a)] Let $f(t_1, \ldots, t_n) \rightarrow r \Leftarrow C$ be
a well-typed and regular program rule with principal type: 
$f: (\tau_1, \ldots,$ $ \tau_n) \rightarrow \tau \in {\it FS}$.
Let $V$ and  $\sigma_{t}$ be an environment and
a type substitution respectively, such that 
$t_i \in {\it Term}_{\Sigma_{\perp}}^{\tau_i\sigma_{t}}(V)$,
$1 \leq i \leq n$. Then
$r \in {\it Expr}_{\Sigma_{\perp}}^{\tau\sigma_{t}}(V)$;

\item[(b)] Let $c(t_1, \ldots, t_n) \approx d(s_1, \ldots, s_m)$ be
a well-typed  regular equation such that $c$ and $d$ have 
principal types:
$c: (\tau_1, \ldots, \tau_n) \rightarrow \tau,
d : (\tau'_1, \ldots, \tau'_m) \rightarrow \tau \in {\it DC}$
(up to variants).
Let $V$ be an environment and $\sigma_{t}
\in {\it TSub}$ such that 
$t_i \in {\it Term}_{\Sigma_{\perp}}^{\tau_i\sigma_{t}}(V)$,
$1 \leq i \leq n$. Then $d(s_1, \ldots, s_m) \in
{\it Term}_{\Sigma_{\perp}}^{\tau\sigma_{t}}(V)$. \hfill \square 
\end{des}
\end{lemma} 

\begin{proof}
 (a)  (respect. (b)) follows from the well-typedness of
  $f(t_1, \ldots, t_n) \rightarrow r \Leftarrow C$
  (respect. $c(t_1, \ldots, t_n) \approx d(s_1, \ldots, s_m)$),
  Lemmas \ref{lema1} and \ref{lema2} and  
  $dvar(r)$  $\subseteq \bigcup_{i=1}^n dvar(t_i)$
  (respect. $dvar(c(t_{1}, \ldots, t_{n}))
   = dvar(d(s_1, \ldots, s_m))$).    
\end{proof}

 \medskip

The next lemma extends the previous one, and ensures
that well-typedness is not only  preserved by type instantiation
 but also by well-typed data instantiation. The complete proof can
 be found in Appendix \ref{demos}.

\begin{lemma}[Type preservation by instances] \label{instancias}

\begin{des} 
\item[(a)] Let $f(t_1, \ldots, t_n) \rightarrow r \Leftarrow C$ be
      a well-typed and regular defining rule with principal type: 
      $f: (\tau_1, \ldots, \tau_n)$ $ \rightarrow \tau \in {\it FS}$. 
      Let $V$ and $\sigma=(\sigma_{t},\sigma_{d})$ be an environment and
      a substitution, respectively. 
      If $t_i\sigma_{d} \in {\it Term}_{\Sigma_{\perp}}^{\tau_i\sigma_{t}}(V)$\/, 
            $1 \leq i \leq n$,
             then $r\sigma_{d} \in {\it 
Expr}_{\Sigma_{\perp}}^{\tau\sigma_{t}}(V)$\/.

\item[(b)] Let $c(t_1, \ldots, t_n) \approx d(s_1, \ldots, s_m)$
      be a well-typed   regular axiom  such that
      $c$ and $d$ have principal types:
      $c: (\tau_1, \ldots,$ $ \tau_n) \rightarrow \tau,
      d: (\tau'_1, \ldots, \tau'_m) \rightarrow \tau \in {\it DC}$
      (up to variants).
      Let $V$ and $\sigma=(\sigma_{t},\sigma_{d})$ be an environment and
      a substitution, respectively. If $t_i\sigma_{d} \in {\it 
Term}_{\Sigma_{\perp}}^{\tau_i\sigma_{t}}(V)$\/, 
         $1 \leq i \leq n$\/,
        then $d(s_1, \ldots, s_m)\sigma_{d} 
        \in {\it Term}_{\Sigma_{\perp}}^{\tau\sigma_{t}}(V)$\/. \hfill \square  
\end{des}

\end{lemma}

Finally, here we have the theorem which ensures the type preservation 
result we were looking for.

\begin{theorem}[Type preservation]
\label{conservacion}
Let $\programa$ be a well-typed stro\-ngly regular program.
Let $V$ be an environment. If $e \rightarrow_{{\cal P}} e'$ and
$e \in {\it Expr}^{\tau}_{\Sigma_{\perp}}(V)$
then $e' \in {\it Expr}^{\tau}_{\Sigma_{\perp}}(V)$\/, for all $\tau 
\in \tipos$\/. \hfill \square

\end{theorem}

\begin{proof}
  The proof proceeds by induction on the structure
     of the $\bpc$ derivation 
       associated to $e \rightarrow_{{\cal P}} e'$. We analyze the last
       inference rule applied in such a proof. For rules
       \textsf{(B)} and \textsf{(RF)} the result is trivial. For
       rules \textsf{(TR)} and \textsf{(MN)} it is enough to apply induction hypothesis.
       It remains to prove rules \textsf{(MUT)} and \textsf{(R)}.

       \begin{des}
       
       \item[\textsf{(MUT)}.] Then  $e  = c(t_1, \ldots, 
       t_n)\sigma_{d}$,
       $e'  = d(s_1, \ldots, s_m)\sigma_{d}$ and $e \aproximado e' \in 
       [{\cal C}]_{\ecuacional}$, for some data substitution
        $\sigma_{d} \in {\it DSub}_{\perp}$ being safe for 
        $c(t_1, \ldots, t_n)$\/.  Suppose that
       $c : (\tau'_1, \ldots, \tau'_n)$ $ \rightarrow \tau',
       d: (\tau''_1, \ldots, \tau''_m) \rightarrow \tau' \in {\it DC}$
       (up to renaming). Since
       $e \in {\it Term}^{\tau}_{\Sigma_{\perp}}(V)$,  then there 
       exists 
       $\sigma_{t} \in {\it TSub}$  such that
       $\tau  = \tau'\sigma_{t}$\/ and
        $t_{i} \sigma_{d} \in 
       {\it Term}^{\tau'_{i}\sigma_{t}}_{\Sigma_{\perp}}(V)$, $1 \leq 
       i \leq n$. From Lemma \ref{instancias} (b), 
       it holds, for the substitution $(\sigma_{t},\sigma_{d})$, that   
       $d(s_1, \ldots, s_m)\sigma_{d}
       \in {\it Term}^{\tau'\sigma_{t}}_{\Sigma_{\perp}}(V)$\/.

      \item[\textsf{(R)}.] Then  $e  = f(t_1, \ldots, 
       t_n)\sigma_{d}$, $e'  = e''\sigma_{d}$ and  $f(t_{1}, \ldots, t_{n})\sigma_{d}  \rightarrow  
       e''\sigma_{d} \Leftarrow C\sigma_{d}  \in 
       [{\cal R}]_{\rightarrow}$, for some data substitution
        $\sigma_{d} \in {\it DSub}_{\perp}$.
        Suppose that
       $f : (\tau'_1, \ldots, \tau'_n)$ $ \rightarrow \tau'  \in {\it 
       FS}$. Since
       $e \in {\it Term}^{\tau}_{\Sigma_{\perp}}(V)$,  then there 
       exists  $\sigma_{t} \in {\it TSub}$  such that
       $\tau  = \tau'\sigma_{t}$ and $t_{i} \sigma_{d} \in 
       {\it Term}^{\tau'_{i}\sigma_{t}}_{\Sigma_{\perp}}(V)$, $1 \leq 
       i \leq n$. 
        From Lemma \ref{instancias} (a), 
       it holds, for the substitution $(\sigma_{t},\sigma_{d})$, that   
       $e'' \sigma_{d}
       \in {\it Expr}^{\tau'\sigma_{t}}_{\Sigma_{\perp}}(V)$\/. 
 \end{des}
 \end{proof}

Note that this type preservation theorem does not hold   for non-regular 
axioms neither collapsing regular axioms, as the following example shows.

\begin{exmp} \label{r3} Let us consider the signature $\Sigma$\/ from 
Example \ref{tiposecuaciones} (1)
and the empty environment $V$\/. Assuming the non-regular axiom
${\it Suc}(x) \approx {\it Suc}(y)$\/, we obtain ${\it Suc}({\it Zero})
\rightarrow_{{\cal P}} {\it Suc}({\it True})$\/, where 
${\it Suc}({\it Zero}) \in {\it Term}_{\Sigma}^{\it Nat}(V)$\/
but ${\it Suc}({\it True}) \not \in {\it Term}_{\Sigma}^{\it Nat}(V)$\/.
Taking the collapsing regular axiom $x \approx {\it Suc}(x)$\/, we get
${\it True} \rightarrow_{{\cal P}} {\it Suc}({\it True})$\/, where
${\it True} \in {\it Term}_{\Sigma}^{\it Bool}(V)$\/ but
${\it Suc}({\it True}) \not \in {\it Term}_{\Sigma}^{\it 
Bool}(V)$\/. \hfill \blacksquare  
\end{exmp}

\section{Model-theoretic Semantics}
\label{modelodeterminos}

In this section  we will present a model-theoretic semantics, showing  also
its relation  to  the rewriting calculi from Section~\ref{calculos}. 
First, we recall some basic notions from  the theory of semantic 
domains~\cite{Sco82,GS90}.

A {\em poset} with bottom $\perp$\/ is any set $S$\/ partially ordered by
$\sqsubseteq$\/, with least element $\perp$\/.  $\textsf{Def}(S)$\/ denotes the
set of all maximal elements $u \in S$\/, also called {\em totally defined}.
Assume $X \subseteq S$\/. $X$\/ is a {\em directed set} iff for all $u,v \in
X$\/ there exists $w \in X$\/ s.t. $u,v \sqsubseteq w$\/. $X$\/ is a {\em cone}
iff $\perp \in X$\/ and $X$\/ is  downwards closed w.r.t. $\sqsubseteq$\/.
$X$\/ is an {\em ideal} iff $X$\/ is a directed cone. We write   
${\cal C}(S)$\/ and ${\cal I}(S)$\/ for  the  sets  of 
cones  and ideals  of  $S$,  respectively. ${\cal I}(S)$ ordered by set
inclusion $\subseteq$\/ is a poset with bottom $\{\perp\}$\/, called the
{\em ideal completion} of $S$\/. Mapping each $u \in S$ into the principal
ideal $\langle u \rangle = \{ v \in S \mid v \sqsubseteq u \}$\/ gives an order
preserving embedding.

A poset $C$ with bottom is a {\em complete partial order} (in 
short, {\em cpo}) iff $C$ has a least upper bound $\bigsqcup C$ (also called {\em 
limit}) for every directed set $D \subseteq C$. An element $u \in C$ 
is called {\em finite} if and only if whenever $u \sqsubseteq 
\bigsqcup D$ for a non-empty directed $D$, there exists $x \in D$ 
such that $u \sqsubseteq x$\/.
It is known that, for any poset with 
bottom $S$,  
${\cal I}(S)$\/ is the least cpo containing $S$\/. Moreover, ${\cal I}(S)$ is 
an algebraic cpo whose finite 
elements correspond to  the principal ideals $\langle x \rangle$, 
$x \in S$\/; see for instance \cite{Mol85}. In particular, elements
$x \in \textsf{Def}(S)$ generate finite and total elements $\langle x 
\rangle$ in the ideal completion.

As in \cite{GHLR99}, we will  use
posets instead  of algebraic cpo's. Such posets will
provide only finite semantic values. The ideal completion of $S$ 
might supply the missing infinite values, but in fact finite values 
are enough for describing the semantics of our programs.
To  represent  non-deterministic lazy 
functions, we use models with posets as carriers, interpreting function  symbols 
as monotonic mappings from elements to cones. 
For given posets $D$\/ and $E$\/, we define the set of all
{\em non-deterministic functions} from $D$\/ to $E$\/ as

 \vspace*{-0.20cm}

\[ {[D \rightarrow_{nd} E]}= \{f: D \rightarrow {\cal C}(E) 
       \ \mid \ \forall u,u' \in D: (u \sqsubseteq_D u' \Rightarrow
                   f(u) \subseteq f(u'))\} \]

\noindent
and the set of all {\em deterministic functions} from $D$\/ to $E$\/ as

\vspace*{-0.20cm}

\[ {[D \rightarrow_{d} E]}= \{f \in {[D \rightarrow_{nd} E]} 
       \ \mid \ \forall u \in D: f(u) \in {\cal I}(E)\} \]

\smallskip 

When given some fixed arguments, a deterministic function $f$  will  return a directed
set of partial values.  Hence, after performing an
ideal completion, deterministic functions become continuous
mappings between algebraic cpos.  On the other hand, a non-deterministic function $f$ 
returns cones, which are the same as elements of Hoare's 
powerdomain~{n}\cite{Sco82,Win85}. Therefore, after performing an ideal 
completion, non-deterministic functions  become continuous functions 
taking values in a powerdomain.

Moreover, any non-deterministic function $f$ can be extended
to a monotonic mapping
$f^*: {\cal C}(D) \rightarrow {\cal C}(E)$ defined
as $f^*(C)= \bigcup_{c \in C} f(c)$\/. Abusing of
notation, we will identify $f$ with its extension
$f^*$ in the sequel.

We are now prepared to introduce our algebras, combining
ideas from \cite{GHLR99,Smo89}.

\begin{definition}[Polymorphically typed algebras]
\label{algebras}
Let $\Sigma$ be
a polymorphic signature.
A Polymorphically Typed algebra ($PT$-algebra) $\algebra$ has the following
structure:

\vspace*{-0.25cm}

{\small
\[ \begin{array}{llll}
\algebra &=& \langle D^{\algebra},T^{\algebra}, \relacion ,
\{K^{\algebra}\}_{K \in  {\it TC}},
\{c^{\algebra} \}_{c \in  {\it DC}},
\{f^{\algebra} \}_{f \in  {\it FS}}
\rangle
\end{array} \]
}

\vspace*{-0.50cm}

\noindent
where: 

\begin{des}
\item[(1)] $D^{\algebra}$ (data universe) is a poset with partial order $\orden$ and
            bottom element $\botom$, and $T^{\algebra}$ (type universe) is a set;

\item[(2)] $\relacion \subseteq D^{\algebra} \times T^{\algebra}$ is a binary
        relation such that for all ${\ell} \in T^{\algebra}$, the extension
        of ${\ell}$ in $\algebra$\/, defined as
        ${\cal E}^{\algebra}({\ell})
          =
      \{u \in D^{\algebra} \ \mid \ u \relacion {\ell}\}$ is a cone in $D^{\algebra}$\/;

\item[(3)] For each $K \in {\it TC}^{n}$\/, 
        $K^{\algebra}:(T^{\algebra})^n \rightarrow T^{\algebra}$ (simply
         $K^{\algebra} \in T^{\algebra}$ if $n=0$\/);

\item[(4)] for  all $c: (\tau_1, \ldots, \tau_n) \rightarrow \tau \in
    {\it DC}_{\perp}$\/, $c^{\algebra} \in {[(D^{\algebra})^n \rightarrow_{d} D^{\algebra}]}$
    satisfies: For all $u_1, \ldots, u_{n} \in D^{\algebra}$, there exists
      $v \in D^{\algebra}$ such that
      $c^{\algebra}(u_1, \ldots, u_n)= \langle v\rangle$\/. Moreover, 
      if $u_1, \ldots, u_{n} \in   \textsf{Def}(D^{\algebra})$ then   
       $v \in   \textsf{Def}(D^{\algebra})$\/;

\item[(5)] for all $f: (\tau'_1, \ldots, \tau'_m) \rightarrow \tau' \in
    {\it FS}$, $f^{\algebra} \in 
   {[(D^{\algebra})^m \rightarrow_{nd} D^{\algebra}]}$\/.  \hfill \square

\end{des}

\end{definition}

Some comments may help to understand this definition. Items (4) and 
(5) mean that constructors and function symbols are interpreted as 
continuous operations (when moving to the ideal completion). Moreover 
item (4) requires that data constructors are interpreted as 
deterministic operations which preserve finite and total elements in 
the ideal completion. As in \cite{Smo89}, $\relacion$ represents the 
membership  relation between data and types. Item (2) requires the 
extensions of types to be cones, which is a natural condition. In 
particular, $\perp^{_\algebra}$ must belong to all types, which is 
consistent with $\perp$'s principal type declaration $\perp: 
\rightarrow \alpha$.

In order to interpret expressions in an algebra $\algebra$
we use {\em valuations} $\eta = (\eta_{t},\eta_{d})$\/, where
$\eta_{t} : {\it TVar} \rightarrow T^{\algebra}$\/ is a {\em type 
valuation} and $\eta_{d} : {\it DVar} \rightarrow D^{\algebra}$\/
is a {\em data valuation}. $\eta_{d}$\/ is called {\em totally defined} 
iff $\eta_{d}(x) \in \textsf{Def}(D^{\algebra})$\/,
for all $x \in {\it DVar}$\/; and $\eta_{d}$\/ is called {\em safe} for
a given $t \in \terminosp$
iff $\eta_{d}(x) \in \textsf{Def}(D^{\algebra})$\/, for all $x \in dvar(t)$
s.t. $x$ has more than one occurrence in $t$\/. ${\it Val}(\algebra)$\/
denotes the set of all valuations over $\algebra$\/.

For a given $\eta = (\eta_{t}, \eta_{d}) \in {\it Val}(\algebra)$\/,
{\em type denotations} 
 $\C \tau \J^{\algebra} \eta =_{{\it Def}} \C \tau \J^{\algebra} \eta_{t} 
 \in T^{\algebra}$\/ and {\em expression denotations}
 $\C e \J^{\algebra}\eta =_{{\it Def}} \C e \J^{\algebra}\eta_{d}
 \in  {\cal C}(D^{\algebra})$\/ are defined recursively as follows:

\begin{des}

\item $\C \alpha \J^{\algebra}\eta_{t} = \eta_{t}(\alpha)$\/,  where
         $\alpha \in {\it TVar}$;
 \item $\C K(\tau_1, \ldots, \tau_n) \J^{\algebra}\eta_{t} =
        K^{\algebra}(\C \tau_1 \J^{\algebra}\eta_{t}, \ldots, \C \tau_n 
       \J^{\algebra}\eta_{t})$, where $K \in {\it TC}^n$ and
  $\tau_i \in \tipos$, $1 \leq i \leq n$\/; 
\item $\C \perp \J^{\algebra}\eta_{d} = \{ \botom \}$;
\item $\C x \J^{\algebra}\eta_{d} =
      \langle \eta_{d}(x)\rangle $,  where $x \in {\it DVar}$\/;
\item $\C h(e_1, \ldots, e_n) \J^{\algebra}\eta_{d} =
        h^{\algebra}(\C e_1 \J^{\algebra}\eta_{d}, \ldots, \C e_n 
       \J^{\algebra}\eta_{d})$, where $h \in {\it DC}^n \cup {\it FS}^n$\/, 
  $e_i \in \expresionesp$, $1 \leq i \leq n$.

\end{des}

As in \cite{GHLR99},  the following result can  be proved easily by 
structural induction. 

\begin{prop}[Properties of denotations]  \label{eval}

\begin{des}

\item[(a)] If $f^{_\algebra}$ is deterministic for every $f 
\in {\it FS}$ occurring in an expression $e$, then $\C e 
\J^{_\algebra} \eta_{d}$ is an ideal;

\item[(b)] For every data term $t$, $\C t 
\J^{_\algebra}\eta_{d}$ is a principal ideal $\langle v \rangle$. 
Moreover,   $u \in \textsf{Def}(D^{\algebra})$ if $t$ is total and $\eta_{d}$ is
totally defined. \hfill \square   

\end{des}

\end{prop}

We are particularly interested in those $PT$-algebras that are well-behaved
w.r.t. types.

\begin{definition}[Well-typed $PT$-algebras and valuations]

\begin{des} 
\item $\algebra$\/ is {\em well-typed} 
if  for all $h: (\tau_1, \ldots , \tau_n) \rightarrow \tau \in 
{\it DC}_{\perp} \cup {\it FS}$\/ and for every type valuation $\eta_{t}$\/,
it holds that $h^{\algebra}({\cal E}^{\algebra}(\C \tau_1 \J^{\algebra} \eta_{t}),
 \ldots, 
{\cal E}^{\algebra}(\C \tau_n \J^{\algebra} \eta_{t}))
\subseteq {\cal E}^{\algebra} (\C \tau\J^{\algebra} \eta_{t})$\/;

\item $\eta = (\eta_{t},\eta_{d}) \in {\it Val}(\algebra)$\/ 
 is {\em well-typed} w.r.t. an environment $V$\/ iff 
for every $x:\tau \in V$, it holds that
$\eta_{d}(x) \in {\cal E}^{\algebra}(\C \tau \J^{\algebra}\eta_{t})$.  \hfill \square
\end{des}
\end{definition}

The next auxiliary lemma is needed to prove
some of the results presented later. It can be proved easily by
structural induction.

\begin{lemma}[Substitution lemma]
\label{auxiliar}
Let $\eta=(\eta_{t},\eta_{d})$ be a valuation over a ${\it PT}$-algebra
$\algebra$. For any $\tau \in \tipos$, $e \in \expresionesp$
and substitution $\sigma=(\sigma_{t},\sigma_{d})$, it holds that
$\C \tau \J^{\algebra}\eta_{t_{\sigma_{t}}} = \C \tau\sigma_{t} \J^{\algebra} \eta_{t}$ 
and $\C e \J^{\algebra} \eta_{d_{\sigma_{d}}}  = \C e\sigma_{d} \J^{\algebra} 
\eta_{d}$, where $\eta_{\sigma}=(\eta_{t_{\sigma_{t}}},\eta_{d_{\sigma_{d}}})$ is the uniquely determined
valuation that satisfies:
$\eta_{t_{\sigma_{t}}}(\alpha)=\C \alpha\sigma_{t} \J^{\algebra}\eta_{t}$,
for all $\alpha \in {\it TVar}$, and
$\eta_{d_{\sigma_{d}}}(x) =d$, for all $x \in {\it DVar}$,
                            where $\langle  d\rangle =\C
                            x\sigma_{d} \J^{\algebra}\eta_{d}$.  \hfill \square

\end{lemma}

We can prove that
expression denotations behave as expected w.r.t. well-typed algebras and
valuations.  

\newpage

\begin{prop}
\label{compatibles}
Let $V$ be an environment. Let $\algebra$ be a 
well-typed $PT$-algebra and
$\eta=(\eta_{t},\eta_{d}) \in {\it Val}(\algebra)$
well-typed w.r.t. $V$\/. For all
$e \in {\it Expr}_{\Sigma_{\perp}}^{\tau}(V)$\/, 
$\C e \J^{\algebra} \eta_{d} \subseteq
{\cal E}^{\algebra}(\C \tau \J^{\algebra}\eta_{t})$\/. \hfill \square   
\end{prop}

\begin{proof}
By structural induction on $e$. For $e=\perp$ or
  $e=c \in {\it DC}^0$, the result follows from the
  well-typedness of $\algebra$ and Lemma \ref{auxiliar}.
  For $e=x \in {\it DVar}$, the well-typedness of $\eta$ entails the
  result. For $e=h(e_{1}, \ldots, e_{n})$, $h:(\tau'_{1}, \ldots, 
  \tau'_{n}) \rightarrow \tau' \in {\it DC} \cup {\it 
  FS}$\/, there exists $\sigma_{t} \in {\it TSub}$ such that
  $\tau =  \tau' \sigma_{t}$ and $e_{i} \in {\it 
  Term}^{\tau'_{i}\sigma_{t}}_{\Sigma_{\perp}}(V)$\/, $1 \leq i \leq 
  n$.  By induction 
  hypothesis, $\C e_{i} \J^{_\algebra} \eta_{d} \subseteq {\cal 
  E}^{_\algebra}(\C \tau'_{i}\sigma_{t} \J^{_\algebra}\eta_{t})$, $1 
  \leq i \leq n$.
    From Lemma \ref{auxiliar} we have that ${\cal 
  E}^{_\algebra}(\C \tau'_{i}\sigma_{t} \J^{_\algebra}\eta_{t})=
  {\cal 
  E}^{_\algebra}(\C \tau'_{i} \J^{_\algebra}\eta_{t_{\sigma_{t}}})$, 
  $1 \leq i \leq n$.
  The well-typedness of $\cal A$ entails 
  $h^{_\algebra}(\C e_{1} \J^{_\algebra} \eta_{d}, \ldots,
  \C e_{n} \J^{_\algebra} \eta_{d}) \subseteq {\cal 
  E}^{_\algebra}(\C \tau' \J^{_\algebra}\eta_{t_{\sigma_{t}}})$. 
  Now, the result follows from  Lemma \ref{auxiliar} and definition 
  of denotation.  
 \end{proof}

Next, we define the notion of {\em model}. Note that reduction/approximation
is interpreted as inclusion, while joinability is
interpreted as existence of a common maximal approximation.

\begin{definition}[Models of a program]
Let $\algebra$ be a $PT$-algebra. We define:

\begin{des}
 
\item[(i)] ${\cal A}$ satisfies a reduction statement $e 
\rightarrow e'$ under a valuation $\eta=(\eta_{t},\eta_{d})$ (noted by 
$(\algebra,\eta_{d}) \modelo e \rightarrow e')$ iff 
$\C e' \J^{\algebra}\eta_{d}  \subseteq \C e \J^{\algebra}\eta_{d}$;

\item[(ii)] ${\cal A}$ satisfies a joinability statement $e 
\lazo e'$ under a valuation $\eta=(\eta_{t},\eta_{d})$ (noted by  
     $(\algebra,\eta_{d}) \modelo e \lazo e'$) iff        
      $\C e \J^{\algebra}\eta_{d} \cap \C e' \J^{\algebra}\eta_{d} 
      \cap   \textsf{Def}(D^{\algebra}) \not =
      \emptyset$. Furthermore, we say that ${\cal A}$ satisfies a set 
      $C$ of joinability statements under a valuation 
      $\eta=(\eta_{t},\eta_{d})$ (noted by  
     $(\algebra,\eta_{d}) \modelo  C$) iff $(\algebra,\eta_{d}) 
     \modelo e \lazo e'$, for all $e \lazo e' \in C$;

\item[(iii)] $\algebra$ satisfies a defining rule
    $l \rightarrow r \Leftarrow C$ iff
     every $\eta=(\eta_{t},\eta_{d}) \in {\it Val}({\algebra})$ such that
      $(\algebra,\eta_{d}) \modelo C$ verifies that
       $({\algebra},\eta_{d}) \modelo l \rightarrow r$;

\item[(iv)] $\algebra$ satisfies an equation $s \approx t$ 
      iff for every $\eta=(\eta_{t},\eta_{d}) \in {\it Val}(\algebra)$\/:
       $\C s \J^{\algebra}\eta_{d}  \supseteq \C t
          \J^{\algebra}\eta_{d}$ if $\eta_{d}$ is safe for
           $s$ and
            $\C t \J^{\algebra}\eta_{d}  \supseteq \C s
          \J^{\algebra}\eta_{d}$ if $\eta_{d}$ is safe for $t$\/;

\item[(v)] Let $\programa$ be a program. $\algebra$ is a model
       of ${\cal P}$ (noted by $\algebra \modelo {\cal P}$) 
      iff $\algebra$ satisfies every defining rule in ${\cal R}$ 
      (noted by $\algebra \models {\cal R}$) and 
      every equation in ${\cal C}$ (noted by $\algebra \models {\cal 
      C}$)\/.  \hfill \square
      
\end{des}
\end{definition}

The rest of the section is devoted to the construction of
free term models, which allows us to prove soundness and completeness
of the rewriting calculi from Sect. \ref{calculos}.

\begin{definition}[Free term models]
Given a program $\programa$
and an environment $V$, we build the term model
$\modterm$ as follows:

\begin{des}

\item  \textsf{Data universe:} Let $X$ be the set of all data variables 
occurring in $V$. Then the {\em data universe} of $\modterm$ is 
$\dominio$, where 

\vspace*{-0.10cm}

\[ {\it Term}_{\Sigma_{\perp}}(X) =_{{\it Def}} 
\{t \in \terminosp \mid {\it dvar}(t) \subseteq X\} \]

\vspace*{-0.10cm}

\noindent
For all $t \in {\it Term}_{\Sigma_{\perp}}(X)$,   ${[t]}$ denotes the 
equivalence class $\{t' \in {\it Term}_{\Sigma_{\perp}}(X) \ \mid$
$ \ t \equivalente
           t'\}$\/;

\item \textsf{Type universe:}  Let $A$ be the set of type variables 
occurring in $V$. Then
   the type universe of $\modterm$ is 
    ${\it T_{\it TC}}(A)=_{{\it Def}} \{ \tau \in \tipos \ \mid \ 
    tvar(\tau) \subseteq A\}$\/;

\item For all ${[t]} \in \dominio$,
     $\tau \in {\it T}_{\it TC}(A)$\/, we define
      $[t] :^{\modterm} \tau$
       iff $t \in {\it Term}_{\Sigma_{\perp}}^{\tau}(V)$\/;

\item For all $K \in {\it TC}^n$, $\tau_i
    \in {\it T}_{\it TC}(A)$, $1 \leq i \leq n$\/:  $K^{\modterm}(\tau_1, \ldots, \tau_n)
        = K(\tau_1, \ldots, \tau_n)$\/;

\item For all $c  \in {\it DC}^n$\/, 
      $[t_i] \in \dominio$, $1 \leq i \leq n$:  $c^{\modterm}([t_1], \ldots , [t_n])
                               = \langle
      [c(t_1, \ldots, t_n)]\rangle$;

\item  For all $f \in {\it FS}^n$\/,  
       $[t_i] \in \dominio$, $1 \leq i \leq n$:
      
   \vspace*{-0.25cm}

  \[ f^{\modterm}([t_1],  \ldots , [t_n]) =
        \{[t] \in \dominio   \ \mid \  f(t_1, \ldots, t_n)
                   \rightarrow_{{\cal P}} t\} \]

\vspace*{-0.10cm}

\item $\perp^{\modterm} = [\perp]$ is the bottom
   element, whereas the partial order is defined as follows:
For all $[s],[t] \in \dominio$,
      $[s] \sqsupseteq^{\modterm} [t]$ iff $s \aproximado t$\/.
\hfill \square
      
 \end{des}
 
\end{definition}

The following theorem ensures that $\modterm$ is a well-defined 
algebra in presence of a strongly regular set of equational axioms. The 
complete proof of the theorem can be found in Appendix \ref{demos}.

\begin{theorem}[$\modterm$ is a well-typed $PT$-algebra] \label{ape1} 
Given a program $\programa$ where ${\cal C}$ is
strongly regular and well-typed, it holds that
$\modterm$ is a $PT$-algebra. Moreover, if all rules
in ${\cal R}$ are regular and well-typed then $\modterm$ is a 
well-typed $PT$-algebra.  \hfill \square
\end{theorem}

The relationship 
between semantic validity in $\modterm$\/ and $\gpc$-derivability 
(which allows us to prove the adequateness theorem below) can be
characterized as follows:

\begin{lemma}[Characterization lemma]
\label{validez}
Consider a program {\small $\programa$}  such that
${\cal C}$ is strongly regular and well-typed.
Let {\small ${[\sigma]} = 
(\sigma_{t},{[\sigma_{d}]}) \in {\it Val}(\modterm)$}\/ be a 
valuation,
represented by a substitution $\sigma = (\sigma_{t},\sigma_{d})$\/.
Then for all  $e,a,b \in {\it Expr}_{\Sigma_{\perp}}(X)$\/,
$t \in {\it Term}_{\Sigma_{\perp}}(X)$\/:

\begin{des}
\item[(a)] $[t] \in \C e \J^{\modterm}[\sigma_{d}]$  if and only if
            $e\sigma_{d} \rightarrow_{{\cal P}} t$;
\item[(b)]  $(\modterm, [\sigma_{d}]) \modelo e \rightarrow 
                 t$ if and only if $e\sigma_{d} \rightarrow_{{\cal P}} 
                 t\sigma_{d}$;
\item[(c)]  $({\modterm},[\sigma_{d}]) \modelo a \lazo b$ if and only if
              $a\sigma_{d} \lazo_{{\cal P}} b\sigma_{d}$\/. \hfill \square 

\end{des}
\end{lemma}

\begin{theorem}[Adequateness of $\modterm$]
\label{adecuacion}
Let $\programa$ be a program such that 
${\cal C}$ is strongly regular and well-typed. Then:

\begin{des}
\item[(1)] $\modterm \modelo {\cal P}$.
\item[(2)] For any $\varphi \equiv  e \rightarrow t$ or
           $\varphi \equiv e \lazo e'$\/, where
           $e,e' \in {\it Expr}_{\Sigma_{\perp}}(X)$ and
           $t \in  {\it Term}_{\Sigma_{\perp}}(X)$\/,       
           the following
       statements are equivalent:  
   
   \begin{des}
    \item[(2.1)] $\varphi$ is derivable in $\gpc$
           (or equivalently, in $\bpc$\/);
    \item[(2.2)] $(\algebra,\eta_{d}) \modelo \varphi$, for all
      $PT$-algebra $\algebra$ such that
      $\algebra \modelo {{\cal P}}$ and for all totally
      defined data substitution $\eta_{d}$;  

   \item[(2.3)] $(\modterm,[id]) \modelo \varphi$,
    where $id$ is the identity partial data substitution
    defined as
    $id(x)=x$\/, for all $x \in X$.  \hfill \square

\end{des}
\end{des}
\end{theorem}

\newpage

\begin{proof}
To prove (1), we need to prove that
  $\modterm \modelo {\cal C}$ and $\modterm \modelo
  {\cal R}$\/.

\noindent
Given $s \approx t \in {\cal C}$ and $\sigma_{d}$ safe for $s$,
$s\sigma_{d} \rightarrow_{{\cal P}} t\sigma_{d}$ holds because of
$s\sigma_{d} \ecuacional t\sigma_{d} \in [{\cal C}]_{\ecuacional}$. By 
Lemma \ref{validez} (b) we get $(\modterm, [\sigma_{d}]) 
\modelo s \rightarrow t$, i.e., $ \C s \J^{\modterm}[\sigma_{d}] 
\supseteq \C t \J^{\modterm}[\sigma_{d}]$. Similarly, assuming that 
    $[\sigma_{d}]$ is safe
  for $t$\/, it can be proved that 
  $\C s \J^{\modterm}[\sigma_{d}] \subseteq \C t 
    \J^{\modterm}[\sigma_{d}]$\/. Hence $\modterm \models {\cal C}$.

\noindent
Consider now
a defining rule $f(t_1, \ldots, t_n) \rightarrow r \Leftarrow C$
and a data valuation $[\sigma_{d}]$ over $\modterm$ such that
$({\modterm},[\sigma_{d}]) \modelo C$.  
If $({\modterm},[\sigma_{d}]) \modelo C$ then,
Lemma \ref{validez} (c) entails that
for all $a \lazo b \in C$,
$a\sigma_{d} \lazo_{\cal P} b\sigma_{d}$.
On the other hand, for any $[t] \in \C r \J^{\modterm}[\sigma_{d}]$,
Lemma \ref{validez} (a) entails  
$r\sigma_{d} \rightarrow_{\cal P} t$\/.
Applying the $\gpc$-rule \textsf{(OR)} with the
instance program rule $f(t_1, \ldots, t_n)\sigma_{d} \rightarrow
r\sigma_{d} \Leftarrow C\sigma_{d}$, we conclude that
$f(t_1, \ldots, t_n)\sigma_{d} \rightarrow_{\cal P} t$.
From Lemma \ref{validez} (a), we get 
$[t] \in \C f(t_1, \ldots ,t_n) \J^{\modterm}[\sigma_{d}]$\/, i.e.
$\C r \J^{\modterm}[\sigma_{d}]  \subseteq 
\C f(t_1, \ldots ,t_n) \J^{\modterm}[\sigma_{d}]$. Hence, $\modterm 
\models {\cal R}$\/.

\medskip

\noindent
(2.1) $\Rightarrow$ (2.2).  This can be proved by induction 
on the length of a $\gpc$-proof for $\varphi$. The assumption that 
$\eta_{d}$ is totally defined is needed when dealing with inference 
rule {\sf (J)}. See Theorem 5.1 in \cite{GHLR99} for a similar proof.

\medskip

\noindent
(2.2) $\Rightarrow$ (2.3). From (1),
${\modterm} \modelo {\cal P}$. From (2.2),
it holds that $({\modterm},[id])$  $\modelo \varphi$, since $[{\it id}]$ 
is totally defined.

\medskip

\noindent
(2.3) $\Rightarrow$ (2.1). This follows from Lemma 
\ref{validez} (b) and (c), taking ${\it id}$ 
for $\sigma_{d}$. 
\end{proof}

Theorem \ref{adecuacion} implies soundness and completeness of the 
rewriting calculi  w.r.t. semantic validity in all models. 
Moreover, the theorem also says that validity in the term model 
$\modterm$ characterizes validity in all models. For this reason,  
$\modterm$ can be regarded as the intended 
(canonical) model of the program $\cal P$\/. More precisely, any given 
$f \in {\it FS}^{n}$\/, $n \geq 0$\/, will denote a deterministic function iff 
$f^{\modterm}([t_{1}], \ldots, [t_{n}])$ is an ideal for all $t_{i} 
\in {\it Term}_{\Sigma_{\perp}}(X)$\/, $1 \leq i \leq n$\/. This property is
undecidable 
in general, but some 
decidable sufficient conditions are known which work quite well in 
practice; see e.g. the sufficient non-ambiguity conditions 
in \cite{GHR93}.

Ignoring algebraic constructors in ACRWL, there is a clear analogy 
between $\modterm$ and so-called $\cal C$-semantics \cite{FLMP93} for 
Horn clause programs. Moreover, Horn clause logic programs correspond 
to CRWL-programs \cite{GHLR99} (and thus also to ACRWL-programs) 
composed of boolean functions. For such programs, it is easily checked 
that $\modterm$ indeed corresponds to the $\cal C$-semantics. By a 
construction similar to that of $\modterm$, using the poset of ground 
partial data terms as carrier, we could obtain also an analogon of 
the least Herbrand model semantics for Horn clause logic 
programming. However, even ignoring equational axioms, $\modterm$ 
bears more interesting information due to Theorem \ref{adecuacion}.

To conclude this section, we present a categorical characterization 
of $\modterm$ as the free model of $\cal P$\/, generated by the set 
of variables contained in the environment $V$\/. We will use only very 
elementary notions from category theory; see e.g. \cite{Pie91}.
First of all, we need a suitable notion of homomorphism which follows 
the idea of {\em loose element-valued homomorphism}, in Hussmann's 
terminology; see \cite{Hus92,Hus93}.

\begin{definition}[Homomorphism]
\label{homomorfismos}
Let $\algebra$ and ${\cal B}$ be two $PT$-algebras.
A {\em homomorphism} $h: \algebra \rightarrow {\cal B}$
is any pair of mappings $(h_t,h_d)$\/, where
$h_t : T^{\algebra} \rightarrow T^{\cal B}$ and
$h_d \in {[D^{\algebra} \rightarrow_d D^{\cal B}]}$ 
which satisfies the following conditions:

\begin{des}
\item[(1)]  $h_{t}$ preserves type constructors: For all $K \in {\it TC}^n$, $\ell_1, \ldots \ell_n
     \in  T^{\algebra}$\/, $h_t(K^{\algebra}(\ell_1$ $,\ldots \ell_n))= K^{\cal B}(h_t(\ell_1),
           \ldots, h_t(\ell_n))$\/;

\item[(2)] $h_{d}$ is element valued: For all
      $u \in D^{\algebra}$, there is $v \in D^{\cal B}$ such
      that $h_d(u)= \langle v \rangle$\/;

\item[(3)] $h_d$ is strict: $h_d(\perp^{\algebra}) =
                   \langle \perp^{\cal B} \rangle$\/;

\item[(4)] $h_{d}$ preserves data constructors: For all 
$c \in {\it DC}^n$\/, $u_i \in D^{\algebra}$, $1 \leq i \leq n$\/:
$h_d(c^{\algebra}(u_1, \ldots, u_n))=
   c^{{\cal B}}(h_d(u_1), \ldots, h_d(u_n))$\/;

\item[(5)] $h_{d}$ loosely preserves defined functions: For all $f \in {\it FS}^n$\/, $u_i \in 
D^{\algebra}$, $1 \leq i \leq n$\/: 
$h_d(f^{\algebra}(u_1, \ldots, $ $u_n)) \subseteq
  f^{{\cal B}}(h_d(u_1), \ldots, h_d(u_n))$\/.

\end{des}

Moreover, $h$ is called a {\em well-typed} homomorphism
if and only if $h_d({\cal E}^{\algebra}(\ell))
\subseteq {\cal E}^{{\cal B}}(h_t(\ell))$ for
all $\ell \in T^{\algebra}$\/. \hfill \square 
\end{definition}

${\it PT}$-algebras of signature $\Sigma$ are the
objects of a category ${\it PTAlg}_{\Sigma}$ whose
arrows are the homomorphisms from
Definition \ref{homomorfismos}. The models of any
given program $\programa$ determine a full subcategory
${\it Mod}_{{\cal P}}$ of ${\it PTAlg}_{\Sigma}$\/. We can
prove the following theorem, whose complete proof is given in 
Appendix \ref{demos}.

\begin{theorem}[$\modterm$ is free] \label{modelolibre}
Let $\programa$ be a program such that
${\cal C}$ is strongly regular and well-typed.
$\modterm$ is freely generated by $V$ in ${\it Mod}_{{\cal P}}$\/, 
that is,
given any $\algebra \modelo {{\cal P}}$ and any
$\eta=(\eta_{t},\eta_{d}) \in {\it Val}({\algebra})$\/ such that 
$\eta_{d}$ is totally defined, 
there exists a unique homomorphism
$h:\modterm \rightarrow {\algebra}$
extending $\eta$\/, i.e. 
such that  
$h_t(\alpha)=\eta_{t}(\alpha)$, for all $\alpha \in A$ 
and
$h_d([x])=\langle  \eta_{d}(x)\rangle$\/,  
for all $x \in X$\/.
Moreover, if ${\algebra}$ and $\eta$ are well-typed
then $h$ is a well-typed homomorphism. \hfill \square  

\end{theorem}

The intuitive meaning of Theorem \ref{modelolibre} is that $\modterm$ 
behaves as the ``least term algebra that is a model of $\cal P$''. An 
alternative characterization of $\modterm$ as the least fixpoint of a 
continuous transformation that maps term algebras to term algebras is 
also possible, as shown in \cite{MP97} for the restriction of 
our framework to an untyped language with free data constructors.

\section{A Lazy Narrowing Calculus for Goal Solving}
\label{goalsolving}

This section presents a {\em Lazy Narrowing Calculus based on
Equational Constructors} ($\calculo$ for short). This calculus provides a goal solving 
procedure that
combines lazy narrowing  (in the spirit of \cite{GHLR99,techanus})
with unification modulo a set of equational axioms ${\cal C}$
(in the line of \cite{JK91,Soc94}).  Differently 
to \cite{GHLR99,techanus} 
(where data constructors are free) we require the introduction of mutation
rules (as in \cite{JK91}) for applying equational axioms to
data constructors.  With respect to \cite{JK91,Soc94}
we need the incorporation of {\em narrowing } for applying
program rules.

$\calculo$ is a quite general and expressive framework for declarative programming, based on 
algebraic 
data constructors and non-deterministic lazy functions. Nevertheless, 
there is still a big gap between our current presentation of lazy 
narrowing and an implemented system. In fact, our narrowing 
calculus $\calculo$ is not intended as an operational model, but 
rather as an abstract description of goal solving that provides a very 
convenient basis for soundness and completeness proofs, while ignoring 
control issues and implementation details.

As in \cite{GHLR99}, goals are finite conjunctions of  
approximation and joinability statements whereas 
solutions will be partial data substitutions such
that the goal affected by such a substitution is provable in
$\gpc$\/.  Due to technical reasons that will become apparent later, we  
divide $\calculo$ computations in two main phases, as in \cite{techanus}.
The first phase transforms an
initial goal $G$ into a quasi-solved goal $G'$ (only
containing variables) by applying the goal transformation rules for 
$\rightarrow$ and $\igual$ presented in Subsection 
\ref{reglasdetransformacion}.
The second phase takes
the resulting $G'$ and using  variable elimination rules, transforms it into
a solved goal which represents a solution in the sense of
Definition \ref{solucion} below.
Each transformation step using  either a $\rightarrow$ or $\lazo$ 
rule  is noted as $G \reduce G'$ whereas $G \reducevar G'$
represents  a transformation step using variable elimination rules.
A {\em derivation} for a goal is a finite sequence  of $\reduce$-steps
(named $\reduce$-derivation)
followed by a finite sequence of $\reducevar$-steps (named
a $\reducevar$-derivation). {\em FAIL}
represents an irreducible inconsistent goal used to write
{\em failure rules}.
Of course, since we work with static types, $\calculo$ will 
preserve types in the case of a well-typed admissible goal and program
(see Theorem \ref{completitud2}). As notation,
${\it dvar}(L)$ stands for the set of data variables occurring in
$L$\/, where $L$ is either a goal, a multiset of joinability/approximation
statements, a program rule or an equational axiom.

\subsection{Admissible Goals}

The next definition introduces formally the notion of (well-typed)
{\em admissible goal}. Admissible goals must  fulfill a number of 
technical requirements needed to achieve the effect of lazy 
unification with {\em sharing} during goal solving.  Example 
\ref{sharing1} below will illustrate the treatment of sharing in 
$\calculo$.

\begin{definition}[Well-typed admissible goals] 
\label{objetivoadmisible}
Let $\programa$ be
a program. An admissible goal $G$ for ${\cal P}$ has the
structure $G \equiv  \exists \tupla \cdot S \Box P \Box E$\/,
where:

\begin{des}
\item $\textsf{evar}(G) =_{{\it Def}}  \tupla$ is called the set of existential variables;

\item $S$ is a system of equations in solved form; i.e.
   $S$ has the form  $x_1 = s_1, \ldots,$  $x_n = s_n$\/,
   where $s_i \in \terminos$\/, $1 \leq i \leq n$\/,
   and $x_{i}$ occurs exactly once in the whole goal, 
   $1 \leq i \leq n$\/;

\item $P \equiv   e_1 \rightarrow t_1, \ldots ,e_k \rightarrow t_k$
      is a multiset of approximation statements. The set
      $\textsf{pvar}(P) =_{{\it Def}} \bigcup_{i=1}^k {\it dvar}(t_i)$
      is called the set of produced variables;

\item $E \equiv  e_1 \lazo e'_1, \ldots, e_m \lazo e'_m$ is
      a multiset of joinability statements.
      
      Additionally, any goal $G$ must satisfy the following conditions:  
\begin{des}

\item[\textsf{(LIN)}] $(t_1, \ldots, t_k)$ is linear;
\item[\textsf{(EX)}] $\textsf{pvar}(P) \subseteq \textsf{evar}(G)$\/, i.e. all produced
   variables are existentially quantified;
\item[\textsf{(NCYC)}] the transitive closure of the relation
 $\gg$ defined as: $x \gg y$ iff there exists $1 \leq i \leq k$ such
 that $x \in {\it dvar}(e_i)$ and $y \in {\it dvar}(t_i)$\/,
 must be irreflexive (i.e. a strict partial order);
\item[\textsf{(SOL)}] ${\it dvar}(S) \cap \textsf{pvar}(P)=\emptyset$\/,
     i.e. the solved part does not contain produced
     variables.
\end{des}
\end{des}

\noindent
$G$ is well-typed iff there exists an environment $V$ such that
for all $e \diamondsuit e' \in S \cup P \cup E$\/,
$\diamondsuit \in \{\rightarrow, \lazo, =\}$\/, there exists
$\tau \in \tipos$ such
that $e,e' \in {\it Expr}^{\tau}_{\Sigma}(V)$\/.  We will write
$\textsf{env}(G)$ for the collection of all environments $V$ such 
that $G$ is well typed w.r.t. $V$. As we will see, well-typedness of 
goals is preserved by $\calculo$, as long as the program is also 
well-typed.   \hfill \square
\end{definition}

In the following, {\em initial goals} will be admissible
goals of the form $\Box \Box E$ whereas
{\em quasi-solved goals}  will  be admissible goals 
such that for all $e \rightarrow t \in P$ and $e' \lazo e'' \in E$ it
holds that $e,t,e',e'' \in {\it DVar}$\/.
Finally, {\em goals in solved form} will be also admissible goals
with the following structure: $\exists \tupla  \cdot S \Box \Box$\/.
It is easy to check that solved goals with
$S \equiv  x_1=s_1, \ldots, x_n=s_n$\/, determine an associated
{\em answer data substitution} $\sigma_{d_{S}}$ defined as 
$\sigma_{d_{S}}(x_i)=s_i$\/,
for all $1 \leq i \leq n$ and $\sigma_{d_{S}}(x)=x$ for
all $x \not = x_i$\/, which is idempotent.

Some comments on the structure of admissible goals may be helpful. Intuitively, 
each equation $x=s$ in the solved part $S$ denotes a computed answer for $x$\/. As we will
show later, if an admissible goal $G$ has a solution $\sigma_{d}$, our lazy
narrowing calculus is able to transform $G$ into
a solved goal $G' \equiv   \exists \tupla  \cdot S' \Box \Box$\/
in such a way that the solved system $S'$\/, viewed as a data
substitution $\sigma_{d_{S'}}$, denotes a solution for $G$ more 
general (modulo the finite set of equational axioms $\cal C$)
than $\sigma_{d}$\/.   Irreflexivity of $\gg$  allows  to  avoid  occur-checks  in  some  of  our 
transformations. Remark that all produced variables
are existentially quantified because such variables are used
to compute intermediate results. Furthermore, since
$(t_1, \ldots , t_k)$ is linear, produced variables are
only ``produced'' once.  The $\calculo$ treatment of approximation 
statements $e \rightarrow t \in P$ takes care of the lazy unification 
of $e$ and $t$, using narrowing with rewrite rules in $\cal R$ and 
mutations with equations in $\cal C$. Moreover, statements $e 
\rightarrow t$ are handled in such a way that the effect of sharing is 
achieved. Recall that in our framework sharing is not only convenient 
for the sake of efficiency, but also necessary for soundness, due 
to {\em call time choice} semantics for non-determinism.

 More concretely,  
the effect of  sharing will be  emulated by means of 
approximation statements of the form $e \rightarrow x$ occurring in  
$P$.  The idea is that $e \rightarrow x$ will never propagate a 
binding of $x$ to $e$, unless $e$ is a data term, but will behave as a 
suspension.  In case that the value of variable $x$ is demanded (or 
computed) somewhere in the rest of the goal, a suitable $\calculo$ 
transformation will ``awake'' $e \rightarrow x$ to trigger the 
evaluation of $e$. Let us see a simple example.

\newpage

\begin{exmp} \label{sharing1}  Consider the rewrite rules for 
\textsf{coin}, \textsf{double} and \textsf{plus} given in the 
Introduction.  Suppose that we want to solve the goal
$G \equiv \Box \Box {\it double(coin)} \lazo R$. As we know, this goal 
has two correct solutions, namely $R=0$ and $R=2$, but not $R=1$. Let 
us sketch how $\calculo$ would in fact compute the two correct 
solutions and avoid the incorrect one.

From the goal ${\it double(coin)} \lazo R$, and applying  
the rule $\textsf{Narro\-wing}_{\igual}$ given in  
Subsection \ref{reglasdetransformacion}, we 
can get the new goal $G_{1} \equiv \exists x \cdot 
\Box {\it coin} \rightarrow x \Box 
{\it plus}(x,x)$ $\lazo R$. Imagine that we would allow to propagate the 
binding $x/{\it coin}$. In such a case, we would get the new goal 
$G_{2} \equiv \Box \Box  R \lazo  {\it plus}({\it coin},{\it coin})$.  Now, 
applying again narrowing using the rule ${\it plus}(1,0) \rightarrow 1$, we 
would get the goal $G_{3} \equiv \Box \Box {\it coin} \rightarrow 
1, {\it coin} \rightarrow 0 \Box  R \lazo 1$. From $G_{3}$, according 
to the rules for $\textsf{coin}$, we would be able to compute the 
solution $R=1$, 
which is known to be unsound in our setting.

In order to avoid this kind of unsound computations, what we do is to forbid 
propagations of bindings such as 
$x/{\it coin}$.  From ${\it plus}(x,x) $ $ \lazo R$ it is not yet obvious 
that the value of $x$ is demanded. Therefore, the computation can only 
proceed by applying  $\textsf{Narrowing}_{\lazo}$ to narrow ${\it 
plus}(x,x)$. This can be attempted with any of the four rewrite rules 
for \textsf{plus}. The second and third one lead to failure, 
while the other two give rise to the two new goals

\vspace*{-0.10cm}

\[ \begin{array}{lll}
G_{2} \equiv \exists x \cdot 
\Box {\it coin} \rightarrow x, x \rightarrow 0, x \rightarrow 0 \Box R 
\lazo 0, \\
G_{3} \equiv \exists x \cdot 
\Box {\it coin} \rightarrow x, x \rightarrow 1, x \rightarrow 1 \Box R 
\lazo 2
\end{array} \]

\vspace*{-0.10cm} 

\noindent
respectively. By binding $x$ to $0$ in $G_{2}$ and $x$ to $1$ in 
$G_{3}$ (rule 
{\em {\sf Imitation+Decom\-position}}$_{\rightarrow}$)  and applying 
$\textsf{Decomposition}_{\rightarrow}$, these goals become:

\vspace*{-0.10cm}

\[ \begin{array}{lll}
G'_{2} \equiv \exists x \cdot x=0
\Box {\it coin} \rightarrow  0 \Box R 
\lazo 0, \\
G'_{3} \equiv \exists x \cdot x=1
\Box {\it coin} \rightarrow 1  \Box R 
\lazo 2
\end{array} \]

\vspace*{-0.10cm}

\noindent
which can be solved by narrowing $\textsf{coin}$ with the rule 
$\textsf{Narrowing}_{\rightarrow}$, leading to the expected solutions.
\hfill \blacksquare 
\end{exmp}

\subsection{Transformation Rules for $\calculo$}
\label{reglasdetransformacion}

Some of the $\lazo$ and $\rightarrow$ goal transformation rules 
described below, related to {\em mutation}, use the 
``linearization''  ${\cal C}_{\rightarrow}$ of $\cal C$\/. Formally, 
${\cal C}_{\rightarrow}$ is obtained from $\cal C$ by replacing each 
strongly regular equational axiom $c(t_1, \ldots,t_n) \approx d(s_1, \ldots, s_m)$
by  the rewriting rules $c(t'_1, \ldots, t'_n) \rightarrow d(s_1, \ldots, s_m)
 \Leftarrow C_1$ and  $d(s'_1, \ldots, s'_m) \rightarrow c(t_1, \ldots, t_n) 
 \Leftarrow C_2$\/,
where $c(t'_1, \ldots, t'_n)$ and $C_1$ are calculated as follows:
For each variable
$x$ in $c(t_1, \ldots, t_n)$ occurring 
$k>1$ times, we replace each $j$-th occurrence of
$x$\/, $2 \leq j \leq k$\/, by a fresh variable $y_j$ adding
in  $C_1$ the joinability statements $x \lazo y_j$\/.
In the same way $d(s'_1, \ldots, s'_m)$
and $C_2$
can be calculated from $d(s_1, \ldots, s_m)$\/. For instance,
for the strongly regular equation $c(x,x,x,y) \approx d(y,y,y,x)$
we get the rewriting rules $c(x,x_1,x_2,y) \rightarrow d(y,y,y,x)
        \Leftarrow x \lazo x_1, x \lazo x_2$ and
$d(y,y_1,y_2,$ $x) \rightarrow c(x,x,x,y) \Leftarrow
      y \lazo y_1, y \lazo y_2$\/.

Using rule instances from $[{\cal C}]_{\rightarrow}$ is equivalent to 
using inequalities from $[{\cal C}]_{\ecuacional}$ as explained in
Definition \ref{calin} above. This new view allows a more uniform 
presentation of goal solving, since algebraic data constructors and 
defined functions have now  similar rewrite rules. Rules \textsf{(MUT)} 
and \textsf{(OMUT)} of Definition \ref{calculitos} must be modified as 
follows:

 \medskip

\noindent
\(\begin{array}{lll}
\textsf{(MUT)} \ \ {\cal C}\mbox{-}{\sf Mutation:} &
\left.
\begin{array}{cc}
C \\
 \hline
s \rightarrow t
\end{array} \right.  & \mbox{if $s \rightarrow t
                            \Leftarrow C \in [{\cal C}]_{\rightarrow}$}
\end{array} \)

\medskip

\noindent
\( \begin{array}{lllll}
\textsf{(OMUT)} \ \ {\sf Outer\ {\cal C}\mbox{-}Mutation:} &
\left.
\begin{array}{cc}
 \ldots, e_i \rightarrow t_i, \ldots, C, s \rightarrow t    \\
 \hline
 c(\etupla_{n}) \rightarrow t
\end{array} \right.
\end{array} \)

\smallskip

~~~~~~~~~~~~~~~~~~~~~~~~~~~~~~~~~~~~~~~~if $t \not = \perp$\/,
$c(\ttupla_{n}) \rightarrow s \Leftarrow C \in [{\cal 
C}]_{\rightarrow}$

\medskip
        
\noindent
where $[{\cal C}]_{\rightarrow} =  \{ (s \rightarrow t
           \Leftarrow C)\sigma_{d} \ \mid 
              \  s \rightarrow t \Leftarrow C \in {\cal C}_{\rightarrow},
                \sigma_{d} \in DSub_{\perp}\}$.

\medskip

In the sequel we will always assume this new version of the rewriting 
calculi. The next proposition ensures that the rewrite rules in ${\cal 
C}_{\rightarrow}$ are well-typed.  The complete proof has been
moved to the Appendix \ref{demos}

\begin{prop}[Preservation of well-typedness by ``linearization''] 
\label{linearizacion} 
Let $c(\ttupla_{n})
\approx d(\stupla_{m})$ be a well-typed strongly regular
equation. Let $c(\tprimatupla_{n}) \rightarrow d(\stupla_{m})
\Leftarrow C_1$ be one  rule
obtained by applying ``linearization''  to  the equational axiom.
Then there exist an environment $V$ and type variants
$c:(\tau_1, \ldots, \tau_n) \rightarrow \tau$ and
$d:(\tau'_1, \ldots, \tau'_m) \rightarrow \tau$ of the principal
types of $c$ and $d$ respectively, such that
$c(\tprimatupla_{n}),d(\stupla_{m})
\in {\it Term}_{\Sigma}^{\tau}(V)$ and $C_{1}$ is well-typed w.r.t. 
$V$.\hfill \square  
\end{prop}

In the description of the transformation rules given below, the notation 
$c(\etupla_{n})$ stands for $c(e_1, \ldots, e_n)$\/, $c \in {\it 
DC^n}$\/, $e_i \in \expresiones$\/, $1 \leq i \leq n$\/. Analogously,
$f(\etupla_{n})$ is a
shorthand for   $f(e_1, \ldots, e_n)$, where
$f \in {\it FS}^n$ and $e_i \in \expresiones$\/, $1 \leq i \leq n$\/.
Furthermore, all bracketed equations
$[x=s]$ occurring in $S$  mean that $x=s$ only occurs
in $S$ if $x \not \in \textsf{pvar}(P)$\/. We regard 
conditions $e \lazo e' \in E$ as symmetric for the purpose of applying 
goal transformations.  Note that no particular strategy is
assumed to select a particular part of $G \equiv   \exists \tupla \cdot S \Box P \Box E$
to be transformed by one of the possible goal transformation rules.

\subsubsection{Transformation rules for $\igual$}

\noindent   
$\textsf{Decomposition}_{\igual}$:

\noindent
~~$\exists \tupla \cdot S \Box P \Box c(\etupla_{n}) \lazo
                    c(\eprimatupla_{n}) , E
     \reduce \exists \tupla \cdot S \Box P \Box  e_1 \lazo e'_1, 
     \ldots, e_{n} \lazo e'_{n},E$
     
\smallskip
           
\noindent
$\textsf{Mutation}_{\lazo}$:

\noindent        
~~$\exists \tupla \cdot S \Box P \Box
           c(\etupla_{n}) \lazo e' , E
     \reduce \exists \xtupla,
     \tupla \cdot S \Box e_{1} \rightarrow t_{1},\ldots, e_n \rightarrow 
     t_n,  P \Box
      C, s \lazo e',  E$
    
\smallskip
      
where ${\it Eq}: c(\ttupla_{n}) \rightarrow s \Leftarrow C$ is a 
variant of a rule in ${\cal C}_{\rightarrow}$
with  $\stackrel{\_}{x}={\it dvar}(Eq)$   

fresh variables.
 
\smallskip

\noindent
$\textsf{Imitation+Decomposition}_{\lazo}$:

\noindent 
~~$\exists \tupla \cdot S \Box P \Box x \lazo c(\etupla_{n}), E 
\reduce$

\noindent
~~~~~~~~~~~~$\exists \xtupla_n, \tupla \cdot {[x = c(\xtupla_{n})]}, (S \Box P \Box
   x_{1} \lazo e_{1},\ldots ,x_n \lazo e_n, E)[x/c(\xtupla_{n})]$
   
\smallskip

where  $\xtupla_{n}$  are fresh variables.
 
\smallskip

\newpage

\noindent
$\textsf{Imitation+Mutation}_{\lazo}$: 

\noindent
~~$\exists \tupla \cdot S \Box P \Box x \lazo 
c(\etupla_n), E
     \ \reduce \ \exists
    \stackrel{\_}{z}, \xtupla_{m}, \tupla \cdot
    {[x=d(\xtupla_{m})]},$
     
\noindent
~~~~~~~~~~~~$(S \Box x_{1} \rightarrow t_{1}, \dots, x_m \rightarrow 
    t_m,  P \Box
      C, s \lazo c(\etupla_{n}),  E) [x/d(\xtupla_{m})]$
      
\smallskip
      
If $c,d$  are  algebraic constructors of the same datatype,  where
${\it Eq}: d(\ttupla_{m}) \rightarrow s \Leftarrow C$   

is a variant 
of a rule in ${\cal C}_{\rightarrow}$, 
$\stackrel{\_}{z}={\it dvar}(Eq)$ and $\xtupla_{m}$   are fresh 
 variables.
 
\smallskip

\noindent
$\textsf{Narrowing}_{\lazo}$:

\noindent
~~$\exists \tupla \cdot S \Box P \Box f(\etupla_{n}) \lazo e', E
      \ \reduce \ \exists \stackrel{\_}{x}, \tupla \cdot
  S  \Box e_{1} \rightarrow t_{1}, \ldots, e_n \rightarrow t_n , P \Box C,
       r \lazo e', E$
       
\smallskip
    
where ${\it Rul}:f(\ttupla_{n}) \rightarrow r \Leftarrow C$  
is a variant of a rule in  ${\cal R}$  with 
$\stackrel{\_}{x}={\it dvar}({\it Rul})$  

fresh variables.

\subsubsection{Transformation Rules for $\rightarrow$}

Before presenting the transformation rules for $\rightarrow$,
we need to introduce the  concept of {\em demanded variable}, i.e.,  
a variable which requires the evaluation of an expression
in order to be unified with the result of such evaluation.

\begin{definition}[Demanded variables]
\label{variabledemandada} 
Let $G \equiv  \exists \tupla \cdot S \Box P \Box E$ be an admissible goal.
A variable $x $ $\in {\it dvar}(G)$ is demanded iff  
there exists a sequence of approximation statements in $P$ of 
the form $x_{0} \rightarrow x_{1}, x_{1} \rightarrow x_{2}, \ldots, 
$ $x_{k-2} \rightarrow x_{k-1},
x_{k-1} \rightarrow x_{k}$, such that $x_{0} = x$ and $x_{k} 
\lazo e \in E$ or $e \lazo x_{k} \in E$. $k=0$ is possible, in which 
case $x \lazo e \in E$ or $e \lazo x \in E$\/, for some $e \in 
\expresiones$\/.

In the following $\textsf{dmvar}(G)$ will denote the set of demanded
variables
in $G$\/. \hfill \square
\end{definition}
 
 As we will see below   in Definition \ref{solucion},
 any solution   (partial data substitution $\sigma_{d}$)
 for $G$ must   guarantee the existence of $\gpc$-proofs for
 all joinability and approximations statements in       $G$     affected
 by $\sigma_{d}$\/. Due to the semantics of joinability and approximation
 statements (see Theorems \ref{equivalencia} and \ref{adecuacion}),     solutions must compute totally
 defined values for demanded variables. Thus, in
 statements of the form $f(\etupla_{n}) \rightarrow x$\/,
 $f \in {\it FS}^n$ or $c(\etupla_{n}) \rightarrow x$\/,        $c \in {\it     DC}^n$ 
 with $x \in \textsf{dmvar}(G)$\/,  the evaluation of
 $f(\etupla_{n})$ or    $c(\etupla_{n})$, respectively, is 
 needed (see transformation rules $\textsf{Mutation}_{\rightarrow}$, 
 $\textsf{Imitation}_{\rightarrow}$  
 and $\textsf{Narrowing}_{\rightarrow}$  below).
 Otherwise ($x$ is not demanded) such   evaluation is delayed
until the application of other goal transformation  produces a 
non-variable binding for $x$, or causes $x$ to become demanded, or 
causes $x$ to disappear from the rest of the goal, in which case 
$f(\etupla_{n}) \rightarrow x$ (respect. $c(\etupla_{n}) \rightarrow 
x$) can be eliminated by using $\textsf{Elimination}_{\rightarrow}$. 
As we have discussed above, these mechanisms achieve the effect of 
sharing. Moreover, it is because we avoid to process such 
approximation statements eagerly that we can speak of {\em lazy 
narrowing}. As in some other related works such as 
\cite{techanus,MOI96,GHLR99} by {\em laziness} we  mean that our 
narrowing calculus has the ability to delay the unification of parameter 
expressions with the left-hand sides of rewrite rules.  This is needed 
for completeness, since both {\em innermost} and {\em outermost}
narrowing are known to be incomplete \cite{You89}. Nevertheless,  
we do not claim that $\calculo$ computations perform only needed steps. 
On the contrary, $\calculo$ computations with redundant steps are possible, partly 
because of unnecessary mutation transformations and partly due to 
other reasons; see discussion in \cite{GHLR99}, Sect. 8. Actual 
programming languages based on our ideas should implement 
refinements of $\calculo$ which avoid unneeded computations, in particular, redundant mutations.
 A first attempt in this direction, limited to a language with multisets and 
 arbitrary free data constructors, has been presented in \cite{ALR98}.  The concept of 
{\em needed narrowing} \cite{AEH94,LLR93}, based on Huet and Levy's theory of needed 
reductions \cite{HL79,HL91} gives a strategy which avoids unneeded 
narrowing steps for so called {\em inductively sequential} rewrite 
rules. However, to our best knowledge there is no theory of needed 
reductions which can be applied to rewriting/narrowing modulo 
equational axioms for data constructors. Even for the case of multisets, the 
notion of needed reduction becomes unclear. For instance, matching a simple pattern such as
$\mi 0 \mid xs \md$ requires an unpredictable amount of evaluation for the 
matching expression.

The transformation rules for $\rightarrow$ are the following:

 \medskip
  
\noindent
$\textsf{Decomposition}_{\rightarrow}$:

\noindent
~~$\exists \tupla \cdot S \Box c(\etupla_{n}) \rightarrow c(\ttupla_{n}), P \Box E
  \ \reduce \ \exists \tupla \cdot S \Box e_{1} \rightarrow t_{1}, \ldots, 
  e_n \rightarrow t_n ,
            P \Box E$

\smallskip    

\noindent
$\textsf{Mutation}_{\rightarrow}$:

\noindent
~~$\exists \tupla \cdot S \Box c(\etupla_{n}) \rightarrow t, P \Box  E
     \ \reduce
    \ \exists \stackrel{\_}{x}, \tupla \cdot S \Box e_{1} \rightarrow 
    t_{1},
     \ldots, e_n \rightarrow t_n,  s \rightarrow t,P \Box C, E$
     
\smallskip

If $t \not \in {\it DVar}$ or
$t \in \textsf{dmvar}(G)$\/,  where 
${\it Eq}: c(\ttupla_{n}) \rightarrow s
       \Leftarrow C$ is a variant of a rule

in ${\cal C}_{\rightarrow}$, with $\stackrel{\_}{x}={\it dvar}({\it Eq})$  fresh variables.
 
\smallskip

\noindent
$\textsf{Imitation+Decomposition}_{\rightarrow}$:

\noindent
~~$\exists \tupla \cdot S \Box x \rightarrow 
c(\ttupla_{n}),P \Box E \ \reduce$

\noindent
~~~~~~~~~~~~$\exists \xtupla_n, \tupla \cdot  {[x = c(\xtupla_{n})]}, (S \Box
   x_{1} \rightarrow t_{1},\ldots, x_n \rightarrow t_n, P \Box 
   E)[x/c(\xtupla_{n})]$
   
\smallskip
 
where  $\xtupla_{n}$  are fresh 
variables.
 
\smallskip

\noindent
$\textsf{Imitation+Mutation}_{\rightarrow}$:

\noindent
~~$\exists \tupla \cdot S \Box x \rightarrow
      c(\ttupla_{n}),P \Box E  \ \reduce$
\ $\exists \stackrel{\_}{z},\xtupla_{m},\tupla \cdot
    {[x=d(\xtupla_{m})]},$
      
\noindent
~~~~~~~~~~~~$(S \Box x_{1} \rightarrow s_{1},
     \ldots, x_m \rightarrow s_m, s \rightarrow c(\ttupla_{n}),
      P \Box C,E)  [x/d(\xtupla_{m})]$
      
\smallskip

If  $c,d$  are algebraic data constructor  of the same datatype,  
where   

${\it Eq}: d(\stupla_m) \rightarrow s \Leftarrow C$ is a variant of a rule 
in ${\cal C}_{\rightarrow}$, 
and $\stackrel{\_}{z}={\it dvar}(Eq)$, 

and $\xtupla_{m}$  are fresh 
variables.

\smallskip

\noindent
$\textsf{Imitation}_{\rightarrow}$:

\noindent
~~$\exists x, \tupla \cdot S \Box c(\etupla_{n})
      \rightarrow x,P \Box E \ \reduce
\ \exists \xtupla_n,\tupla \cdot  S \Box (e_{1} \rightarrow x_{1},\ldots, 
e_n  \rightarrow x_n,P \Box E)[x/c(\xtupla_{n})]$

\smallskip
      
If $x \in \textsf{dmvar}(G)$\/,   where
$\xtupla_{n}$  are fresh variables.

\smallskip 

\noindent
$\textsf{Elimination}_{\rightarrow}$:

\noindent
~~$\exists x, \tupla \cdot  S \Box e \rightarrow x,P
      \Box E \ \reduce \ \exists \tupla  \cdot S \Box P \Box E$
      
\smallskip      

If $x \not \in {\it dvar}(P \Box E)$\/.
 
\smallskip

\noindent
$\textsf{Narrowing}_{\rightarrow}$:

\noindent
~~$\exists  \tupla  \cdot S \Box f(\etupla_{n})
     \rightarrow t, P \Box E \ \reduce \  \exists  \stackrel{\_}{x},
      \tupla \cdot S  \Box e_{1} \rightarrow t_{1},\ldots, e_n \rightarrow t_n,r
      \rightarrow t, P \Box C,E$
      
If $t \not \in {\it DVar}$ or $t \in \textsf{dmvar}(G)$\/,  
 where ${\it Rul}: f(\ttupla_{n}) \rightarrow r
       \Leftarrow C$  is a variant of a rule 

in ${\cal R}$, and  $\stackrel{\_}{x}={\it dvar}({\it Rul})$   are fresh variables.

\subsubsection{Failure Rules}

The {\em failure rules}  below
should be applied before the rest of $\reduce$-rules in order to detect
failures as soon as possible. The set $\textsf{svar}(e)$ in rule
\textsf{Cycle} denotes the set of {\em safe variables} occurring
in $e$\/, i.e. the set of variables $x$  
such that $x$ occurs in $e$ at some position whose ancestor positions are all 
occupied by {\em free} constructors. In \cite{GHLR99}, a different notion of safe 
variable is used: ``$x$ is safe in $e$ iff $x$ occurs in $e$ at some position 
whose ancestor positions are all occupied by constructors''. This 
notion wouldn't lead to a correct \textsf{Cycle} rule in a language with 
algebraic constructors. For instance, in our
framework, if $c(a) \approx a \in {\cal C}$ then
$x \lazo c(x)$ has a solution $x=a$\/.

 \medskip

\noindent
$\textsf{Conflict}_{\lazo}$:

\noindent
~~$\exists \tupla \cdot S \Box P \Box c(\etupla_{n})
           \lazo d(\etupla'_{m}), E \ \reduce \ {\it FAIL}$
           
\smallskip

If $c \not  = d$ and $c,d$  are free constructors, or $c$  is free
and $d$  is algebraic 

or $c$  is algebraic and $d$  is free.
     
\medskip

\noindent
\textsf{Cycle:}

\noindent
~~$\exists \tupla \cdot S \Box P \Box  E \reduce \ {\it FAIL}$
 
 \smallskip
 
If $E$  contains a {\em  variable cycle}, i.e.,  a sequence of strict 
equalities of the form: 

\vspace*{-0.10cm}

\[  x_{0} \igual e_{1}[x_{1}], x_{1} \igual e_{2}[x_{2}], \ldots,
     x_{n-2} \igual e_{n-1}[x_{n-1}], x_{n-1} \igual e_{n}[x_{0}], \]
     
\vspace*{-0.10cm}

where $n \geq 1$, $x_{i} \in {\it DVar}$, $0 \leq i \leq n-1$,  and
$e_{i}[x_{i}] \in \expresiones-{\it DVar}$, 

$1 \leq i \leq n$, represents an expression in which  $x_{i} \in \textsf{svar}(e_{i})$.

\medskip

\noindent
$\textsf{Conflict}_{\rightarrow}$:

\noindent
~~$\exists \tupla \cdot  S \Box c(\etupla_{n})
           \rightarrow d(\ttupla_{m}),P \Box E \ \reduce \ {\it FAIL}$ 
           
\smallskip

If $c \not  = d$ and $c,d$  are free constructors, or $c$  is free
and $d$  is algebraic 

or $c$  is algebraic and $d$  is free.

\medskip

The rule \textsf{Cycle} above is a generalization of that presented in
\cite{GHLR99}.  In \cite{GHLR99}   an   ``occur-check''  failure is 
detected by the rule: $ \textsf{Cycle}^{_*}: \ \exists \tupla 
\cdot S \Box P \Box x \lazo e, E \reduce {\it FAIL}$, if
$e \not \in {\it DVar}$ and $x \in \textsf{svar}(e)$. In our setting, 
such a rule is not enough. For instance, $\textsf{Cycle}^{_*}$ applied to 
the goal  $G \equiv \Box \Box x \lazo {\it Suc}(y), y \lazo {\it 
Suc}(x)$ would be unable to detect failure, whereas $\textsf{Cycle}$ is. 
Note that no $\calculo$ transformation can convert $G$ into the goal 
{\small $G' \equiv \Box \Box x \lazo {\it Suc(Suc(x))}$} $, y \lazo {\it 
Suc(x)}$, where $\textsf{Cycle}^{*}$ would suffice to detect the 
failure.
     
\subsubsection{Variable Elimination Rules}
\label{rev}

\noindent
\textsf{Produced variable Elimination:}

\medskip

 \noindent
~~$\exists y, \tupla \cdot  S \Box x \rightarrow y, P
\Box E \reducevar \exists \tupla \cdot S \Box (P \Box  E) [y/x]$

\smallskip
 
\noindent
\textsf{Identity:}

\noindent
~~$\exists \tupla \cdot  S \Box P \Box x \lazo x,E
        \reducevar \exists \tupla \cdot S \Box P \Box  E$
        
\smallskip
        
If $x \not \in \textsf{pvar}(P)$\/. 

\smallskip

\newpage

\noindent
\textsf{Non-produced variable elimination}:

\noindent
~~$\exists \tupla  \cdot S \Box P \Box x \lazo y,E
        \reducevar \exists \tupla \cdot  x=y, (S \Box P \Box 
        E)[x/y]$

If $x,y \not \in \textsf{pvar}(P)$.

\medskip

Differently to \cite{GHLR99}, $\reduce$
needs {\em don't know choice} in the application of transformation
rules\footnote{Of course don't know non-determinism also
appears in the selection of $\cal C$-equations and program rules to be applied.}.
The reason is the incorporation of equational axioms for algebraic constructors.
More precisely, when a statement $c(\etupla_n) \lazo c(\eprimatupla_n)$ (respect.
$c(\etupla_n) \rightarrow c(\eprimatupla_n)$), where $c$ is algebraic,
has to be reduced it is  not known in advance which transformation rule 
($\textsf{Decomposition}_{\diamondsuit}$ or $\textsf{Mutation}_{\diamondsuit}$, where $\diamondsuit \in 
\{\igual,\rightarrow\}$) will  succeed.
For instance, considering
$c(a) \approx c(b) \in {\cal C}$\/, where $a$ and $b$ are
free constant symbols, we get that $c(a) \lazo c(b)$
must be reduced using $\textsf{Mutation}_{\igual}$ but $c(a) \lazo c(a)$ should
be reduced using $\textsf{Decomposition}_{\igual}$. However, in both cases, both
rules are applicable. Some other times, frequently in presence
of variables, both rules are able to capture a solution, even
the same. For instance, consider the goal $\Box \Box c(a) \lazo c(x)$
which clearly has as possible solutions $\{x=a\}$ and $\{x=b\}$\/.
Using $\textsf{Decomposition}_{\lazo}$
and $\textsf{Imitation+Decomposition}_{\lazo}$ we get the solution $\{x=a\}$\/.
Now, applying $\textsf{Mutation}_{\lazo}$ to $c(a)$ with $c(a) \rightarrow c(b)
\in {\cal C}_{\rightarrow}$, $\textsf{Decomposition}_{\rightarrow}$ and
$\textsf{Imitation+Decomposition}_{\lazo}$ we get another solution $\{x=b\}$\/.
Furthermore, applying $\textsf{Mutation}_{\igual}$ to $c(x)$ with
$c(a) \rightarrow c(b) \in {\cal C}_{\rightarrow}$ we can again
capture the solution $\{x=a\}$ computed previously. This shows that 
$\calculo$ can compute repeated solutions,
something undesirable from a practical point of view, which shows  its 
practical weakness.

  Another problem is that mutations with respect to `` symmetric'' $\cal     
  C$-equations  such as $\mi x,y | {\it zs}     \md     \approx \mi     y,x     |{\it 
  zs} \md$      can     lead to infinite $\calculo$-derivations. This difficulty 
  can be avoided in an  implementation, as we have explored     in \cite{ALR98} 
  for the  particular case of the datatype multiset.    There are also some     
  known techniques      to alleviate the problem of     computing redundant     
  solutions for set unification problems \cite{AD95,AD97}.
  At the level  of arbitrary equational data constructors (even  under 
  the restriction to strongly regular equations), it seems      quite hard 
  to design more efficient      narrowing calculi, without sacrificing
  completeness. We will come back to this point in      the     next section.

\subsection{Solutions}

In order to establish the soundness and completeness of $\calculo$\/, 
we must define the notion of {\em solution}. Our definition refers to 
the rewriting calculi from Section \ref{calculos}. However, 
Theorem \ref{adecuacion} shows that solutions can be also characterized in 
terms of the free term models $\modterm$\/.

\begin{definition}[Solution] 
\label{solucion}
 Let $\programa$ be a program.
Let $G \equiv   \exists \tupla \cdot S \Box P \Box E$ be an admissible
goal for ${\cal P}$ and $\sigma_{d} \in {\it DSub}_{\perp}$\/.
We say that $\sigma_{d}$ is a solution of $G$ iff

\begin{des}
\item[\textsf{(TOT)}] $x\sigma_{d} \in \terminos$ is a total data term 
         for all $x \in {\it DVar}-\textsf{pvar}(P)$\/;
\item[\textsf{(EQ)}] $x\sigma_{d} =  s\sigma_{d}$ for all $x=s \in S$\/;
\item[\textsf{(GORC)}] For all $e \lazo e' \in E$ and $e'' \rightarrow t \in P$
      there exists a $\gpc$-proof of $e\sigma_{d} \lazo e'\sigma_{d}$ and
      $e''\sigma_{d} \rightarrow t\sigma_{d}$ respectively. The multiset composed
      of all such $\gpc$-proofs will be called a {\em witness} ${\cal M}$ for
      $G$ and $\sigma_{d}$\/.
\end{des}

\noindent
In the following $\textsf{Sol}(G)$ will denote the set of all solutions
for $G$\/. \hfill \square

\end{definition}

Solutions for goals are {\em partial data substitutions}.
This is because of the presence of produced variables. Such
variables, which are not present in initial goals,
can appear (existentially quantified) in intermediate goals of a
computation. Since they
occur in right-hand sides of approximation statements, they
serve to express approximation and thus may need to be given only
partial values. 
For instance, consider the functions
$\textsf{duo}: {\it Set}(\alpha) \rightarrow {\it Bool}$ and
$\textsf{om}: \rightarrow \alpha$ defined in Example \ref{conjuntos}.
Consider now 
the admissible goal $G  \equiv \Box \Box duo(\{om\}) \lazo {\it True}$ which
has the empty substitution as solution. By applying
$\textsf{Narrowing}_{\lazo}$ using the
program rule variant ${\it duo}(\{x_1,y_1\}) \rightarrow {\it True}$\/, we get
$G' \equiv   \exists x_1,y_1 \cdot \Box \{om\} \rightarrow \{x_1,y_1\}
\Box {\it True} \lazo {\it True}$\/.
Clearly, any solution $\sigma_{d}'$ for $G'$ must map $x_1$ and $y_1$
(produced variables) into $\perp$\/. Otherwise there is no witness
for $\{om\} \rightarrow \{x_1,y_1\}\sigma_{d}'$\/.  Notice that for initial
goals (where only $E$ is present), solutions
are {\em total data substitutions} due to condition \textsf{(TOT)} in
Definition \ref{solucion}.

In spite of algebraic data constructors, condition \textsf{(EQ)} in Definition \ref{solucion}
requires syntactic identity. The reason is that the solved part $S$ of a goal 
represents a partially computed answer substitution. Moreover,  $\calculo$ enumerates
non deterministically all the possible solutions to any admissible
goal, even those being equivalent modulo ${\cal C}$\/.  This is 
because the transformation rules  $\textsf{Imitation}_{\rightarrow}$, 
$\textsf{Imitation+Decomposition}_{\diamondsuit}$ and
$\textsf{Imitation+Mutation}_{\diamondsuit}$, where $\diamondsuit \in 
\{\rightarrow,\lazo\}$, have been designed purposefully to mimic the 
proof steps of any given solution witness. In particular, given a 
solution $\sigma_{d}$ for a goal $G$, $\calculo$ can choose to 
propagate bindings $x/t$ in such a way that the new resulting goal 
$G'$ has a solution $\sigma'_{d}$ such that $x \sigma_{d}$ is 
syntactically identical to $t \sigma'_{d}$. The following example 
will clarify this idea. 

\begin{exmp} \label{raro} Consider an admissible goal
of the form $G \equiv   \exists \tupla \cdot S \Box x \rightarrow c(\ttupla_n),
P \Box E$ having a solution $\sigma_{d}$\/. Let us analyze the
witness $\cal M$ associated to $G$ and $\sigma_{d}$\/. If the
$\gpc$-proof in ${\cal M}$ for $x\sigma_{d} \rightarrow c(\ttupla_{n})\sigma_{d}$
uses:

\begin{des}
\item[\textsf{(1)}] $\textsf{(DC)}$ as last inference step, then it  
holds that $x\sigma_{d} = c(\stupla_n)$ and ${\cal M}$ contains 
subproofs for $s_{i} \rightarrow t_{i}\sigma_{d}$\/, $1 \leq i \leq n$\/. By applying the
        propagation rule  $\textsf{Imitation+Decomposition}_{\rightarrow}$,
        the resulting goal has a solution $\sigma_{d}'$ defined
        as:  $\sigma_{d}'(x_i)$ $=s_i$\/, $1 \leq i \leq n$\/,
         so that  $\sigma_{d}'=\sigma_{d}[\backslash \{x_1, $ $\ldots, x_n\}]$ and
          $x\sigma_{d}'$ is syntactically equal to
         $c(\xtupla_n) \sigma_{d}'$\/;

\item[\textsf{(2)}] \textsf{(OMUT)} as last inference step, then
           $x\sigma_{d}=d(\stupla_m)$ and there exists
     ${\it Eq}:d(s'_1, \ldots, $ $s'_m) \rightarrow s'' \Leftarrow C
     \in [{\cal C}]_{\rightarrow}$ such that the $\gpc$-proof
         for $x\sigma_{d} \rightarrow c(\ttupla_n)\sigma_{d}$ in
       $\cal M$ contains subproofs
    for $s_i \rightarrow s'_i$\/,
     $C$ and $s'' \rightarrow c(\ttupla_n)\sigma_{d}$ respectively.
   Trivially there exists a fresh variant ${\it Eq'}: d(s^*_1, \ldots,
s^*_m) \rightarrow s^* \Leftarrow C^*$ in ${\cal C}_{\rightarrow}$
such that ${\it Eq} = {\it Eq'}\sigma_{d_{0}}$ for some
$\sigma_{d_{0}}\in {\it DSub}_{\perp}$\/. By applying the propagation
rule $\textsf{Imitation+Mutation}_{\rightarrow}$ with ${\it Eq'}$\/,
the resulting goal has as solution $\sigma_{d}'$ defined as
$\sigma_{d}'(x_i)  =  s_i$\/, $1 \leq i \leq m$\/,
$\sigma_{d}'(x)   =  \sigma_{d_{0}}(x)$\/, for all $x \in {\it dvar(Eq')}$
and $\sigma_{d}'(x)  =  \sigma_{d}(x)$ otherwise, where $\sigma_{d}'$
verifies that $x\sigma_{d}$ is syntactically equal to
$d(\xtupla_m)\sigma_{d}'$\/. \hfill \blacksquare  
\end{des}

\end{exmp}

Similar considerations motivate the design of all the transformation
rules for $\calculo$. They are chosen to enable a completeness proof 
(Lemma \ref{progreso}) that relies
on a multiset ordering for witnesses (Definition \ref{ordenmult}), as we will see 
in next section.

\section{Soundness and Completeness of $\calculo$} 
\label{completeness}

In this section we establish the soundness and completeness of 
$\calculo$ w.r.t. the notion of solution given in 
Definition \ref{solucion} above. First,
we present a correctness lemma which ensures that $\reducevar$-steps 
preserve quasi-solved goals, $\reduce$-steps 
preserve admissibility of goals and  
fail only in case of unsatisfiable goals
and $\reduce$,  $\reducevar$-steps do not introduce new solutions.
The proof proceeds by inspecting all $\reduce$ and $\reducevar$
transformation rules and can be found in Appendix \ref{demos}.

\begin{lemma}[Correctness lemma]
\label{lemadecorreccion} 
Let $\programa$ be a program
with ${\cal C}$ strongly regular. Let $G$ be an admissible goal.
Then:

\noindent
${\bf (Invariance_{1})}$ If $G \reduce G'$ and $G$ is admissible, then $G'$ is 
           admissible;

\noindent
${\bf (Invariance_{2})}$ If $G \reducevar G'$ and $G$ is quasi-solved then $G'$ is 
           quasi-solved;

\noindent
${\bf (Correctness_{1})}$ If $G \reduce {\it FAIL}$ then $\textsf{Sol}(G)=\emptyset$\/;

\noindent
${\bf (Correctness_{2})}$ If $G \reduce G'$ or $G \reducevar G'$ and
     $\sigma_{d}' \in \textsf{Sol}(G')$ then there exists
     $\sigma_{d} \in \textsf{Sol}(G)$ 
with    $\sigma_{d}=\sigma_{d}'[\backslash 
(\textsf{evar}(G) \cup \textsf{evar}(G'))]$\/. \hfill \square

\end{lemma}

From Lemma \ref{lemadecorreccion} we can easily  prove the following
correctness theorem.

\begin{theorem}[Correctness of $\calculo$]
\label{correccion}
Let $\programa$ be a program with ${\cal C}$ strongly regular.
Let $G$ be an initial goal and $G'$ a quasi-solved goal such that
$G \reducederi G' \reducevarderi  G'' \equiv  \exists \tupla \cdot S \Box \Box$\/. Then
$\sigma_{d_{S}} \in \textsf{Sol}(G)$\/.  \hfill \square
\end{theorem}

\begin{proof}
First note that $\sigma_{d_{S}}$ is trivially a solution 
for the goal $G'' \equiv  \exists \tupla  \cdot S \Box \Box$.  On the other hand, 
suppose that the derivation $G \reducederi G' \reducevarderi  G''$ 
has the form:

\vspace*{-0.10cm}

\[ G \equiv G_{0} \reduce  G_{1} \reduce \ldots
   \reduce  G_{i} \equiv G' \reducevar  G_{i+1} \reducevar  \ldots 
   \reducevar  
   G_{n} \equiv G'' \]

\vspace*{-0.10cm}

\noindent
Now, if we repeatedly apply backwards 
item ${\bf (Correctness_{2})}$ of Lemma \ref{lemadecorreccion},  we 
have that there exists a solution $\sigma_{d}$ of $G$ such that 
$\sigma_{d} = \sigma_{d_{S}}[\backslash  \bigcup_{i=0}^{_n}\textsf{evar}(G_{i}))]$.  But noting that 
$\textsf{evar}(G) = \emptyset$ and ${\it dvar}(G) \cap \bigcup_{i=0}^{_n}\textsf{evar}(G_{i})
= \emptyset$, we can conclude that $\sigma_{d} = \sigma_{d_{S}}[{\it 
dvar}(G)]$. But then, since $\sigma_{d_{S}}$ is a total data 
substitution, we have that $\sigma_{d_{S}} \in \textsf{Sol}(G)$. 
\end{proof}

We address now the question of completeness of $\calculo$\/. 
Given a solution $\sigma_{d}$ of a goal $G$ we need to ensure the 
existence of some {\em terminating} sequence of $\calculo$-transformations, 
leading to a goal in solved form whose associated data substitution is 
more general than $\sigma_{d}$, modulo the finite set $\cal C$ of 
equational axioms\/. The completeness proof relies on a 
multiset ordering for witnesses (defined in Definition \ref{solucion}).
The definition of this ordering
is borrowed from \cite{GHLR99}.

\begin{definition}[Multiset ordering for proofs]
\label{ordenmult}
Let  $\programa$ be a program  and
${\cal M} =  \mi \Pi_1, \ldots,  $ $\Pi_n \md$\/,
${\cal M'} =  \mi \Pi'_1, \ldots, \Pi'_m \md$
multisets of $\gpc$-proofs of approximation and
joinability statements. We define
${\cal M} \triangulo {\cal M'} \Leftrightarrow
\mi |\Pi_1|, \ldots, |\Pi_n| \md \menor \mi
|\Pi'_1|, \ldots, |\Pi'_m| \md$\/,
where $|\Pi|$ is the size (i.e. the number of inference steps)
of $\Pi$, and $\menor$ is the multiset extension \cite{DM79}
of the usual ordering over the natural numbers. \hfill \square 
\end{definition}

\noindent
Then, in order to prove that $\calculo$ is complete, we can
argue as follows: 
Given any non-quasi-solved admissible goal
$G \equiv   \exists \tupla \cdot S \Box P \Box E$ and
$\sigma_{d} \in \textsf{Sol}(G)$  with witness $\cal M$\/, there exists a 
$\reduce$-transformation
rule $T$ such that $G \reduce G'$ by applying $T$ and
 $G'$ has a solution
$\sigma_{d}'$  with witness ${\cal M}' \triangulo {\cal M}$\/. Note that this 
holds for 
Example \ref{raro}, since in item \textsf{(1)}  $\cal M'$ loses one application of the
$\gpc$-rule \textsf{(DC)} 
whereas in item \textsf{(2)}  $\cal M'$ loses one application of the $\gpc$-rule
\textsf{ (OMUT)}. Analyzing
all the possible forms of an admissible goal and reasoning as
suggested by Example \ref{raro},
we can  state the following progress lemma, whose complete proof can be 
found in Appendix \ref{demos}.

\begin{lemma}[Progress lemma for $\reduce$]
\label{progreso} 
Let $\programa$ be a program where $\cal C$ is strongly regular.
Let $G$ be a non quasi-solved admissible goal (different from
FAIL and such that no failure rules can be applied to it), $\sigma_{d} 
\in \textsf{Sol}(G)$ and $\cal M$ a witness
for $G$ and $\sigma_{d}$\/. Then, there exists a $\reduce$-transformation
rule $T$ such that $G \reduce G'$ using $T$ and:

\begin{des}
\item there exists $\sigma_{d}' \in \textsf{Sol}(G')$ such that
       $\sigma_{d}= \sigma_{d}'[\backslash
       (\textsf{evar}(G) \cup \textsf{evar}(G'))]$\/;
\item there exists $\cal M'$ a witness for $G'$ and $\sigma_{d}'$ such that
     ${\cal M'} \triangulo {\cal M}$\/;
\item if $G$ and $\cal P$ are well-typed then for all $V \in 
\textsf{env}(G)$, there exists $V' \supseteq V$ such that $V' \in 
\textsf{env}(G')$. \hfill \square  
\end{des}
\end{lemma}

Using the progress lemma, we can prove now the following
completeness result for $\reduce$\/:

\begin{theorem}[Completeness of $\reduce$]
\label{completitud} 
Let $\programa$ be a program with ${\cal C}$ strongly regular,
$G$ an initial goal and $\sigma_{d} \in \textsf{Sol}(G)$. Then
there exist a quasi-solved goal $G'$ and
$\sigma_{d}' \in \textsf{Sol}(G')$ such that
$G \reducederi G'$ and
$\sigma_{d}=\sigma_{d}'[dvar(G)]$. Furthermore, if
$G$ and ${\cal P}$ are well-typed then for any environment $V \in 
\textsf{env}(G)$, there exists an environment $V' \supseteq V$ such 
that $V' \in \textsf{env}(G')$.  \hfill \square
\end{theorem}

\begin{proof}
Thanks to Lemma \ref{progreso}, it is possible to
 build a $\reduce$-derivation:
 $G \equiv  G_0 \reduce G_1$ $ \reduce G_2$  $\reduce $ $\ldots$ for
 which there exist $\sigma_{d} \equiv \sigma_{d_{0}},\sigma_{d_{1}}, 
 \sigma_{d_{2}},\ldots$ and
${\cal M} \equiv {\cal M}_0, {\cal M}_1,$  ${\cal M}_2, \ldots$ such that
$\sigma_{d_{i}} \in \textsf{Sol}(G_i)$, $x\sigma_{d_{i}} =x\sigma_{d_{i-1}}$,
for all $x \in {\it DVar}-(\textsf{evar}(G_{i-1}) \cup \textsf{evar}(G_i))$,
${\cal M}_i$ is a witness for $G_i$
and $\sigma_{d_{i}}$ and
${\cal M}_i \triangulo {\cal M}_{i-1}$. Since $\triangulo$
is a well-founded ordering, such a derivation is finite
and finishes with a quasi-solved goal $G_n \equiv G'$.
Furthermore, since $G$ has no existential variables it holds that
$\sigma_{d} = \sigma_{d_{n}}[dvar(G)]$. The last part of the theorem follows from
the well-typedness of  $G$\/, ${\cal P}$ and the last item in
 Lemma \ref{progreso}. 
\end{proof}

Note that $\reduce$-rules involving algebraic data
constructor and propagating bindings have two versions.
The first one is a standard imitation whereas the second
one combines imitation of the outermost constructor in some ${\cal 
C}$-equation with mutation via that ${\cal C}$-equation.
With these rules the termination of $\reduce$ is
ensured since the selection
of the transformation rule for getting $G'$ from $G$ depends
very directly on the witness associated to the given solution (as seen in  
Example \ref{raro} and in the proof of Lemma \ref{progreso}).
In Appendix \ref{apendice} we present an alternative narrowing 
calculus, which results from the calculus LNC from 
\cite{GHLR99} by adding mutation rules in the spirit of  \cite{JK91}. 
This  alternative is less indeterministic than $\calculo$. On the 
negative side, redundant and/or diverging computations due to 
mutation transformations are still possible.

As for soundness and completeness, the calculus shown in  
 Appendix \ref{apendice} is obviously correct in the sense of Theorem 
 \ref{correccion}. Unfortunately, we have been unable to prove the 
 analogon of the progress lemma \ref{progreso}. The reason is that 
 propagations of a binding $x/t$ can cause the witness for the new 
 goal to include a big number of mutation steps, thus preventing it 
 to decrease w.r.t. the multiset ordering. For instance,
consider the initial goal $G \equiv   \Box \Box
x\lazo b, c(x) \lazo c(a)$, where $c$ is a free data constructor
and $a \approx b \in {\cal C}$\/. Let $\sigma_{d}=\{x/a\}$ be
a solution for $G$\/. Assume
that instead of having our rule $\textsf{Imitation+Mutation}_{\lazo}$,
we had the binding rule \textsf{Binding} presented in
Appendix \ref{apendice}. Applying this
rule to $G$ we get $G' \equiv   x=b \Box \Box c(b) \lazo c(a)$\/.
Considering Definition \ref{solucion} but replacing \textsf{(EQ)}
by equality modulo ${\cal C}$ we get that
$G'$ has two possible solutions: $\sigma_{d}'=\{x/a\}$ or
$\sigma_{d}''=\{x/b\}$\/. Assume now that
${\cal M}$ (a witness for $G$ and $\sigma_{d}$)
has a $\gpc$-proof $\Pi$ for $c(a) \lazo c(a)$ consisting of:

\begin{des}
\item[(1)] one application of rule \textsf{(J)} for proving
          $c(a) \lazo c(a)$ by proving that
          $c(a) \rightarrow c(a)$;
\item[(2)] one application of rule \textsf{ (DC)} for proving that
         $c(a) \rightarrow c(a)$ by proving
         that $a \rightarrow a$\/. Since (1) requires
         to prove $c(a) \rightarrow c(a)$ twice,
         then in fact we have two applications
         of \textsf{ (DC)};
\item[(3)] one application of \textsf{ (DC)} for proving that
         $a \rightarrow a$. Again by (2) we have two applications
         of \textsf{ (DC)}.

\end{des}

Thus $\Pi$ has 5 inference steps. But
any $\gpc$-proof $\Pi'$ for $c(a) \lazo c(b)$ has more inference steps
than $\Pi$\/. That is, there is no witness
for $G'$ decreasing in $\triangulo$\/. We conjecture that the
goal solving calculus given in Appendix \ref{apendice} is also
complete, but we were unable to find a termination ordering
for the completeness proof. To
prove completeness of this calculus is an interesting
open problem, since its behaviour is less wildly indeterministic.
Therefore, it is much better suited as a basis for implementations.

Let us now present several results related to $\reducevar$-rules.  For that, we define the following
well-founded ordering,  useful for proving that
any $\reducevar$-derivation always terminates 
(see Lemma \ref{progreso1}).

\begin{definition}[Order for quasi-solved goals]
Let $G \equiv \exists \tupla \cdot S \Box P \Box E$ and
$G' \equiv  \exists \tupla'  \cdot S' \Box P' \Box E'$ be quasi-solved goals.
We say that $G'  \triangleright G$ iff $n < m$, where $n$ and $m$ are
the number of approximation and joinability statements
occurring in
$P' \Box E'$ and $P \Box E$ respectively.  \hfill \square

\end{definition}

Finally we state the lemma which ensures
the termination of $\reducevar$ along with
the preservation of types, quasi-solved goals and
solutions. The proof is easy  and follows by inspection
of the $\reducevar$-rules.
The notation $\sigma_{d} =_{{\cal C}} \sigma'_{d}[X]$, $X 
\subseteq {\it DVar}$, means that $x\sigma_{d} \equivalente 
x\sigma'_{d}$, for all $x \in X$, whereas $\sigma_{d} \leq_{{\cal 
C}} \sigma'_{d}[X]$ means that there exists $\sigma''_{d} \in {\it 
DSub}_{\perp}$ such that $x\sigma'_{d} \equivalente x \sigma_{d} 
\sigma''_{d}$, for all $x \in X$.

\begin{lemma}[Progress lemma for $\reducevar$]
 \label{progreso1} Let
$\programa$ be a program with ${\cal C}$ strongly regular.
Let $G$ be a quasi-solved but not solved goal (different from FAIL).
Then there exists a $\reducevar$-transformation rule $T$ such
that $G \reducevar G'$ using $T$ and:

\begin{des}
\item there exists $\sigma_{d}' \in \textsf{Sol}(G')$ such that 
            $\sigma'_{d} \leq_{{\cal C}} \sigma_{d}$;
\item $G' \triangleright G$;
\item if ${\cal P}$ and $G$ are well-typed then for any $V \in 
\textsf{env}(G)$ it holds that $V \in \textsf{env}(G')$.  \hfill \square

\end{des}

\end{lemma} 

\begin{proof}
Firstly we prove that if $G$ is quasi-solved (not in solved 
form) and different from {\em FAIL} then there exists a $\reducevar$-rule
 applicable to $G$\/. We have two possibilities: If $P$ is not
 empty, then $P$ contains an approximation statement of the form $x 
 \rightarrow y$ and we can apply the rule
 \textsf{Produced variable elimination}. Otherwise ($P$ is empty), it 
 holds that 
 for all $x \lazo y \in E$ we have that
 $x,y \not \in \textsf{pvar}(P)$. Then we can apply
 either rule \textsf{Identity} or rule \textsf{Non-produced variable
 elimination}. The proof of the first item of the theorem proceeds by 
 inspecting all variable elimination rules.
 For rule \textsf{Identity}, it is enough to take $\sigma'_{d}$ equal 
 to $\sigma_{d}$. Then it is trivial that $\sigma'_{d} \in 
 \textsf{Sol}(G')$ and $\sigma'_{d} \leq \sigma_{d}$. For \textsf{Produced variable elimination}, we 
 define $\sigma'_{d}$ as: $\sigma'_{d}(y)=w$, where $w$ is a fresh 
 variable, and  $\sigma'_{d}(z)=\sigma_{d}(z)$ otherwise. Finally, for \textsf{Non-produced variable elimination} it is known that $\sigma_{d}(x) 
 \lazo \sigma_{d}(y)$ is $\gpc$-provable. Then, there is $t \in 
 \terminos$ such that the approximation statements $\sigma_{d}(x) 
 \rightarrow t$ and $\sigma_{d}(y) \rightarrow t$ are 
 $\gpc$-provable. Let us define $\sigma'_{d}$ as follows: 
 $\sigma'_{d}(x)=\sigma'_{d}(y)=t$, and 
 $\sigma'_{d}(z)=\sigma_{d}(z)$ otherwise. From 
 Theorem \ref{equivalencia} \textsf{(c)}, it holds that $\sigma_{d}(x) 
 \equivalente \sigma_{d}(y) \equivalente t = \sigma'_{d}(x)=\sigma'_{d}(y)$. 
 Then $\sigma_{d} =_{{\cal C}} \sigma'_{d}$ and trivially 
 $\sigma'_{d}$ is a solution for $G'$.
 The rest of the  lemma is straightforward. 
\end{proof}

Using this lemma we get:

\begin{theorem}[Completeness of $\reducevar$]
\label{completitud1}
Let $\programa$ be a program with ${\cal C}$ strongly regular.
Let $G$ be a quasi-solved goal and $\sigma_{d} \in \textsf{Sol}(G)$.
There exists a solved goal $\exists \tupla  \cdot S \Box \Box$
such that $G \reducevarderi \exists \tupla \cdot  S \Box \Box$ and
$\sigma_{d_{S}} \leq_{{\cal C}} \sigma_{d}$\/. \hfill \square
 
\end{theorem}

\begin{proof}
From Lemma \ref{progreso1} it is possible
to build a $\reducevar$-derivation $G \equiv  
G_0 \reducevar G_1$ $ \reducevar$ $G_2 \ldots$ such that
$G_{i+1}  \triangleright G_i$  and for which there exist
$\sigma_{d}  \equiv \sigma_{d_{0}},\sigma_{d_{1}}, \sigma_{d_{2}}, \ldots$, verifying
$\sigma_{d_{i}} \in \textsf{Sol}(G_i)$ and
$\sigma_{d_{i+1}} \leq _{{\cal C}} \sigma_{d_{i}}$.
Since $\triangleright$ is a well-founded ordering, the derivation
at hand is finite, ending in some goal $G_n \equiv  \exists \tupla 
\cdot S \Box \Box$. Hence, it holds that $\sigma_{d_{n}} \leq_{{\cal 
C}} \sigma_{d}$. Now, the result follows if we prove that 
$\sigma_{d_{S}} \leq \sigma_{d_{n}}$. 
Consider $x \in {\it DVar}$. If $x=s \in S$, then
$x\sigma_{d_{n}} = s \sigma_{d_{n}} = 
x \sigma_{d_{S}} \sigma_{d_{n}}$\/. If $x$ does not occur in $S$ then 
$x \sigma_{d_{n}} = x \sigma_{d_{S}} 
\sigma_{d_{n}}$. Hence the result holds. 
\end{proof}

Note that the theorem above establishes that  $\sigma_{d_{S}} 
\leq_{{\cal C}} \sigma_{d}$.  In general, $\sigma_{d_{S}} 
\leq  \sigma_{d}$  cannot be guaranteed, because of the 
variable elimination rule \textsf{Non-produced variable elimination}.  
Let us illustrate this 
by means of a simple 
example.

\begin{exmp} Consider the quasi-solved goal $G \equiv \Box \Box x 
\lazo y$ for a program $\programa$ such that the equational axiom $a 
\approx b$ belongs to $\cal C$. Then the data substitution 
$\sigma_{d}$ defined as $\sigma_{d}(x)=a$, $\sigma_{d}(y)=b$ and 
$\sigma_{d}(z)=z$ otherwise, is 
clearly a solution for $G$. In particular, note that we can prove
$ x\sigma_{d}  \lazo y\sigma_{d}$ in $\gpc$ by using the  
rule $\textsf{(J)}$, proving previously
$x \sigma_{d} \rightarrow a$ (by using $\textsf{(DC)}$) and 
$y\sigma_{d} \rightarrow a$ (by using $\textsf{(OMUT)}$ with 
the oriented equational axiom $b \rightarrow a$).

Now, the unique possible applicable rule to $G$ is the rule \textsf{Non-produced variable elimination}, which transforms $G$ into the 
solved goal $G' \equiv x = y \Box \Box$.   Clearly $\sigma_{d_{S}} 
\leq  \sigma_{d}$ does not holds, but $\sigma_{d_{S}} 
\leq_{{\cal C}} \sigma_{d}$  holds, because of the substitution 
$\sigma'_{d}$ defined as $\sigma'_{d}(y)=a$, $\sigma'_{d}(z)=z$ 
otherwise. In fact, for this $\sigma'_{d}$ we get $x \sigma_{d} = a = 
x \sigma_{d_{S}} \sigma'_{d}$, $y \sigma_{d} = b \equivalente a = y 
\sigma_{d_{S}} \sigma'_{d}$, $z \sigma_{d} = z = z 
\sigma_{d_{S}}\sigma'_{d}$, for $z \not = x,y$. \hfill \blacksquare   

\end{exmp}

From Theorems \ref{completitud} and \ref{completitud1} we get
our final completeness result.

\begin{theorem}[Completeness of $\calculo$]
\label{completitud2}
Let $\programa$ be a program with ${\cal C}$ strongly regular.
Let $G$ be an initial goal and $\sigma_{d} \in \textsf{Sol}(G)$. There exist a quasi-solved goal
$G'$ and a solved goal $G'' \equiv \exists \tupla \cdot S \Box \Box$
such that $G \reducederi G' \reducevarderi G''$ and
$\sigma_{d_{S}} \leq_{{\cal C}} \sigma_{d}[{\it dvar}(G)]$. Furthermore,
if $G$ and ${\cal P}$ are well-typed then
$G\sigma_{d_{S}}$ is well-typed. \hfill \square

\end{theorem}  

\begin{proof}
From Theorem \ref{completitud} it holds that there exist a 
quasi-solved goal $G'$ and $\sigma'_{d} \in \textsf{Sol}(G')$ such that 
$G \reducederi G'$ and $\sigma_{d} = \sigma'_{d}[{\it dvar}(G)]$. 
From Theorem \ref{completitud1} it holds that there exists a goal 
$G'' \equiv  \exists \tupla \cdot S \Box \Box$ in solved form such that $G' 
\reducevarderi G''$ and $\sigma_{d_{S}} \leq_{{\cal C}} 
\sigma'_{d}$.  Hence,  the result follows trivially.

The well-typedness of $G\sigma_{d_{S}}$  follows from Theorem 
\ref{completitud} and Lemma \ref{progreso1}. The proof can be 
reasoned as follows: Let $V$ be an environment such that $V \in {\sf 
env}(G)$. Then, Theorem \ref{completitud} ensures that there exists an 
environment $V'$ extending $V$ such that $V' \in \textsf{env}(G')$, 
i.e., $G'$ is well-typed w.r.t. $V'$. Now, applying repeatedly Lemma 
\ref{progreso1}, we have that $V' \in \textsf{env}(G'')$, i.e., $G''$ is 
well-typed w.r.t. $V'$. But note that since $V'$ extends $V$, then 
$V' \in \textsf{env}(G)$. On the other hand, for all $x=t \in S$, it is 
known that there exists a common type $\tau$ for $x$ and $t$ in $V'$. 
Hence, the effect of applying $\sigma_{d_{S}}$ to $G$ is to replace 
variables by terms which have the same type (in $V'$) that the variables which 
replace, i.e., $V' \in \textsf{env}(G\sigma_{d_{S}})$. 
\end{proof}

\section{Conclusions}
\label{conclusiones}

We have presented a general framework for functional logic  
programming  with {\em algebraic polymorphic datatypes}, whose 
data constructors can be controlled
by a specified set of equational  axioms. 
Defined functions are lazy and possibly non-deterministic.
The combination of all these features together is not found in other 
related works we are aware of
 \cite{JP89, Jay92, Leg94, DOPR91, DR93, 
DOPR96, DPR96, Mes92, Mes93, CDE+99, KKV95, BKK+96,
DFI+98, Llo99, MM95, HS90, GHSST92}.

Following the CRWL approach \cite{GHLR99}, we have given  rewriting 
calculi and models which provide an adequate  declarative  
semantics  for  our programs.  This   is   shown   by   the   existence   of   free   models   for 
programs (Theorem \ref{modelolibre}), the adequateness of the rewriting calculi 
w.r.t.  models  (Theorem \ref{adecuacion}),   and   type preservation   results 
(Theorems \ref{conservacion}, \ref{ape1} and  \ref{modelolibre}).
We have also presented a narrowing calculus for goal solving (named 
$\calculo$), proving soundness (Theorem \ref{correccion}), completeness 
(Theorem \ref{completitud2}) 
and well-typedness of computed answers (Theorem \ref{completitud2}). 
$\calculo$ is not intended as an operational model, but rather as an 
abstract description of goal solving that provides a  convenient 
basis  for soundness and completeness results, while ignoring  
control issues and implementation details.

In the near future, we plan to implement the instance of our framework given by 
the equational axioms for multisets, which is expected to allow for many 
optimizations w.r.t. the general case. A first step in this direction can be 
found in \cite{ALR98}. We also aim at enriching our framework with constraints, 
coming from a constraint system given as a suitable extension of the equational 
axioms for the data constructors. For instance, for sets and multisets we could 
introduce disequality and membership constraints, in analogy to \cite{DR93,Leg94}.
For the particular case of multisets, the enriched framework has been 
successfully developed by the first author in her Phd thesis \cite{Are98}. The 
extension of this work towards a general scheme for declarative programming with 
constraints over arbitrary algebraic datatypes, is left for future research. 

 \bigskip
 
\noindent
{\bf Acknowledgments:}   The authors 
are  indebted to their colleagues 
A. Gil-Luezas and F.J. L\'opez-Fraguas for their
support to the development of this work.  The constructive criticisms 
of several anonymous referees have helped to improve an older version 
of the paper.

\bibliography{biblio}

\newpage
 
\appendix

\section{Another lazy narrowing calculus based on Equational Constructors}
\label{apendice}

Consider the following goal solving calculus resulting of adding
\textsf{Mutation} rules to the lazy narrowing calculus 
from~\cite{GHLR99}:

\medskip

\noindent
{\bf Transformation rules for $\lazo$\/:} \\

\noindent
{\small
 \( \begin{array}{lllll}
\textsf{Decomposition}_{\lazo}:
                 \ \exists \tupla \cdot S \Box P \Box c(\etupla_n) \lazo
                    c(\eprimatupla_n) , E
      \reduceap  \exists \tupla \cdot  S \Box P \Box  \ldots,e_i \lazo e'_i,
 \ldots, E \\

\textsf{Mutation}_{\lazo}:
        \ \exists \tupla  \cdot S \Box P \Box
           c(\etupla_n) \lazo e' , E
     \reduceap   \\

\ \ \ \ \ \ \ \ \ \ \ \ \ \ \ \ \ \ \ \ \ \ \exists \stackrel{\_}{x},
     \tupla \cdot  S \Box \ldots, e_i \rightarrow t_i, \ldots, P \Box
      C, s \lazo e',  E \\
\ \ \ \ \ \ \ \ \ \mbox{where  $e' \not \in {\it DVar}$ and
    ${\it Eq}: c(\ttupla_n) \rightarrow s \Leftarrow C$
      is a variant of a rule in ${\cal C}_{\rightarrow}$} \\
\ \ \ \ \ \ \ \ \ \mbox{with $\stackrel{\_}{x}={\it dvar}(Eq)$ fresh variables.} 
\end{array}\)

\noindent
\( \begin{array}{llll}
\mbox{\sf Identity:}  \ \exists \tupla \cdot S  \Box P \Box x \lazo x , E
    \reduceap  \exists \tupla \cdot  S \Box P \Box  E \\

\ \ \ \ \ \ \ \ \ \mbox{if $x \not \in {\sf pvar}(P)$.} \\

\mbox{\sf Binding:} \ \exists \tupla \cdot S \Box P \Box  x \lazo s, E 
    \reduceap  \exists \tupla \cdot x = s, (S \Box P \Box E)\sigma_{d} \\
\ \ \ \ \ \ \ \ \ \mbox{if $s \in \terminos$ , $x \not =  s$, $x \not \in  {\sf pvar}(P)$ and
                                          $dvar(s) \cap {\sf
    pvar}(P)=\emptyset$,} \\
\ \ \ \ \ \ \ \ \ \mbox{where $\sigma_{d}=\{x/s\}$.} \\

\textsf{Imitation}_{\lazo}: \ \exists \tupla \cdot S \Box P \Box x \lazo 
   c(\etupla_n), E  \reduceap \\

\ \ \ \ \ \ \ \ \ \ \ \ \ \ \ \ \ \ \ \ \ \ \exists \xtupla_n, \tupla \cdot x = c(\xtupla_n), (S \Box P \Box 
   \ldots ,x_i \lazo e_i, \ldots, E)\sigma_{d} \\
\ \ \ \ \ \ \ \ \ \mbox{if $c(\etupla_n) \not \in
   \terminos$ or ${\it dvar}(c(\etupla_n)) \cap {\sf pvar}(P)
\neq \emptyset$\/, $x \not \in {\sf pvar}(P)$} \\
\ \ \ \ \ \ \ \ \ \mbox{where $\sigma_{d}=\{x/c(\xtupla_n)\}$ and $\xtupla_n$ fresh variables.}\\
\textsf{Narrowing}_{\lazo}: \ \exists \tupla \cdot S \Box P \Box
f(\etupla_n) \lazo e', E
      \reduceap  \\
\ \ \ \ \ \ \ \ \ \ \ \ \ \ \ \ \ \ \ \ \ \ \exists \xtupla, \tupla \cdot 
  S  \Box \ldots, e_i \rightarrow t_i,\ldots , P \Box C,
       r \lazo e', E\\
\ \ \ \ \ \ \ \ \ \mbox{where ${\it Rul}=f(\ttupla_n) \rightarrow r 
            \Leftarrow C$ is a variant of a rule in ${\cal R}$} \\

\ \ \ \ \ \ \ \ \ \mbox{with $\xtupla={\it dvar}({\it Rul})$ fresh variables.}

\end{array} \) \\
}

\medskip

\noindent
{\bf Transformation rules for $\rightarrow$\/:} \\

\medskip

\noindent
{\small
\( \begin{array}{llll}
\textsf{Decomposition}_{\rightarrow}: \ \exists \tupla \cdot S \Box c(\etupla_n) \rightarrow
   c(\ttupla_n), P \Box E  \reduceap 
  \exists \tupla \cdot  S \Box \ldots, e_i \rightarrow t_i ,\ldots, P \Box E\\
\textsf{Mutation}_{\rightarrow}:
 \ \exists \tupla \cdot  S \Box c(\etupla_n) \rightarrow t, P \Box  E
     \reduceap
    \ \exists \xtupla, \tupla \cdot S \Box
     \ldots, e_i \rightarrow t_i, \ldots, s \rightarrow t,P \Box C, E \\
\ \ \ \ \ \ \ \ \ \mbox{where $t \not \in {\it DVar}$ and  
     ${\it Eq} : c(\ttupla_n)
                  \rightarrow
                     s \Leftarrow C$ is a variant
                     of a rule in ${\cal C}_{\rightarrow}$} \\
\ \ \ \ \ \ \ \ \ \mbox{with $\xtupla={\it dvar}({\it Eq})$ fresh
      variables.} \\ 

\mbox{\sf Output binding:} \ \exists \tupla \cdot S \Box x \rightarrow t, P 
        \Box E  \reduceap  \exists \tupla \cdot  [x = t], 
          (S \Box P \Box E)\sigma_{d} \\
 \ \ \ \ \ \ \ \ \ \mbox{if $t \not \in {\it DVar}$,  where $\sigma_{d}=\{x/t\}$.} \\
  
\mbox{\sf Input binding:} \ \exists x, \tupla \cdot  S \Box t \rightarrow x, 
    P \Box E  \reduceap   \exists \tupla \cdot S \Box 
   (P \Box E)\sigma_{d} \\
\ \ \ \ \ \ \ \ \ \mbox{if $t \in \terminos$, 
       where $\sigma_{d}=\{x/t\}$}\\

 \mbox{\sf Input imitation:}\ \exists x, \tupla \cdot  S \Box c(\etupla_n) 
      \rightarrow x,P \Box E  \reduceap 
\ \exists \xtupla_n,\tupla \cdot  S \Box (\ldots, e_i \rightarrow x_i,
      \ldots, P \Box E)\sigma_{d} \\
\ \ \ \ \ \ \ \ \ \mbox{if  $c(\etupla_n) \not \in  
   \terminos$, $x \in {\sf dmvar}(E)$, where
           $\sigma_{d}=\{x/c(\xtupla_n)\}$,} \\

\ \ \ \ \ \ \ \ \ \mbox{$\xtupla_n$ fresh variables.}\\

\textsf{Elimination}_{\rightarrow}:
 \  \exists x, \tupla \cdot  S \Box e \rightarrow x,P
      \Box E  \reduceap  \exists \tupla \cdot    S \Box P \Box E  \ \ \ \ \ \ \ \ \ \mbox{if $x \not \in {\it dvar}(P \Box E)$\/.} \\

\textsf{Narrowing}_{\rightarrow}:
 \ \exists  \tupla \cdot  S \Box f(\etupla_n)
     \rightarrow t, P \Box E  \reduceap   \exists  \xtupla,
      \tupla \cdot S  \Box \ldots, e_i \rightarrow t_i,\ldots,r
      \rightarrow t, P \Box C,E \\
\ \ \ \ \ \ \ \ \ \mbox{if  $t \not \in {\it DVar}$ or  $t \in {\sf dmvar}(G)$\/, where
       ${\it Rul}: f(\ttupla_n) \rightarrow r
       \Leftarrow C$ is a variant of a rule} \\ 
\ \ \ \ \ \ \ \ \ \mbox{in ${\cal R}$ with $\xtupla={\it dvar}({\it Rul})$
 fresh variables\/.}
\end{array} \) \\
}
\medskip

\noindent
{\em Failure rules} are  the same that those for $\calculo$.
Considering the new definition of solution:

\begin{definition}[Solution]
 Let $\programa$ be a program.
Let $G \equiv   \exists \tupla \cdot  S \Box P \Box E$ be an admissible
goal for ${\cal P}$ and $\sigma_{d} \in {\it DSub}_{\perp}$\/.
We say $\sigma_{d}$ is a solution of $G$ iff

\begin{des}
\item $x\sigma_{d} \in \terminos$ for all $x \in {\it DVar}-\textsf{pvar}(P)$\/;
\item $x\sigma_{d}  \equivalente s\sigma_{d}$ for all $x=s \in S$\/;
\item For all $e \lazo e' \in E$ and $e'' \rightarrow t \in P$
      there exists a $\gpc$-proof of $e\sigma_{d} \lazo e'\sigma_{d}$ and
      $e''\sigma_{d} \rightarrow t\sigma_{d}$ respectively. 
   \hfill \square
\end{des}
\end{definition}

\noindent
We ensure that
the lazy narrowing calculus above is sound
in the sense of Theorem~\ref{correccion} and conjecture that possibly
verifies the following  completeness theorem:

\begin{conj}

Let $\programa$ be a program where ${\cal C}$ is strongly regular,
$G$ an initial goal and $\sigma_{d} \in \textsf{Sol}(G)$. Then there exists a solved form
$\exists \tupla \cdot  S \Box \Box$ such that $G \reduceap^* \exists 
\tupla  \cdot S \Box \Box$ and
$\sigma_{d_{S}} \leq_{\cal C} \sigma_{d}[dvar(G)]$.  Furthermore if ${\cal P}$ and $G$ and
well-typed then  $G\sigma_{d_{S}}$ is well-typed. \hfill \square
\end{conj}

\section{Proofs}
\label{demos}


\begin{proof}[Proof of  Theorem \ref{equivalencia}]

\noindent
\textsf{(a $\Rightarrow$)}. The result holds because
  any step within a $\gpc$-proof can be easily replaced
 by one or several $\bpc$-steps. This is trivial for
\textsf{(B)}, \textsf{(RR)}, \textsf{(DC)} and \textsf{(J)} rules.
A \textsf{(OMUT)}-step can be replaced by several
$\bpc$-steps, according to the following scheme:

\begin{center}
\textsf{(TR)} $\frac{\mbox{{\sf (MN)}} 
  \frac{\mbox{$e_1 \rightarrow t_1, 
\ldots, e_n \rightarrow t_n$}}{\mbox{$c(e_1, \ldots, e_n) \rightarrow c(t_1, \ldots, t_n)$}} 
          \qquad 
  \mbox{\sf (TR)} \frac{\mbox{\sf (MUT)} 
\frac{}{\mbox{$c(t_1, \ldots, t_n) \rightarrow s$}} \qquad \frac{}{\mbox{$s 
\rightarrow t$}}}
{\mbox{$c(t_1, \ldots, t_n) \rightarrow t$}} 
}{\mbox{$c(e_1, \ldots, e_n) \rightarrow t$}}$
\end{center}

\noindent
where $c(t_1, \ldots, t_n) \ecuacional s \in {[{\cal C}]}_{\ecuacional}$\/.
Analogously, a \textsf{(OR)}-step can be simulated in
$\bpc$ as follows:
 
\begin{center}
\textsf{(TR)} $\frac{\mbox{{\sf (MN)}} \frac{\mbox{$e_1 \rightarrow t_1, 
\ldots, e_n \rightarrow t_n$}}{\mbox{$f(e_1, \ldots, e_n) \rightarrow 
f(t_1, \ldots, t_n)$}} \qquad \mbox{\sf (TR)} \frac{\mbox{\sf (R)} 
\frac{\mbox{$C$}}{\mbox{$f(t_1, \ldots, t_n) \rightarrow r$}} \qquad 
\frac{}{\mbox{$r \rightarrow t$}}}{\mbox{$f(t_1, \ldots, t_n) \rightarrow t$}}
}{\mbox{$f(e_1, \ldots, e_n) \rightarrow t$}}$
\end{center}

\noindent
where $f(t_1, \ldots, t_n) \rightarrow r \Leftarrow C
\in {[{\cal R}]}_{\rightarrow}$\/. \\

\noindent
\textsf{(a $\Leftarrow$)}.  Due to the inference rule {\sf (J)}, it is 
enough to prove that every $\bpc$-provable approximation statement $e 
\rightarrow t$ is also $\gpc$-provable. We reason by induction on the 
length of the given $\bpc$-derivation. if $t = \perp$, then $e 
\rightarrow t$ can be derived by rule ${\sf (B)}$. If $e$ is some 
variable $x$, then $t$ must be also $x$ (because $x \rightarrow t$ 
with $t \not = x$ cannot be proved in $\bpc$ with $\cal C$ being 
strongly regular) and $x \rightarrow x$ can be derived by rule {\sf 
(RR)}. Otherwise, we can assume $e = h(e_{1}, \ldots, e_{n})$ for 
some $h \in {\it DC}^n \cup {\it FS}^n$. Now we can distinguish three 
cases: 

\smallskip

\begin{des}
\item[(i)] $h = f \in {\it FS}^n$. Then, from the $\bpc$-proof of $e 
\rightarrow t$ (of length $l$, say) we obtain a sequence of rewrite 
steps:

\vspace*{-0.10cm}

\[ f(e_{1}, \ldots, e_{n}) \rightarrow^{*} f(t_{1}, \ldots, t_{n}) 
\rightarrow r \rightarrow^{*} t \]

\vspace*{-0.10cm}

\noindent
where each step  applies either a rewrite rule of the form $e 
\rightarrow \perp$, or a rewrite rule from $[{\cal R}]_{\rightarrow}$, 
or a rewrite rule of the form $s' \rightarrow s$ such that $s' 
\ecuacional s \in [{\cal C}]_{\ecuacional}$. In particular, the 
rewrite step $f(t_{1}, \ldots, t_{n}) \rightarrow r$ will correspond 
to some rewrite rule $f(t_{1}, \ldots, t_{n}) \rightarrow r \Leftarrow 
C \in [{\cal R}]_{\rightarrow}$. By induction hypothesis, we can 
assume that $e_{i} \rightarrow t_{i}$ ($1 \leq i \leq n$), $C$ and $r 
\rightarrow t$ are $\gpc$-derivable (because they have $\bpc$-proofs 
of size less than $l$). Then we can conclude that $f(e_{1}, \ldots, 
e_{n}) \rightarrow t$ is $\gpc$-derivable by applying ${\sf (OR)}$.

\item[(ii)] $h=c \in {\it DC}^n$, $t=c(t_{1}, \ldots, t_{n})$ and 
the $\bpc$-proof of $e \rightarrow t$ determines a sequence of rewrite 
steps:

\vspace*{-0.10cm}

\[ c(e_{1}, \ldots, e_{n}) \rightarrow^{*}  c(t_{1}, \ldots, t_{n})\]

\vspace*{-0.10cm}

\noindent
where each step uses some of the rewrite rules mentioned in (i), 
applied at some position strictly below the root. Then, we can use 
the induction hypothesis to find $\gpc$-proofs for $e_{i} \rightarrow 
t_{i}$ ($1 \leq i \leq n$), and apply rule ${\sf (DC)}$ to conclude 
that $c(e_{1}, \ldots, e_{n}) \rightarrow c(t_{1}, \ldots, t_{n})$ 
is $\gpc$-derivable.

\item[(iii)] $h = c \in {\it DC}^n$ and the $\bpc$-proof of $e 
\rightarrow t$ determines a sequence of rewrite steps

\vspace*{-0.10cm}

\[ c(e_{1}, \ldots, e_{n}) \rightarrow^{*} c(t_{1}, \ldots, t_{n}) 
  \rightarrow  s \rightarrow^{*} t \]

\vspace*{-0.10cm}

\noindent
where each step uses some of the rewrite rules mentioned in (i), and 
in particular, the step $c(t_{1}, \ldots, t_{n}) \rightarrow s$ is 
such that $c(t_{1}, \ldots, t_{n}) \ecuacional s \in [{\cal 
C}]_{\ecuacional}$. By induction hypothesis, we can assume the 
existence of $\gpc$-proofs for $e_{i} \rightarrow t_{i}$ ($1 \leq i 
\leq n$) and $s \rightarrow t$. Then, we can apply rule {\sf (OMUT)} 
to conclude that $c(e_{1}, \ldots, e_{n}) \rightarrow t$ is 
$\gpc$-derivable.

\end{des}

\smallskip

\noindent
\textsf{(b)}. It is straightforward from the structures of the 
 inequational calculus  presented in Definition \ref{calin}
     and the $\bpc$ calculus. 
     
\smallskip

\noindent
\textsf{(c $\Rightarrow$)}. If $s \lazo t$ is $\bpc$-provable, 
then there 
exists
      $t' \in \terminos$ such that $s \rightarrow t'$ and
      $t \rightarrow t'$ are $\bpc$-provable. From \textsf{(b)}, we get that
      $s \aproximado t'$ and $t \aproximado t'$. From 
Proposition \ref{caracteristicas} \textsf{(c)},
      it  follows that $s,t \in \terminos$ and $s \equivalente t' 
      \equivalente t$\/. 
      
\smallskip

\noindent
\textsf{(c $\Leftarrow$)}. From $s \equivalente t$ and item \textsf{(b)}, we get
    that $s \rightarrow s$ and $t \rightarrow s$ are $\bpc$-provable. 
    As
    $s \in \terminos$, we can apply \textsf{(J)} to get the result.
\end{proof}

\medskip
    
    
\begin{proof}[Proof of Lemma \ref{instancias}]

\noindent
\textsf{(a)}. Assume that $\{x_1, \ldots, x_m\} =_{{\it Def}}
  \{x \in \bigcup_{i=1}^n dvar(t_i) \ \mid
  \ x\sigma_{d} \not = x\}$\/. Suppose that
   $x_i\sigma_{d} = t'_i$\/, $1 \leq i \leq m$\/.
  Since $t_i\sigma_{d} \in {\it Term}_{\Sigma_{\perp}}^{\tau_i\sigma_{t}}(V)$
    then, applying $n$ times Lemma \ref{lema5}, there
   exist $\tau'_i$, $1 \leq i \leq m$, such that 
    $t'_i \in  {\it Term}_{\Sigma_{\perp}}^{\tau'_i}(V)$ and
    $V[x_1:\tau'_1, \ldots , x_m:\tau'_m] \deduce t_i :\tau_i\sigma_{t}$,
   $1 \leq i \leq n$\/. From Lemma \ref{lema3} \textsf{(a)}, it  follows
  that $V[x_1:\tau'_1, \ldots , x_m:\tau'_m] \deduce r:\tau\sigma_{t}$.
   Now, from Lemma \ref{lema4},
  $r\sigma_{d} \in {\it Expr}_{\Sigma_{\perp}}^{\tau\sigma_{t}}(V)$. 
  
 \smallskip

\noindent
\textsf{(b)}.  Assume that $y_1, \ldots, y_k$ are all the variables
   of $c(t_1, \ldots, t_n)$ such that $y_j$, $1 \leq j \leq k$,
   occurs $n_j > 1$ times in $c(t_1, \ldots, t_n)$\/. Consider
   new variables $y_{ij}$, $1 \leq i \leq k$, $2 \leq j \leq n_i$\/.
   Let $c(t'_1, \ldots, t'_n)$ be the term resulting  by
   replacing in $c(t_1, \ldots, t_n)$ each $p$-th occurrence
   of $y_i$ by $y_{ip}$, $1 \leq i \leq k$,
   $2 \leq p \leq n_i$\/. For $\sigma_{d}' \in {\it DSub}_{\perp}$
   defined as:

     \[ \begin{array}{lllll}
      z\sigma_{d}' & = &
     \left\{
     \begin{array}{llll}
      y_i\sigma_{d} & z = y_{ij}, 1 \leq i \leq k, 2 \leq j \leq n_i \\
      z\sigma_{d}  & {\it otherwise}
     \end{array} \right.
      \end{array} \]

\noindent
we have that $c(t'_1, \ldots, t'_n)\sigma_{d}' =
c(t_1, \ldots, t_n)\sigma_{d}$ and $d(s_{1}, \ldots, s_{m}) 
\sigma_{d} = d(s_{1}, \ldots,$ $ s_{m}) \sigma_{d}'$\/. 

\smallskip

Assume now that $\{x_1, \ldots, x_l\} =_{{\it Def}}
  \{x \in \bigcup_{i=1}^n dvar(t'_i) \ \mid
  \ x\sigma_{d}' \not = x\}$\/. Suppose that
   $x_i\sigma_{d}' = t''_i$\/, $1 \leq i \leq l$\/.
  Since $t_i'\sigma_{d}' = t_i\sigma_{d} \in {\it 
Term}_{\Sigma_{\perp}}^{\tau_{i}\sigma_{t}}(V)$
    then, applying $n$ times Lemma \ref{lema5}, there
   exist $\tau''_i$, $1 \leq i \leq l$, such that
    $t''_i \in  {\it Term}_{\Sigma_{\perp}}^{\tau''_i}(V)$ and
    $V[x_1:\tau''_1, \ldots , x_l:\tau''_l] \deduce t'_i :\tau_i\sigma_{t}$,
   $1 \leq i \leq n$\/. On the other hand,  note that
   all variables $y_{ij}$, $y_{i}$, $1 \leq i \leq k$,
   $2 \leq j \leq n_{i}$, are annotated in
   $V[x_1:\tau''_1, \ldots , x_l:\tau''_l]$ with the same
   type-annotation. Therefore, we have also that
   $V[x_1:\tau''_1, \ldots , x_l:\tau''_l] \deduce t_i 
   :\tau_i\sigma_{t}$\/.
  Now, Lemma \ref{lema3} \textsf{(b)} ensures that
  $V[x_1:\tau''_1, \ldots , x_l:\tau''_l] \deduce
  d(s_1, \ldots, s_m) :\tau\sigma_{t}$. Finally,
  Lemma \ref{lema4} entails
$d(s_1, \ldots, s_m)\sigma_{d}' \in
{\it Term}_{\Sigma_{\perp}}^{\tau\sigma_{t}}(V)$. Now, the result 
follows from $d(s_1, \ldots, s_m)\sigma_{d}'= d(s_1, \ldots, 
s_m)\sigma_{d}$\/. 
\end{proof}

\medskip


\begin{proof}[Proof the Theorem \ref{ape1}]

\noindent
Let us prove that $\modterm$ verifies all items in   
Definition \ref{algebras}.  
From Proposition \ref{caracteristicas}  we have that $\dominio$ is a poset with partial
order $\sqsubseteq^{\modterm}$ and bottom element 
$\perp^{\modterm}$\/, i.e. item \textsf{(1)} holds. 

\medskip

Item \textsf{(2)}: Let us prove that 
${\cal E}^{\modterm}(\tau)$ is 
a cone in $\dominio$, for all $\tau \in T_{{\it TC}}(A)$\/.
Consider $[t_1],[t_2] \in \dominio$ such that
$[t_1] \in {\cal E}^{\modterm}(\tau)$ (i.e. $V \deduce t_{1}:\tau$) and
$[t_2] \sqsubseteq^{\modterm} [t_1]$\/. Then
$t_1 \aproximado t_2$, that is,
from Theorem \ref{equivalencia},
$t_1 \rightarrow_{\cal P} t_2$\/. Now, from
the proof of Theorem \ref{conservacion}, and assuming $\mathcal{C}$
strongly regular and well-typed, we get 
$t_2 \in {\it Term}_{\Sigma_{\perp}}^{\tau}(V)$\/,
that is $[t_2] \in {\cal E}^{\modterm}(\tau)$. 

\medskip

Item \textsf{(3)} is trivial. In order to prove items \textsf{(4)} and \textsf{(5)}, we have 
to check firstly that for all $h \in  {\it DC}^{n} \cup {\it FS}^n$,
$h^{\modterm}$ is well-defined, i.e. it does not depend on
the selected representants of the equivalence classes.
Consider elements $t_{i},
s_{i} \in {\it Term}_{\Sigma_{\perp}}(X)$, such that 
$t_{i} \equivalente s_{i}$, $1 \leq i \leq n$. If
$s_{i} \equivalente t_{i}$, $1 \leq i \leq n$, then by 
Theorem \ref{equivalencia} and the $\bpc$-rule
\textsf{(MN)}, it holds that for any $c \in {\it DC}^n$: 
$c(t_{1} \ldots, t_{n}) \equivalente c(s_{1}, \ldots, 
s_{n})$, i.e. $[c(t_{1} \ldots, t_{n})]=[c(s_{1} \ldots, s_{n})]$. 
Hence, $c^{\modterm}([t_{1}], \ldots, [t_{n}]) =
c^{\modterm}([s_{1}], \ldots,$ $ [s_{n}])$\/.
Similarly, for all $f \in {\it FS}^n$,
$f^{\modterm}([t_{1}], \ldots, 
[t_{n}])=\{[t] \ \mid \ f(t_{1}, \ldots, t_{n}) \rightarrow_{{\cal P}} 
t\}$\/. But, from Theorem \ref{equivalencia} and the $\bpc$-rules 
\textsf{(MN)} and \textsf{(TR)}, it holds that $f^{\modterm}([t_{1}], \ldots, 
[t_{n}]) = f^{\modterm}([s_{1}], \ldots, 
[s_{n}])$.

\smallskip
 
The monotonicity of $h^{\modterm}$, for all $h \in {\it 
DC}^n \cup {\it FS}^n$ follows from the following fact: If $[t_{i}] 
\sqsubseteq^{\modterm} [s_{i}]$, $1 \leq i \leq n$, then 
$s_{i} \aproximado t_{i}$.  Theorem \ref{equivalencia} \textsf{(b)} along with
the $\bpc$-rule
\textsf{(MN)}  entail $h(s_{1}, \ldots, s_{n})  \rightarrow_{{\cal P}}
h(t_{1}, \ldots, t_{n})$. If $h \in {\it DC}^n$,  it is clear that
$[h(t_{1}, \ldots, t_{n})] \sqsubseteq^{\modterm} [h(s_{1}, \ldots, 
s_{n})]$, and hence $h^{\modterm}([t_{1}], \ldots, [t_{n}]) \subseteq
h^{\modterm}([s_{1}], \ldots, [s_{n}])$. Otherwise, if $h \in {\it 
FS}^n$ then, for any $[t] \in h^{\modterm}([t_{1}], \ldots,$ $ [t_{n}])$
--by Theorem \ref{equivalencia} together with the $\bpc$-rule \textsf{(TR)}-- 
we have that $h(s_{1}, \ldots, s_{n})$ $ 
\rightarrow_{{\cal P}} t$, i.e. $[t] \in 
h^{\modterm}([s_{1}], \ldots, [s_{n}])$. Hence 
$h^{\modterm}([t_{1}], \ldots, [t_{n}]) \subseteq
h^{\modterm}([s_{1}],$ $ \ldots, [s_{n}])$\/. 

\smallskip

Knowing that $c^{\modterm}$ is monotonic for all $c \in {\it DC}$, 
item  \textsf{(4)} follows from the fact that  
$\textsf{Def}(\modterm)$ $ = \{ [t] \in \dominio
\ \mid \ t  \mbox{ is a total term}\}$\/. This is true because of 
Proposition \ref{caracteristicas} {\sf (c)}. Finally,  
item \textsf{(5)} follows from the monotonicity of $f^{\modterm}$ 
for all $f \in {\it FS}$, and the fact that  
$\{[t] \in \dominio \ \mid \ f(t_1, \ldots, t_n) \rightarrow_{\cal P} t\}$
is a cone. This is true by transitivity of rewriting (rule {\sf (TR)} 
in $\bpc$).

\smallskip

All valuations over the term algebra $\modterm$\/ can be represented
by means of substitutions. Any substitution $\sigma = (\sigma_{t},\sigma_{d})$\/
such that $\sigma_{t} : {\it TVar} \rightarrow {\it T}_{\it TC}(A)$
and $\sigma_{d} : {\it DVar} \rightarrow {\it Term}_{\Sigma_{\perp}}(X)$\/, represents
the valuation $[\sigma] = (\sigma_{t},[\sigma_{d}])$\/, where 
$[\sigma_{d}](x)= [\sigma_{d}(x)]$\/. It is easy to check that 
$\C \tau \J^{\modterm} \sigma_{t} = \tau \sigma_{t}$\/ for all $\tau \in 
{\it T}_{\it TC}(A)$\/, and
$\C t \J^{\modterm} [\sigma_{d}] = \langle [t \sigma_{d} ] \rangle$\/ for all $t \in 
{\it Term}_{\Sigma_{\perp}}(X)$\/. 

\smallskip

Assume now that $\mathcal{P}$ is strongly regular. 
In order to prove that $\modterm$ is well-typed,
suppose  that $c: (\tau_1, \ldots, \tau_n) \rightarrow \tau
\in {\it DC}$ and $[\sigma]=(\sigma_{t},[\sigma_{d}]) \in 
{\it Val}(\modterm)$\/. Consider any
$[t_i] \in {\cal E}^{\modterm}(\tau_i\sigma_{t})$, $1 \leq i \leq n$\/,
then $t_i \in {\it Term}^{\tau_i\sigma_{t}}_{\Sigma_{\perp}}(V)$\/,
$1 \leq i \leq n$. Thus, $c(t_1, \ldots, t_n)
\in {\it Term}^{\tau\sigma_{t}}_{\Sigma_{\perp}}(V)$, and
$[c(t_1, \ldots, t_n)] \in {\cal E}^{\modterm}(\tau\sigma_{t})$\/.
But ${\cal E}^{\modterm}(\tau\sigma_{t})$ is a cone, then we have
that $\langle [c(t_1, \ldots, t_n)] \rangle \subseteq
 {\cal E}^{\modterm}(\tau\sigma_{t})$, that is,
$c^{\modterm}([t_1], \ldots, [t_n]) \subseteq
{\cal E}^{\modterm}(\C \tau \J^{\modterm} \sigma_{t})$. 

\smallskip

Assume now that $f: (\tau_1, \ldots, \tau_n) \rightarrow \tau
\in {\it FS}$\/. Again, it holds that
$f(t_1, \ldots, t_n) \in {\it Term}^{\tau\sigma_{t}}_{\Sigma_{\perp}}(V)$.
From Theorem \ref{conservacion}, it is easy to check
that $\{[t] \ \mid \ f(t_1, \ldots, t_n) \rightarrow_{\cal P} t\}
\subseteq {\cal E}^{\modterm}(\tau\sigma_{t})$, that is,
$f^{\modterm}([t_1], \ldots, [t_n]) \subseteq
{\cal E}^{\modterm}(\C \tau \J^{\modterm} \sigma_{t})$.
\end{proof}

\medskip


\begin{proof}[Proof of  Characterization Lemma  \ref{validez}]

\noindent
\textsf{(a $\Rightarrow$)}. We argue by structural induction on
$e$\/. 

\begin{des}

\item If $e =  \perp$ then, due to the strong regularity of
      ${\cal C}$,  we have that $t =  \perp$. Hence it is enough 
      to apply the $\bpc$-rule \textsf{(B)}.

\item If $e \in  X \cup {\it DC}^0$\/, then  
 $\C e \J^{\modterm}[\sigma_{d}] = \langle  [e\sigma_{d}]\rangle$\/.
 Since $[t] \in \langle  [e\sigma_{d}]\rangle $, then
     $[t] \sqsubseteq^{\modterm} [e\sigma_{d}]$, that is,
     $e\sigma_{d} \aproximado t$.
     The result follows from Theorem \ref{equivalencia} \textsf{(b)}.

\item If $e = c(e_1, \ldots, e_n)$\/, $c \in {\it DC}^n$\/, then
   there exist elements $[s_i] \in \C e_i \J^{\modterm}
      [\sigma_{d}]$, $1 \leq i \leq n$, 
      such that
      $[t] \in c^{\modterm}([s_1], \ldots, [s_n]) =\langle  
       [c(s_1, \ldots, s_n)]\rangle$, i.e. $c(s_{1}, \ldots, s_{n}) 
       $ $\aproximado t$\/. From 
       Theorem \ref{equivalencia} \textsf{(b)}, we have
       that $c(s_1, \ldots, s_n) \rightarrow_{\cal P} t$.
      On the other hand, by induction hypothesis:   
       $e_i\sigma_{d} \rightarrow_{\cal P} s_i$,
      $1 \leq i \leq n$. Applying the $\bpc$-rule \textsf{(MN)}, we have
      that $c(e_1\sigma_{d}, \ldots, e_n\sigma_{d})
       $ $\rightarrow_{\cal P} c(s_1, \ldots ,s_n)$. 
       Then, from the $\bpc$-rule \textsf{(TR)},
     we can conclude $e\sigma_{d} \rightarrow_{\cal P} t$.

\item If $e =  f(e_1, \ldots, e_n)$, then 
there exist elements $[s_i] \in \C e_i \J^{\modterm}[\sigma_{d}]$, $1 
\leq i \leq n$, such that
$[t] \in f^{\modterm}([s_1], \ldots ,[s_n])$, that is,
$f(s_1, \ldots ,s_n) \rightarrow_{\cal P} t$. Then the last $\bpc$-rule applied in this $\bpc$-proof has been
either \textsf{(B)} or \textsf{(R)}. If \textsf{(B)}, then the result follows
trivially. Otherwise, $f(s_1, \ldots, s_n) \rightarrow t
\Leftarrow C \in {[\cal R]}_{\rightarrow}$. 
By induction hypothesis, $e_i\sigma_{d} \rightarrow_{\cal P} s_i$, $1 
\leq i \leq n$\/, 
then, by \textsf{(MN)}, we get that
$f(e_1\sigma_{d}, \ldots, e_n\sigma_{d}) \rightarrow_{\cal P}
f(s_1, \ldots, s_n)$\/. Using now the $\bpc$-rule \textsf{(TR)}, the 
result can be concluded.
 
\end{des}

\noindent
\textsf{(a $\Leftarrow$)}. We proceed by induction
 on the number of inference steps for
the $\gpc$-proof associated to $e\sigma_{d} \rightarrow_{\cal P} t$\/.
For that, let us analyze the last rule applied in such a
$\gpc$-proof.

\begin{des}

\item For rules \textsf{(B)} and \textsf{(RR)} the result is trivial.

\item For rule \textsf{(DC)}, we have that $e\sigma_{d} =  c(e_1, \ldots, e_n)$,
     $t =  c(t_1, \ldots, t_n)$ and
     $e_i \rightarrow_{\cal P} t_i$, $1 \leq i \leq n$\/.
      If $e =  x \in  X$, then the result is trivial.
      Otherwise, $e =  c(e'_1, \ldots, e'_n)$ and
       $e'_i\sigma_{d} = e_i$, $1 \leq i \leq n$. By induction hypothesis,
       we get $[t_i] \in  \C e'_i \J^{\modterm}[\sigma_{d}]$, $1 \leq i 
       \leq n$\/.
       Hence, $[c(t_1, \ldots, t_n)] \in
   \langle [c(t_1, \ldots, t_n)] \rangle =  
   c^{\modterm}([t_1], \ldots, [t_n])  \in
            \C c(e'_1, \ldots ,e'_n) \J^{\modterm}[\sigma_{d}]$\/.
   
\item For rule \textsf{(OMUT)}, we have that $e\sigma_{d} =  c(e_1, \ldots , 
e_n)$, 
$e_i  \rightarrow_{{\cal P}} s_i$\/, $1 \leq i \leq n$,
$s \rightarrow_{{\cal P}} t$\/, for some
$c(s_1, \ldots , s_n) \ecuacional s  \in
{[{\cal C}]}_{\ecuacional}$.  If $e =  x \in  X$ then  
the result is trivial. Otherwise,
$e  =  c(e'_1, \ldots, e'_n)$  and
       $e'_i\sigma_{d} =  e_i$, $1 \leq i \leq n$\/.
       From induction hypothesis,
       $[s_i] \in \C e'_i \J^{\modterm}[\sigma_{d}]$, $1 \leq i \leq n$. Hence:

\medskip

\( \begin{array}{lll} 
\ \ \ s \rightarrow_{\cal P} t  
 \Rightarrow_{\mbox{ (Definition of  $\sqsubseteq^{\modterm}$)}}
  \end{array} \)
   
 \( \begin{array}{lll} 
\ \ \ {[t]} \sqsubseteq^{\modterm} {[s]} 
 \Rightarrow_{\mbox{ ($c(s_1, \ldots, s_n) \rightarrow_{\cal P} s$)}} \\
  \end{array} \)
  
\( \begin{array}{lll} 
\ \ \ {[t]} \sqsubseteq^{\modterm} {[c(s_1, \ldots, s_n)]} 
 \Rightarrow_{\mbox{ (${[t]} \in \langle {[c(s_1, \ldots, s_n)]}
 \rangle$ and def. of $\modterm$)}} \\
  \end{array} \)
  
\( \begin{array}{lll} 
\ \ \  {[t]} \in c^{\modterm}({[s_1]}, \ldots, {[s_n]}) 
 \Rightarrow_{\mbox{ (${[s_i]} \in \C e'_i 
 \J^{\modterm}{[\sigma_{d}]}$\/, $1 \leq i \leq n$)}} \\
  \end{array} \)
  
\( \begin{array}{lll} 
\ \ \  {[t]} \in \C c(e'_1, \ldots, e'_n) \J^{\modterm}{[\sigma_{d}]}
  \end{array} \)
  
\medskip
 
\item For rule \textsf{(OR)}, $e\sigma_{d} =   f(e'_1, \ldots , e'_n)$\/,
$e'_i  \rightarrow_{{\cal P}}  t_i$, $1 \leq i \leq n$,  
$r \rightarrow_{{\cal P}} t$, and $C$ is $\gpc$-provable, 
for some $f(t_1, \ldots, t_n) \rightarrow r \Leftarrow C \in
{[{\cal R}]}_{\rightarrow}$. It holds necessarily that
$e =  f(e_1, \ldots, e_n)$ and $e_i\sigma_{d} = e'_i$, $1 \leq i \leq 
n$.
From induction hypothesis,
$[t_i] \in \C e_i \J^{\modterm}[\sigma_{d}]$, $1 \leq i \leq n$, that
is $f^{\modterm}([t_1], \ldots, [t_n])
\in \C f(e_1, \ldots, e_n) \J^{\modterm}[\sigma_{d}]$\/.
Hence, from $f(t_1, \ldots, t_n) \rightarrow_{\cal P} r$,
 $r \rightarrow_{\cal P} t$ and
the transitivity of $\gpc$,
it holds that $f(t_1, \ldots, t_n) \rightarrow_{\cal P} t$,
that is  $[t] \in  f^{\modterm}([t_1], \ldots,$ $ [t_n])
\subseteq \C f(e_1, \ldots, e_n) \J^{\modterm}[\sigma_{d}]$.

\end{des}

\medskip

\noindent
\textsf{(b $\Rightarrow$)}. If $(\modterm,[\sigma_{d}]) \modelo e 
\rightarrow t$ then $\C e \J^{\modterm}[\sigma_{d}] \supseteq \C t 
\J^{\modterm}[\sigma_{d}]$. , Now, since $\C t 
\J^{\modterm}[\sigma_{d}] = \langle [t\sigma_{d}] \rangle$, we get in 
particular $[t\sigma_{d}] \in \C e \J^{\modterm}[\sigma_{d}]$. By 
{\bf (a $\Rightarrow$)}, we can conclude that $e\sigma_{d } 
\rightarrow_{{\cal P}} t\sigma_{d}$.

\medskip

\noindent
\textsf{(b $\Leftarrow$)}. If  $e\sigma_{d } 
\rightarrow_{{\cal P}} t\sigma_{d}$, then $[t\sigma_{d}] 
\in \C e \J^{\modterm}[\sigma_{d}]$ follows by {\bf (a $\Leftarrow$)}. 
Now, since $\C t 
\J^{\modterm}[\sigma_{d}] = \langle [t\sigma_{d}] \rangle$, we can 
conclude that $\C e \J^{\modterm}[\sigma_{d}] \supseteq \C t 
\J^{\modterm}[\sigma_{d}]$, i.e., $(\modterm,[\sigma_{d}]) \modelo e 
\rightarrow t$.

\medskip

\noindent
\textsf{(c $\Rightarrow$)}. If $(\modterm,[\sigma_{d}]) \modelo a \lazo b$,
      then there exists $[t] \in \textsf{Def}(\dominio)$ such
      that $[t] \in \C a \J^{\modterm}[\sigma_{d}] \cap
      \C b \J^{\modterm}[\sigma_{d}]$. From \textsf{(a $\Rightarrow$)},
      it holds that $a\sigma_{d} \rightarrow_{\cal P} t$ and
      $b\sigma_{d} \rightarrow_{\cal P} t$. From \textsf{(J)}, it holds
       $a\sigma_{d} \lazo_{\cal P} b\sigma_{d}$.  
       
\medskip

\noindent
\textsf{(c $\Leftarrow$)}. If $a\sigma_{d} \lazo_{\cal P} b\sigma_{d}$,
 then there exists $t \in \terminosprima$ such that
 $a\sigma_{d} \rightarrow_{\cal P} t$,
 $b\sigma_{d} \rightarrow_{\cal P} t$\/. From \textsf{(a $\Leftarrow$)},
 it holds that
 $[t] \in \C a \J^{\modterm}[\sigma_{d}] \cap
          \C b \J^{\modterm}[\sigma_{d}]$, that is,
$({\modterm},[\sigma_{d}]) \modelo a \lazo b$.
\end{proof}

\medskip


\begin{proof}[Proof of Theorem  \ref{modelolibre}]

\noindent
Consider $h=(h_t,h_d)$ defined as  $h_t(\tau)= \C \tau \J^{\algebra} 
\eta_{t}$,  for all $\tau \in T_{\it TC}(A)$ and $h_d([t])=\C t \J^{\algebra} 
\eta_{d}$, for all $t \in \terminospprima$\/. By Theorem 
\ref{adecuacion}, we know that $t \equivalente t'$ implies $\C t 
\J^{\algebra}\eta_{d} = \C t' \J^{\algebra}\eta_{d}$. Therefore, 
$h_{d}$ is well defined. Obviously, $h$ extends $\eta$ by definition. 
Let us prove that $h$ is a homomorphism. 

\smallskip 

We prove firstly that $h_{d}$ is monotonic.
Assume that $[t] \sqsubseteq^{\modterm} [s]$. Then $[s] 
\aproximado [t]$, or equivalently (Theorem \ref{equivalencia}), 
$s \rightarrow_{{\cal P}} t$. From Theorem \ref{adecuacion},
$\C t \J^{\algebra}\eta_{d} \subseteq \C s \J^{\algebra}\eta_{d}$,
i.e. $h_{d}([t]) \subseteq h_{d}([s])$\/.

\smallskip

Items \textsf{(1)},\textsf{(2)}  and \textsf{(3)} of Definition \ref{homomorfismos} 
follow from the definition of $h_{d}$ and Proposition \ref{eval} 
{\sf (b)}. To prove item \textsf{(4)}, consider
$c  \in {\it DC}^n$ and
    $[t_i] \in \dominio$, $1 \leq i \leq n$. Then:

\medskip

\noindent
\( \begin{array}{llll}
\ \ \ \  h_d(c^{\modterm}([t_1], \ldots, [t_n])) 
=_{\mbox{(Definition of $c^{\modterm}$)}} 
\end{array} \)

\noindent
\( \begin{array}{llll}
\ \ \ \  h_d(\langle  [c(t_1, \ldots, t_n)]\rangle ) 
=_{\mbox{(Monotonicity of $h_{d}$)}}
\end{array} \)

\noindent
\( \begin{array}{llll}
\ \ \ \  h_d([c(t_1, \ldots, t_n)])  
=_{\mbox{(Definition of $h_{d}$)}} 
\end{array} \)

\noindent
\( \begin{array}{llll}
\ \ \ \ \C c(t_1, \ldots, t_n) \J^{\algebra} \eta_d 
=_{\mbox{(Definition of denotation)} }
\end{array} \)

\noindent
\( \begin{array}{llll}
\ \ \ \    c^{\algebra}(\C t_{1} \J^{\algebra}\eta_{d}, \ldots,
    \C t_{n} \J^{\algebra}\eta_{d}) 
=_{\mbox{(Definition of $h_{d}$)}} 
\end{array} \)

\noindent
\( \begin{array}{llll}
\ \ \ \ c^{\algebra}(h_d([t_1]), \ldots, h_d([t_n]))
\end{array} \)

\medskip

\noindent
Finally, let us consider item {\sf (5)}. Given $f \in {\it FS}^n$ and  
    $[t_i] \in \dominio$, $1 \leq i \leq n$, we have to prove that.
    
\vspace*{-0.10cm}

\[ h_{d}(f^{\modterm}([t_{1}], \ldots, [t_{n}]) \subseteq 
f^{\algebra}(h_{d}([t_{1}]), \ldots, h_{d}([t_{n}])) \]

\vspace*{-0.10cm}

\noindent
By the definitions of $f^{\modterm}$ and $h_{d}$, this is equivalent 
to the following inclusion:

\vspace*{-0.10cm}

\[ \bigcup  \{\C t \J^{\algebra}\eta_{d} \mid [t] \in \dominio, 
f(t_{1}, \ldots, t_{n}) \rightarrow_{{\cal P}} t \}
\subseteq 
f^{\algebra} (\C t_{1} \J^{\algebra}\eta_{d}, \ldots
\C t_{n} \J^{\algebra}\eta_{d})\]

\vspace*{-0.10cm}

\noindent
Now, for each $t$ such that $f(t_{1}, \ldots, t_{n}) 
\rightarrow_{{\cal P}} t$, Theorem \ref{adecuacion} ensures that 
$(\algebra,\eta_{d}) \modelo f(t_{1}, \ldots, t_{n})$ $ \rightarrow t$, 
which means $\C t \J^{\algebra}\eta_{d} \subseteq f^{\algebra}
(\C t_{1} \J^{\algebra}\eta_{d}, \ldots
\C t_{n} \J^{\algebra}\eta_{d})$. Therefore, the inclusion holds.

\medskip

In order to prove that
$h$ is unique, it is enough to assume that there exists another homomorphism
$h'=(h'_{t},h'_{d})$  extending $\eta$ and to conclude that $h = h'$. 
Firstly, let us prove that for any $\tau \in T_{{\it TC}}(A)$, it 
holds that  $h_{t}(\tau) =h'_{t}(\tau)$. We proceed by structural
induction on $\tau$:

\smallskip

\begin{des}

\item $\tau = \alpha \in A$. Since $h'$ and $h$ extend $\eta$ then 
      we have that $h_{t}(\alpha) = h'_{t}(\alpha) = \eta_{t}(\alpha)$. 
      
\item $\tau = K \in {\it TC}^{_0}$.  Then the result follows 
from the definition of homomorphism.
 
\item $\tau = K(\tau_{1}, \ldots, \tau_{n})$, where $K \in {\it 
TC}^{_n}$, $\tau_{i} \in T_{{\it TC}}(A)$, $1 \leq i \leq n$. Then: 

\medskip  

\( \begin{array}{lllll}
\ \ \ \ h_{t}(K(\tau_{1}, \ldots, \tau_{n})) 
=_{\mbox{ (Definition of $h_{t}$)}} 
\end{array} \)

\( \begin{array}{lllll}
\ \ \ \ \C K(\tau_{1}, \ldots, \tau_{n}) \J^{\algebra}\eta_{t} 
=_{\mbox{ (Definition of denotation)}} 
\end{array} \)

\( \begin{array}{lllll}
\ \ \ \ K^{\algebra}(\C \tau_{1} \J^{\algebra}\eta_{t}, \ldots, \C \tau_{n} 
\J^{\algebra}\eta_{t}) 
=_{\mbox{ (Definition of  $h_{t}$)}}
\end{array} \)

\( \begin{array}{lllll}
\ \ \ \ K^{\algebra}(h_{t}(\tau_{1}), \ldots, h_{t}(\tau_{n})) 
=_{\mbox{ (Induction hypothesis)} }
\end{array} \)

\( \begin{array}{lllll}
\ \ \ \ K^{\algebra}(h'_{t}(\tau_{1}), \ldots, h'_{t}(\tau_{n})) 
=_{\mbox{ ($h'$ is homomorphism)} }
\end{array} \)

\( \begin{array}{lllll}
\ \ \ \ h'_{t}(K^{\modterm}(\tau_{1}, \ldots, \tau_{n})) 
=_{\mbox{ (Definition of $\modterm$)}}
\end{array} \)

\( \begin{array}{lllll}
\ \ \ \ h'_{t}(K(\tau_{1}, \ldots, \tau_{n}))
\end{array} \)

\end{des}  

\smallskip

Let us prove  now that for any $t \in {\it Term}_{\Sigma_{\perp}}(X)$, it 
holds that  $h_{d}([t]) =h'_{d}([t])$. We proceed by structural
induction on $t$:

\medskip 

\begin{des}

\item $t = \perp$. Then, since
 $h_d$ and $h'_d$  are strict then $h'_d([\perp]) = h_d([\perp]) = 
 \langle \perp^{\algebra} \rangle$.

\item $t = x \in  X$. Then, since 
  $h'_d$ and $h_d$  extend $\eta$,  it holds that
  $h'_d([x]) = h_d([x]) = \langle \eta_{d}(x) \rangle$\/.
  
 \item $t \in {\it DC}^{_0}$. Then the result follows from the 
       definition of homomorphism.

\item  $t =c(t_1, \ldots, t_n)$, where $c \in {\it DC}^{_n}$, 
$t_{i} \in {\it Term}_{\Sigma_{\perp}}(X)$, $1 \leq i \leq n$. 
Then:   

\medskip

\( \begin{array}{llll}
\ \ \ \ h'_d([c(t_1, \ldots, t_n)]) 
=_{\mbox{ (Monotonicity of $h'_{d}$)}} 
\end{array} \)

\( \begin{array}{llll}
\ \ \ \ h'_d(\langle [c(t_1, \ldots, t_n)] \rangle) 
=_{\mbox{ (Definition of $\modterm$)}}
\end{array} \)

\( \begin{array}{llll}
\ \ \ \ h'_d(c^{\modterm}([t_1], \ldots, [t_n])) 
=_{\mbox{ (Definition of homomorphism)} }
\end{array} \)

\( \begin{array}{llll}
\ \ \ \ c^{\algebra}(h'_d([t_1]), \ldots, h'_d([t_n])) 
=_{\mbox{ (Induction hypothesis)}} 
\end{array} \)

\( \begin{array}{llll}
\ \ \ \ c^{\algebra}(h_d([t_1]), \ldots, h_d([t_n])) 
=_{\mbox{ (Definition of $h_{d}$)}} 
\end{array} \)

\( \begin{array}{llll}
\ \ \ \ c^{\algebra}(\C t_1 \J^{\algebra}\eta_{d}, \ldots, 
         \C t_n \J^{\algebra}\eta_{d})  
=_{\mbox{ (Definition of denotation)}}
\end{array} \)

\( \begin{array}{llll}
\ \ \ \ \C c(t_1, \ldots, t_n)   \J^{\algebra}\eta_{d} 
=_{\mbox{ (Definition of $h_{d}$)}} 
\end{array} \)

\( \begin{array}{llll}
\ \ \ \ h_{d}([c(t_{1}, \ldots, t_{n})]) 
\end{array} \)

\end{des}

\smallskip

 Now, since $h_{t}=h'_{t}$ and $h_{d}=h'_{d}$ it holds that $h=h'$.
It remains to prove the second part of
the theorem. For that, assume that ${\algebra}$
is well-typed. Consider $[t] \in \dominio$ and
$\tau \in T_{{\it TC}}(A)$ such that
$t \in {\it Term}_{\Sigma_{\perp}}^{\tau}(V)$\/.
From Proposition \ref{compatibles}, it holds that
$\C t \J^{\algebra} \eta_{d}
\subseteq {\cal E}^{\algebra}(\C \tau \J^{\algebra}\eta_{t})$,
that is, $h_d([t]) \subseteq {\cal E}^{\algebra}(h_t(\tau))$\/.
\end{proof}

\medskip

    
\begin{proof}[Proof of Proposition \ref{linearizacion}]

Since $c(\ttupla_{n}) \approx d(\stupla_{m})$ is
well typed,   there exist an environment $V'$ and type
variants $c:(\tau_{1}, \ldots, \tau_{n}) \rightarrow \tau$ and
$d:(\tau'_{1}, \ldots, \tau'_m) \rightarrow \tau$
of the type declarations for $c$ and $d$ respectively, such that
$c(\ttupla_{n}),d(\stupla_{m}) \in {\it Term}_{\Sigma}^{\tau}(V')$\/.

\smallskip

Assume that $x_1, \ldots, x_k$ are all the variables occurring
$p_i>1$ times in $c(\ttupla_{n})$, $1 \leq i \leq k$\/.
Suppose that the ``linearization'' process has replaced each $j$-th occurrence
of $x_i$ by a fresh variable $y_{ij}$\/, $2 \leq j \leq p_i$,
$1 \leq i \leq k$\/. Consider the new environment:
$V =  V' \cup \{y_{ij}:\tau^{*}_i \ \mid \ x_i:\tau^{*}_{i} \in V'\}$.
It holds trivially that $c(t'_1, \ldots, t'_n),d(s_1,\ldots, s_m)
\in {\it Term}_{\Sigma}^{\tau}(V)$ and that $C_{1}$ is well-typed. 
 \end{proof}

\medskip


\begin{proof}[Proof of Correctness Lemma \ref{lemadecorreccion}]

\noindent
${\bf (Invariance_{1})}$ By analyzing all $\reduce$-rules and
proving for each one of them that $G'$ (resulting by applying the 
corresponding $\reduce$-rule to $G$) verifies the conditions {\sf 
(LIN)}, {\sf (EX)}, {\sf (NCYC)} and {\sf (SOL)} in 
Definition \ref{objetivoadmisible}.  We will only give
succinct explanations justifying the preservation of admissibility for
the transformation rules for $\lazo$. Similar reasonings can be used to 
prove that  the transformation rules for $\rightarrow$ 
preserve also admissibility.  

\medskip

\noindent
${\sf Decomposition}_{\lazo}$: ${\sf pvar}(P)$, $\tupla$ and $S$  do not 
change and $\gg$ becomes finer. Then $G'$ is admissible.

\smallskip

\noindent
${\sf Mutation}_{\lazo}$: Condition {\sf (LIN)} holds since $\ttupla_{n}$ is linear and 
with fresh variables.  Since all variables ($\bar{x}$) introduced by ${\it Eq}$ are 
existentially quantified, then condition {\sf (EX)} holds. Variables in each 
$t_{i}$, $1 \leq i \leq n$\/, are fresh (hence not appearing in any left-hand side of 
approximation conditions), so no cycle of produced variables can be 
created, i.e. condition {\sf (NCYC)} holds. Finally, condition {\sf (SOL)} 
is trivially satisfied by $G'$,  
since $S$ does not change and all variables in $t_{i}$ are fresh. 

\smallskip

\noindent
${\sf Imitation+Decomposition_{\lazo}}$: If $x \not \in {\sf pvar}(P)$ then 
the application of $ [x/c(\xtupla_{n})]$ 
does not modify the right-hand sides of the approximation statements 
in $P$, i.e. ${\sf pvar}(P)$ does not change. Hence conditions 
{\sf (LIN)} and {\sf (EX)} are verified by $G'$. On the other hand,  
since
$x \not \in {\sf pvar}(P)$ and $\xtupla_{n}$ are fresh variables, then 
condition {\sf (SOL)} holds.  Since no 
produced variables are introduced in the left-hand sides of 
conditions in $P$\/, condition {\sf (NCYC)} is true for $G'$.

If  $x \in {\sf pvar}(P)$\/, condition {\sf (LIN)} holds since 
$\xtupla_{n}$ are fresh variables and $x$ occurs only once as 
produced variable. Similarly, since $\xtupla_{n}$ is existentially 
quantified in $G'$, then condition {\sf (EX)} holds.  Since the 
substitution $[x/c(\xtupla_{n})]$ does not affect to $S$, then condition {\sf (SOL)} is 
verified by $G'$\/. Since $G$ verifies {\sf (NCYC)}, then any cycle in $G'$
must have the form $\ldots \gg x_{i} \gg \ldots$, for some $1 \leq i 
\leq n$\/. But such a cycle can be reproduced in $G$ by replacing 
each $x_{i}$ by variable $x$\/. Hence $G'$ must verify condition {\sf 
(NCYC)}. 

\smallskip

\noindent
${\sf Imitation+Mutation_{\lazo}}$: $\xtupla_{m}$ and 
all variables introduced by ${\it Eq}$ are new and existentially 
quantified. Hence conditions {\sf (LIN)} and {\sf (EX)} hold.  
If $x \not \in {\sf pvar}(P)$ then {\sf (SOL)} holds. Furthermore, the 
right-hand sides of approximation statements are not affected by  the 
substitution $[x/d(\xtupla_{m})]$ and all variables introduced in $G'$ are fresh. Hence, 
$G'$ does not contain cycles (i.e. condition {\sf (NCYC)} holds).

If $x \in {\sf pvar}(P)$, $S$ does not change. Furthermore, since all 
variables introduced in $G'$ are fresh, then condition {\sf (SOL)} is 
true in $G'$\/.  On the other hand, note that variables introduced 
by {\it Eq} do not occur in the left-hand sides of approximation 
statements, i.e. such variables can not generate cycles. Since $G$ 
has no cycles of variables, then a cycle in $G'$  must have the form 
$\ldots \gg x_{i} \gg \ldots$, for some $1 \leq i 
\leq m$\/. But such a cycle can be reproduced in $G$ by replacing 
each $x_{i}$ by variable $x$\/. Hence $G'$ must verify condition {\sf 
(NCYC)}.

\smallskip

\noindent
${\sf Narrowing_{\lazo}}$: Similar to the case ${\sf 
Mutation}_{\lazo}$.

\medskip

\noindent
${\bf (Invariance_{2})}$  We proceed as in ${\bf (Invariance)_{1}}$, 
analyzing all  variable elimination rules.

\medskip

\noindent
{\sf Produced variable elimination:} It holds that ${\sf 
pvar}(G')={\sf pvar}(G)-\{y\}$\/. Hence condition {\sf (LIN)} holds.
Furthermore, $S$ does not  change. Hence condition {\sf (SOL)} is 
verified by $G'$. Since ${\sf evar}(G')={\sf 
evar}(G)-\{y\}$ but $y$ does not occur in $G'$, then condition {\sf 
(EX)}  holds. Finally note that variable $x$ introduced by the 
substitution
$[y/x]$ can not produce a cycle because otherwise, variable $y$ 
would produce a cycle in $G$. Hence {\sf (NCYC)} holds.

\smallskip

\noindent
{\sf Identity:} Trivial.

\smallskip

\noindent
{\sf Non-produced variable elimination:} Since $x \not \in {\sf 
pvar}(P)$ then the propagation $x/y$ does not affect to the right-hand sides 
of $P$\/, i.e. the set of produced variables does not change when 
applying the transformation rule. Hence {\sf (LIN)} holds. $\tupla$ 
does not change and $x,y \not \in {\sf 
pvar}(P)$\/, then {\sf (EX)} and {\sf (SOL)} hold. Finally, since 
$y$  is not a produced variable, then no cycles are produced and {\sf (NCYC)} is verified 
by $G'$\/.

\smallskip

It remains to prove that all approximation statements and joinability 
conditions in $G'$ only contain variables, but this is trivial from 
the   definition of variable elimination rules. 
 
\medskip

\noindent
${\bf (Correctness_{1})}$ We proceed by considering the failure rules, one by one. 
For rules ${\sf Conflict}_{\diamondsuit}$, where $\diamondsuit \in 
\{\lazo, \rightarrow\}$,  the correctness holds straightforwardly
since for any $\sigma_{d} \in {\it DSub}_{\perp}$, the statements
  $c(\etupla_n)\sigma_{d} \lazo d(\eprimatupla_m)\sigma_{d}$ or
  $c(\etupla_n)\sigma_{d} \rightarrow d(\eprimatupla_m)\sigma_{d}$ are not
  $\gpc$-provable. 
  
\smallskip

For rule {\sf Cycle} let us assume that 
  $\sigma_d \in {\sf Sol}(G)$\/.  Then, there exist $\gpc$-proofs for
$x_{n-1}\sigma_d \igual e_{n}[x_{0}]\sigma_d$, $x_{n-2} \sigma_d 
\igual e_{n-1}[x_{n-1}]\sigma_d$, $\ldots$, 
$x_{1} 
\sigma_d \igual e_{2}[x_{2}]\sigma_d$ and $x_{0}\sigma_d 
\igual e_{1}[x_{1}] \sigma_d$.  This sequence of $\gpc$-proofs 
implies that $ 
x_{0}\sigma_d \igual e_{1}[e_{2}$ $[\ldots 
[e_{n-1}[e_{n}[x_{0}]]]]]\sigma_d$ is $\gpc$-provable. Then there exists 
$t \in 
\terminos$  such that $x_{0}\sigma_d \rightarrow t$ and 
$e_{1}[e_{2}[\ldots [e_{n-1}[e_{n}[x_{0}]]]]]\sigma_d 
\rightarrow t$ are both $\gpc$-provable. 
Since $x_{0}$  is a safe variable in
$e_{1}[e_{2}[\ldots [e_{n-1}[e_{n}[x_{0}]]]]]$,  then $x_{0}\sigma_d$ 
 must be a strict subterm of $t$ in some position whose ancestor 
 positions are all occupied by free constructors.  This contradicts 
 the fact that $x_{0}\sigma_d \rightarrow t$ is 
 $\gpc$-provable.

\medskip

\noindent
${\bf (Correctness_{2})}$ The proof proceeds again by inspecting all 
transformation rules for $\calculo$ except for failure rules
and
  checking one by one that $\sigma_{d}$ is a solution for $G$. The 
  whole proof is too large  and does not reveal interesting ideas. 
  Therefore, we will only analyze  
  those rules for $\lazo$ referring to the application of 
  equational axioms in $\cal C$\/. In the rest of the proof, the 
  notation $e \rightarrow_{{\cal P}} t$ (respect. $e \lazo_{{\cal P}} 
  e'$) indicate that $e \rightarrow t$ (respect. $e \lazo e'$) is 
  $\gpc$-provable.
  
  \medskip
  
\noindent
${\sf Mutation}_{\lazo}$:  Consider $\sigma_{d}$ defined as $\sigma_{d}(x)=x$ 
for all $x \in {\it dvar(Eq)}$ and $\sigma_{d}(x)=\sigma'_{d}(x)$ 
otherwise. All items of 
     Definition \ref{solucion} hold trivially, except for
     item {\sf (GORC)}. In order to prove {\sf (GORC)}, it is enough 
     to find a $\gpc$-proof for $c(\etupla_{n})\sigma_{d} \lazo 
     e'\sigma_{d}$ (the rest of 
     approximation/joinability statements in $G$ do not change).
     But, since $\sigma_{d}'$ is a solution for $G'$,  we know:  
     
     \begin{itemize}
     \item[(*)] $e_i\sigma_{d} \rightarrow_{{\cal P}} t_i\sigma'_{d}$,
                 $s\sigma'_{d} \lazo_{{\cal P}} e' \sigma_{d}$ and
           $C\sigma'_{d} $ are $\gpc$-provable.
           
     \end{itemize}
     
     \noindent
     Then:
       
       \smallskip  

      \begin{des}
      \item[(1)] $s\sigma'_{d}  \rightarrow_{{\cal P}} m$, 
                 $e' \sigma_{d} \rightarrow_{{\cal P}} m$, 
                 for some $m \in \terminos$\/;

     \item[(2)] $c(\ttupla_n)\sigma'_{d}  \rightarrow_{{\cal P}} 
     s\sigma'_{d} $, since
                ${\it Eq}\sigma'_{d}  \in [{\cal C}]_{\rightarrow}$ 
                and $C\sigma'_{d}$ is $\gpc$-derivable\/;

     \item[(3)] From (1), (2) and the transitivity of $\gpc$
           (Theorem \ref{equivalencia} {\sf (a)} ensures that
            $\gpc$ and $\bpc$ are equivalent) we get
           $c(\ttupla_{n})\sigma'_{d} \rightarrow_{{\cal P}} m$;
           
      \item[(4)] From (*) and the $\gpc$-rule {\sf (DC)} we have  
            $c(\etupla_n) \sigma_{d} \rightarrow_{{\cal P}} 
            c(\ttupla_n)\sigma'_{d} $;

      \item[(5)] From (3),(4) and the  transitivity of
           $\gpc$ we get $c(\etupla_n) \sigma_{d}
                   \rightarrow_{{\cal P}} m$;
      \item[(6)] From (1),(5) and the $\gpc$-rule {\sf (J)} we can
         build a $\gpc$-proof for
         $c(\etupla_n)\sigma_{d}$ $ \lazo e'\sigma_{d}$\/.

      \end{des}

\smallskip

\noindent     
${\sf Imitation+Mutation}_{\lazo}$: 

\smallskip

If $x \not \in {\sf pvar}(P)$, then consider $\sigma_{d}$ defined
as $\sigma_{d}(z)=z$ for all $z \in {\it dvar(Eq)}$ and
          $\sigma_{d}(z)=\sigma_{d}'(z)$ otherwise.  Conditions
           {\sf (TOT)} and
          {\sf (EQ)} from Definition \ref{solucion} hold from 
          the proper definition of $\sigma_{d}$\/.
          For proving {\sf (GORC)} it is enough to find
          a $\gpc$-proof for $x\sigma_{d} \lazo c(\etupla_n)\sigma_{d}$\/.
          The rest of approximation/joinability statements
          are $\gpc$-provable since $\sigma_{d}(x)=\sigma'_{d}(x)=
          d(\xtupla_m)\sigma'_{d}$\/.
          So, let us find a proof for
          $x\sigma_{d} \lazo c(\etupla_n)\sigma_{d}$\/:
          
          \smallskip
          
          \begin{des}
          \item[(1)] $\sigma_{d}(x_i) \rightarrow t_i\sigma_{d}'$ are
              $\gpc$-provable, then using the $\gpc$-rule {\sf (DC)} we can
              derive $d(\xtupla_m)\sigma_{d}$ $ \rightarrow
                       d(\ttupla_m)\sigma_{d}'$\/.
         \item[(2)] $C\sigma_{d}'$ is $\gpc$-derivable. Then
                  using {\sf (OMUT)} we have also
                  that $d(\ttupla_m)\sigma'_{d} \rightarrow_{{\cal P}} s\sigma_{d}'$\/.
         \item[(3)] $s\sigma_{d}' \lazo c(\etupla_n)\sigma_{d}$ is $\gpc$-derivable,
                   then there exists $m \in \terminos$ such
                   that $s\sigma_{d}' \rightarrow m$ and
                   $c(\etupla_n) \sigma_{d} \rightarrow m$ are both
                   $\gpc$-provable.
         \item[(4)] From (1),(2),(3) and the transitivity of
         $\gpc$\/, it holds that $d(\xtupla_m)\sigma_{d}
                \rightarrow  m$ is $\gpc$-provable. But
                $d(\xtupla_m)\sigma_{d} = \sigma_{d}(x)$, so
                $\sigma_{d}(x) \rightarrow m$ is $\gpc$-provable.
        \item[(5)] From (3), (4) and {\sf (J)} we get finally
                that $\sigma_{d}(x) \lazo c(\etupla_n)\sigma_{d}$ is
                $\gpc$-provable, and thus {\sf (GORC)} holds. 

          \end{des}
          
\smallskip

If $x \in {\sf pvar}(P)$, then it is enough to define 
$\sigma_{d}$ as $\sigma_{d}(z)=z$, for all $z \in {\it 
dvar(Eq)} \cup \xtupla_{m}$, 
$\sigma_{d}(x)=d(\xtupla_{m})\sigma'_{d}$ and 
$\sigma_{d}(z)=\sigma'_{d}(z)$ otherwise, and 
reasoning as done above.  
\end{proof}

\medskip


\begin{proof}[Proof of  Progress Lemma \ref{progreso}]

We analyze all the possible forms of a goal $G \equiv \exists 
\tupla \cdot S \Box P \Box E$ with the properties stated in the lemma.
In order to avoid tedious repetitions, we will  treat in detail only those cases which 
justify the presence of the $\reduce$-rules 
$\textsf{Imitation}_{\rightarrow}$, 
$\textsf{Imitation+Decomposition}_{\rightarrow}$ and 
$\textsf{Imitation+Mutation}_{\rightarrow}$.  
For the rest of the cases,  
we will only mention the $\reduce$-transformation rule $T$
decreasing  ${\cal M'}$\/.

\smallskip

We assume that in each
of the cases below, $G'$ is the goal resulting of applying
$T$ to $G$ and $\sigma_{d}'=  \sigma_{d}$ unless otherwise stated.
As notation $\Pi: \varphi$ indicates that $\Pi$ is a
$\gpc$-proof for $\varphi$ whereas
$(\Pi_1, \ldots, \Pi_n)+R$ stands for the  $\gpc$-proof
composed of $\Pi_1$ followed by $\Pi_2$ $\ldots$
followed by $\Pi_n$ followed by one application
of the $\gpc$-rule $R$\/.

\smallskip

We begin by analyzing the possible forms of joinability conditions in 
$E$\/.

\medskip

\noindent
$\bullet$ $G \equiv  \exists \tupla \cdot S \Box P \Box c(\etupla_{n}) \lazo 
c(\eprimatupla_{n}),E$\/, where $c$ is a free data constructor. 
Then it is enough to choose $T=\textsf{Decomposition}_{\lazo}$\/;

\medskip

\noindent
$\bullet$ $G \equiv  \exists \tupla \cdot S \Box P \Box c(\etupla_{n}) \lazo 
d(\eprimatupla_{m}),E$\/, where $c,d$  are algebraic data constructor. 
Then $T$ must be either $\textsf{Decomposition}_{\lazo}$ or 
$\textsf{Mutation}_{\lazo}$, 
according to the witness ${\cal M}$\/;

\medskip

\noindent
$\bullet$ $G \equiv  \exists \tupla \cdot  S \Box P \Box x \lazo c(\etupla_{n}),E$\/, 
where $c$ is  a free data constructor. Then the transformation rule 
decreasing the witness if \textsf{Imitation+}${\sf De}{\sf com}{\sf po}{\sf si}
{\sf ti\-on}_{\lazo}$;  

\medskip

\noindent
$\bullet$ $G \equiv \exists \tupla \cdot S \Box P \Box x \lazo c(\etupla_{n}),E$\/, 
where $c$ is an algebraic data constructor. 
Then $T$ must be either the rule $\textsf{Imitation+Decomposition}_{\lazo}$ or  
the rule {\small $\textsf{Imitation+Mutation}_{\lazo}$} or the rule $\textsf{Mutation}_{\lazo}$, 
according to the witness ${\cal M}$\/;

\medskip

\noindent
$\bullet$ $G \equiv  \exists \tupla  \cdot S \Box P \Box f(\etupla_{n}) \lazo e',E$\/, 
then $T=\textsf{Narrowing}_{\lazo}$.

\medskip

\noindent
If all joinability conditions in $G$ are different from the analyzed previous 
cases, then:

\[ (1) \ \mbox{ All $e \lazo e' \in E$ are of the form $x \lazo y$\/, 
where $x,y \in {\it DVar}$}\]

We continue now analyzing all the possible forms of approximation 
statements in $P$\/.  

\medskip

\noindent
$\bullet$ $G \equiv \exists \tupla \cdot S \Box c(\etupla_{n}) \rightarrow 
c(\ttupla_{n}),P\Box E$\/, where $c$ is a free data constructor\/. 
Then $T=\textsf{Decomposition}_{\rightarrow}$;

\medskip

\noindent
$\bullet$ $G \equiv  \exists \tupla \cdot S \Box c(\etupla_{n}) \rightarrow 
d(\ttupla_{m}),P \Box E$\/, $c,d$ are algebraic data constructors.
In this case, ${\cal M}$ contains a $\gpc$-proof $\Pi_0$ for
$c(\etupla_{n}) \sigma_{d} \rightarrow d(\ttupla_{m})\sigma_{d}$.
Let us analyze all possible forms of $\Pi_0$:

\smallskip

\begin{des}

\item Assume that $\Pi_0 =  
( \ldots, \Pi_i : e_i\sigma_{d} \rightarrow
             t_i \sigma_{d}, \ldots ) + {\sf (DC)}$,
where $c =  d$ and $n =  m$. Let us take $T=  
\textsf{Decomposition}_{\rightarrow}$\/.
If we take
${\cal M'}=  ({\cal M}-\mi \Pi_0 \md) \cup \mi \Pi_1,
   \ldots, \Pi_n \md$, it holds that ${\cal M'} \triangulo {\cal M}$.
   Trivially, it holds that for all $V \in {\sf env}(G)$ we have that 
   $V \in {\sf env}(G')$. 
   
\item  Assume that $\Pi_0 =  ( \ldots , \Pi_i :
               e_i\sigma_{d} \rightarrow s_i, \ldots,
                \Pi^c : C, \Pi^r : s \rightarrow
                d(t_1, \ldots, t_m)\sigma_{d}) + {\sf (OMUT)}$\/, 
where ${\it Eq}: c(s_1, \ldots , s_n) \rightarrow s \Leftarrow C \in
    [{\cal C}]_{\rightarrow}$. There exists a variant ${\it Eq'}:
    c(s'_1, \ldots, s'_n) \rightarrow
    s' \Leftarrow C'$ of a rule in ${\cal C}_{\rightarrow}$
    such that ${\it dvar(Eq')} \cap {\it dvar}(G)=  \emptyset$ and
    ${\it Eq} =   {\it Eq'} \sigma_{d_{0}}$, for some
    $\sigma_{d_{0}} \in {\it DSub}_{\perp}$. Let us take 
    $T=  \textsf{Mutation}_{\rightarrow}$ with ${\it Eq'}$ and  $\sigma_{d}'$
    defined as $\sigma_{d}'(x)=  \sigma_{d_{0}}(x)$ if 
    $x \in {\it dvar(Eq')}$,
    $\sigma_{d}'(x)=  \sigma_{d}(x)$ otherwise.  
    Note that
    $e_i \sigma_{d}' \rightarrow s'_i \sigma_{d}' = 
    e_i \sigma_{d} \rightarrow s_i$. Analogously
     $C' \sigma_{d}' =   C$ and {\small $s' \sigma_{d}' \rightarrow
     d(t_1, \ldots, t_m) \sigma_{d}'$} $= 
    s \rightarrow d(t_1, \ldots, t_m) \sigma_{d}$.
    Hence ${\cal M'} =   ({\cal M}-\mi \Pi_0 \md) \cup
              \mi \Pi_1, \ldots, \Pi_n, \Pi^c, \Pi^r  \md$
  verifies that ${\cal M'} \triangulo {\cal M}$. 
    
    \smallskip

    \indent
    Let us prove now  the second part of the lemma.
    From Proposition \ref{linearizacion}, there exists an
    environment $V^{*}$  such that $c(\sprimatupla_{n}),
    s' \in {\it Term}_{{\Sigma_{\perp}}}^{\tau}(V^{*})$ and
    $C'$ is well-typed in $V^{*}$,
    where $c: (\tau_{1}, \ldots, \tau_{n}) \rightarrow \tau$
    is a variant of the type declaration associated to $c$. Let $V$ be 
    an environment such that $V \in {\sf env}(G)$. Then, it holds that
     $c(\etupla_{n}),d(\ttupla_{m})
    \in {\it Term}_{{\Sigma_{\perp}}}^{\tau^{*}}(V)$, for
    some $\tau^{_*} \in \tipos$. Furthermore, there must be some
    $\sigma_{t} \in {\it TSub}$ such that $e_{i} \in 
    {\it Term}_{{\Sigma_{\perp}}}^{\tau_{i}\sigma_{t}}(V)$\/,
     $1 \leq i \leq n$\/, and $\tau^{*} =  \tau \sigma_{t}$\/.
    On the other hand, since all variables in ${\it Eq'}$ are new, 
    then we   can choose $V^{*}$ in such a way  that ${\it dvar}(V) \cap
    {\it dvar}(V^{*}) =   \emptyset$. 
    Considering the new environment 
    $V' =  V \cup V^{*}\sigma_{t}$ and Lemma \ref{lema2}, it holds that $C'$ is well-typed
     in such an 
    environment, $e_{i},s'_{i}  \in {\it 
    Term}_{{\Sigma_{\perp}}}^{\tau_{i}\sigma_{t}}(V')$, $1 \leq i \leq n$,
    $s', d(\ttupla_{m}) \in {\it 
    Term}_{{\Sigma_{\perp}}}^{\tau\sigma_{t}}(V')$. Hence $V' \in {\sf 
    env}(G')$.

\end{des}

\medskip

\noindent
$\bullet$ $G \equiv   \exists \tupla \cdot S \Box x \rightarrow 
c(\ttupla_{n}) , P\Box E$\/.
In this case,  ${\cal M}$ contains a $\gpc$-proof $\Pi_0$ for
$x\sigma_{d} \rightarrow c(\ttupla_{n})\sigma_{d}$. The possible forms of 
$\Pi_0$ are:

\smallskip

\begin{des}

\item $\Pi_0 =  ( \ldots, \Pi_i : s_i \rightarrow
             t_i \sigma_{d}, \ldots ) + {\sf (DC)}$,
  where $\sigma_{d}(x)=  c(\stupla_{n})$. Let us choose the 
  transformation rule
  $T=  \textsf{Imitation+}{\sf De\-com\-po\-si\-tion}_{\rightarrow}$. Consider 
  $\sigma_{d}'$ defined as $\sigma_{d}'(x_i)=  s_i$ and
   $\sigma_{d}'(x)=  \sigma_{d}(x)$ otherwise. It holds that
   $\sigma_{d}'(x) =   \sigma_{d}(x)=  c(s_1, \ldots, s_n)
   = c(x_1, \ldots, x_n)\sigma_{d}'$. On the other hand,
   $\sigma_{d}'(x_i) \rightarrow t_i\sigma_{d}' =
   s_i \rightarrow t_i \sigma_{d}$. Furthermore, for all
   $e \in \expresiones$ ($x_i \not \in {\it dvar}(e)$)
   we have that $e[x/c(x_1, \ldots,$ $ x_n)]\sigma_{d}'=
   e\sigma_{d}$. Now, the witness verifying the lemma is
   ${\cal M'}=({\cal M}-\mi \Pi_0 \md) \cup \mi \Pi_1,
   \ldots, \Pi_n \md$\/.

  \smallskip
  
  \indent
  The second part of the lemma proceeds as follows: 
  Assume that $c:(\tau_{1}, \ldots, \tau_{n}) \rightarrow \tau \in 
  {\it DC}$.  Consider an environment $V \in {\sf env}(G)$. Then, it 
  holds that   
  $x$ and $c(\ttupla_{n})$ have a common type in $V$\/, i.e.
  there exists $\sigma_{t} \in {\it TSub}$ such that
  $x:\tau\sigma_{t} \in V$ and $c(\ttupla_{n}) \in {\it 
    Term}_{{\Sigma_{\perp}}}^{\tau\sigma_{t}}(V)$ (i.e.
    $t_{i} \in {\it Term}_{{\Sigma_{\perp}}}^{\tau_{i}\sigma_{t}}(V)$, $1 \leq i \leq n$).
     Let us take $V' =  V [x_{1}:\tau_{1}\sigma_{t}, \ldots, 
     x_{n}:\tau_{n}\sigma_{t}]$.  Noting that  
     $x$ and $c(\xtupla_{n})$ have type 
   $\tau\sigma_{t}$ in $V'$, it is straightforward to check  that $V' 
   \in {\sf env}(G')$.

\item $\Pi_0 =  ( \ldots , \Pi_i :
               s_i \rightarrow l_i, \ldots,
                \Pi^c : C, \Pi^r : s \rightarrow
                c(t_1, \ldots, t_n)\sigma_{d}) + {\sf (OMUT)}$,
         where $\sigma_{d}(x)=d(s_1, \ldots, s_m)$ and ${\it Eq}:
    d(l_1, \ldots , l_m) \rightarrow s \Leftarrow C \in
    [{\cal C}]_{\rightarrow}$. We can find a variant ${\it Eq'}:
    d(l'_1, \ldots, l'_m) \rightarrow
    s' \Leftarrow C'$ of a rule in ${\cal C}_{\rightarrow}$
    such that ${\it dvar(Eq')} \cap ({\it dvar}(G) \cup
    \{x_1, \ldots , x_m\})=\emptyset$ and
    ${\it Eq} = {\it Eq'} \sigma_{d_{0}}$, for some
    $\sigma_{d_{0}} \in {\it DSub}_{\perp}$.
Let us consider $T=\textsf{Imitation+Mutation}_{\rightarrow}$ with ${\it Eq'}$.
 Consider 
     $\sigma_{d}'$ defined as 
     $\sigma_{d}'(x_i) = s_i$, $1 \leq i \leq m$,
    $\sigma_{d}'(x) = \sigma_{d_{0}}(x)$, if $x \in {\it dvar(Eq')}$,
    $\sigma_{d}'(x) = \sigma_{d}(x)$, otherwise.
It is holds that $\sigma_{d}'(x)=\sigma_{d}(x)=d(s_1,\ldots, s_m)
    = d(x_1,\ldots,x_m)\sigma_{d}'$, $\sigma_{d}'(x_i) \rightarrow l'_i\sigma_{d}' =
    s_i \rightarrow l_i$, $C'\sigma_{d}'=C$ and
    $s'\sigma_{d}' \rightarrow c(t_1,\ldots,t_n)\sigma_{d}'=
     s \rightarrow c(t_1, \ldots, t_n)\sigma_{d}$.
    Furthermore, for all $e \in \expresiones$ ($x_i \not \in {\it dvar}(e)$,
    ${\it dvar}(e) \cap {\it dvar(Eq')}= \emptyset$) it holds that
     $e[x/d(x_1, \ldots, x_m)]\sigma_{d}' = e\sigma_{d}$.
    The witness ${\cal M'} = ({\cal M}-\mi \Pi_0 \md) \cup
              \mi \Pi_1, \ldots, \Pi_m, \Pi^c, \Pi^r \md$
 verifies that ${\cal M'} \triangulo {\cal M'}$\/.

\smallskip

\indent
From Proposition \ref{linearizacion},
    there exist an environment $V^*$ and
     a type variant $d: (\tau'_1, \ldots, \tau'_m)
    \rightarrow \tau$ of the principal type of $d$ such that
    $l'_i \in  {\it Term}_{\Sigma}^{\tau'_i}(V^*)$\/, $1 \leq i \leq 
    m$\/,  
    $s' \in {\it Term}_{\Sigma}^{\tau}(V^*)$ and $C'$ is well-typed 
    w.r.t. $V$.
    Since $c$ and $d$ are constructors of the same datatype,
   then there exists a type variant of the principal type of $c$
   of the form $c: (\tau_1, \ldots, \tau_n)
   \rightarrow \tau$\/.  Consider $V \in {\sf env}(G)$. Then it holds 
   that $x:\tau^* \in V$
   and $c(\ttupla_n) \in {\it Expr}_{\Sigma}^{\tau^*}(V)$, for some 
   $\tau^{_*} \in \tipos$\/. We can
   find $\sigma_{t} \in {\it TSub}$ such that $\tau^*=\tau\sigma_{t}$\/.
   From transparency, we have that
   $t_i \in {\it Term}_{\Sigma}^{\tau_i\sigma_{t}}(V)$\/. Furthermore,
   we can choose $\xtupla_m$ and $V^*$ such that
   ${\it dvar}(V) \cap \xtupla_m = \emptyset$,
   ${\it dvar}(V) \cap {\it dvar}(V^*) = \emptyset$ and
   ${\it dvar}(V^*) \cap \xtupla_m = \emptyset$\/. Let us consider
   the environment $V' =  V \cup \{x_i:\tau'_i\sigma_{t}\ \mid \ 1 \leq i \leq m\}
               \cup V^*\sigma_{t}$. It holds that $V' \in {\sf 
               env}(G')$ .
 \end{des} 
 
  \medskip
 
 \noindent
$\bullet$ $G \equiv \exists \tupla \cdot S \Box f(\etupla_{n}) \rightarrow c(\ttupla_{m}),P 
 \Box E$\/. Then $T=\textsf{Narrowing}_{\rightarrow}$.
 
 \medskip

 \noindent
 If all approximation statements in $P$ are different from the 
 previous analyzed cases then:
 
 \[ (2) \ \mbox{ All $e \rightarrow t \in P$ are of the form $e \rightarrow 
 x$} \]

 Moreover, if $P$ is empty then $G$ is quasi-solved due to (1). Otherwise,
 since $G$ is not quasi-solved,  there exists $e \rightarrow 
 x \in P$ such that $e \not \in {\it DVar}$. We choose any $e \rightarrow 
 x \in P$ such that $e \not \in {\it DVar}$ and do the following 
 process:
 
 \medskip
 
 \begin{des}
 
 \item If $x$ does not occur elsewhere in $G$, we stop;
 \item If there is some $u \lazo v \in E$ such that $x = u$ or $x = 
 v$, then we stop;
 \item Otherwise, there is some $e' \rightarrow x' \in P$ with $x 
 \in {\it dvar}(e')$. Then, we  repeat the process with $e' \rightarrow x'$.
 
\end{des}

\medskip

Since $\gg$ is irreflexive, the process  above ends after $m+1$ steps, 
generating the 
following sequence of approximation statements: $e_{0} \rightarrow 
x_{0}, e_{1} \rightarrow x_{1}, \ldots, e_{m} \rightarrow x_{m}$, 
where $e_{0} = e$, $x_{0}=x$ and $x_{i} \in {\it dvar}(e_{i+1})$, $0 
\leq i \leq m-1$. Let us chose the biggest $i$, $0 \leq i \leq m$, 
such that $e_{i} \not \in {\it DVar}$ and $e_{k}$, $i < k \leq m$ is 
a variable (it exists because $e_{0} \not 
\in {\it DVar}$).  Then, if the process above finished because $x_{m}$ 
does not occur elsewhere in $G$, then we can apply the transformation 
rule ${\sf Elimination}_{\rightarrow}$ to $e_{m} \rightarrow x_{m}$. 
Otherwise, the process has finished because $x_{m} \lazo z$ or $z 
\lazo x_{m}$ occurs in $E$. But in such a case $x_{i}$ is a demanded 
variable and either ${\sf Imitation}_{\rightarrow}$ or ${\sf 
Narrowing}_{\rightarrow}$ can be applied to $e_{i} \rightarrow x_{i}$, 
depending on the structure of $e_{i}$. 
\end{proof}

\end{document}